\documentclass[11pt, draftclsnofoot, onecolumn]{IEEEtran}
\IEEEoverridecommandlockouts
\usepackage{amsmath,amsfonts}
\usepackage{acronym}
\usepackage{amsmath}
\usepackage{amssymb}
\usepackage{graphicx}
\usepackage[margin=10pt,font=small,labelfont=bf,textfont=it,justification=centerlast]{caption}
\usepackage{subcaption}
\usepackage{verbatim}
\usepackage{cite}
\usepackage{xcolor}
\usepackage{colortbl}

\definecolor{Red}{HTML}{FF0000}
\definecolor{CornflowerBlue}{RGB}{100,149,237}
\definecolor{GreenYellow}{RGB}{173,255,47}
\newcolumntype{C}[1]{>{\centering\arraybackslash}m{#1}}

\usepackage[normalem]{ulem}

\acrodef{$P_{EM}$}{probability of emulation, or false alarm}
\acrodef{$P_{FA}$}{probability of false alarm}
\acrodef{$P_{MD}$}{probability of missed detection}
\acrodef{$P_{D}$}{probability of detection}

\acrodef{ACF}{autocorrelation function}
\acrodef{ACG}{automatic	gain control}
\acrodef{ACI}{adjacent channel interference}
\acrodef{ACK}{acknowledge}
\acrodef{AcR}{autocorrelation receiver}
\acrodef{ADC}{analog-to-digital converter}
\acrodef{AF}{amplify \& forward}
\acrodef{AFL}{anchor-free localization}
\acrodef{AGNSS}{assisted-GNSS}
\acrodef{AGPS}{assisted GPS}
\acrodef{AI}{artificial intelligence}
\acrodef{AIC}{Akaike information criterion}
\acrodef{AOA}{angle-of-arrival}
\acrodef{AOD}{angle-of-departure}
\acrodef{AOT}{approximate optimum threshold}
\acrodef{AP}{access point}
\acrodef{API}{application programming interface}
\acrodef{ASK}{amplitude shift keying}
\acrodef{ASNR}{accumulated signal-to-noise ratio}
\acrodef{AUB}{asymptotic union bound}
\acrodef{AWGN}{additive white Gaussian noise}
\acrodef{BAN}{body area network}
\acrodef{BAV}{balanced antipodal Vivaldi}
\acrodef{BCH}{Bose Chaudhuri Hocquenghem}
\acrodef{BEP}{bit error probability}
\acrodef{BER}{bit error rate}
\acrodef{BF}{brute force}
\acrodef{BFC}{block fading channel}
\acrodef{BIC}{Bayesian information criterion}
\acrodef{BLUE}{best linear unbiased estimator}
\acrodef{BPAM}{binary pulse amplitude modulation}
\acrodef{BPF}{bandpass filter}
\acrodef{BPPM}{binary pulse position modulation}
\acrodef{bps}{bits per second}
\acrodef{BPSK}{binary phase shift keying}
\acrodef{BPZF}{band-pass zonal filter}
\acrodef{BS}{base station}
\acrodef{BSC}{binary symmetric channel}
\acrodef{BTB}{Bellini-Tartara bound}
\acrodef{c.c.d.f.}{complementary cumulative distribution function}
\acrodef{c.d.f.}{cumulative distribution function}
\acrodef{CAD}{computer-aided design}
\acrodef{CAIC}{consistent Akaike information criterion}
\acrodef{CAP}{continuous aperture phased}
\acrodef{CCF}{cross correlation function}
\acrodef{CCI}{co-channel interference}
\acrodef{CD}{cooperative diversity}
\acrodef{CDMA}{code division multiple access}
\acrodef{CEOT}{channel ensemble optimum threshold}
\acrodef{CEP}{codeword error probability}
\acrodef{CFAR}{constant	 false alarm rate}
\acrodef{ch.f.}{characteristic function}
\acrodef{CH}{cluster head}
\acrodef{CIR}{channel impulse response}
\acrodef{CL}{centroid localization}
\acrodef{CM}{channel model}
\acrodef{CNR}{clutter-to-noise ratio}
\acrodef{CP}{ciclic prefix}
\acrodef{CPR}{channel pulse response}
\acrodef{CR}{channel response}
\acrodef{CRB}{Cram\'{e}r-Rao bound}
\acrodef{CRC}{cyclic redundancy check}
\acrodef{CRLB}{Cram\'{e}r-Rao lower bound}
\acrodef{CS}{clock skew}
\acrodef{CSCG}{circularly symmetric complex Gaussian}
\acrodef{CSI}{channel state information}
\acrodef{CSMA}{carrier sense multiple access}
\acrodef{CSS}{chirp spread spectrum}
\acrodef{CTS}{clear-to-send}
\acrodef{CW}{continuous wave}
\acrodef{DAA}{detect and avoid}
\acrodef{DAB}{digital audio broadcasting}
\acrodef{DBB}{digital base band}
\acrodef{DBPSK}{differential binary phase shift keying}
\acrodef{DCM}{dual-carrier modulation}
\acrodef{DDP}{detected direct path}
\acrodef{DF}{detect \& forward}
\acrodef{DFMS}{monopole dual feed stripline antenna}
\acrodef{DGPS}{differential GPS}
\acrodef{DLL}{delay-locked loop}
\acrodef{DoD}{Department of Defense}
\acrodef{DoF}{degrees of freedom}
\acrodef{DP}{direct path}
\acrodef{DR}{detection rate}
\acrodef{DRT}{distance ratio test}
\acrodef{DS-SS}{direct-sequence spread-spectrum}
\acrodef{DS}{delay spread}
\acrodef{DTR}{differential transmitted-reference}
\acrodef{DVB-H}{digital video broadcasting\,--\,handheld}
\acrodef{DVB-T}{digital video broadcasting\,--\,terrestrial}
\acrodef{e.m.}{electromagnetic}
\acrodef{ECC}{European Community Commission}
\acrodef{ED}{energy detector}
\acrodef{EDR}{energy detector receiver}
\acrodef{EFIM}{equivalent Fisher information matrix}
\acrodef{EIRP}{effective radiated isotropic power}
\acrodef{EKF}{extended Kalman filter}
\acrodef{ELP}{equivalent low-pass}
\acrodef{EM}{electromagnetic}
\acrodef{EMCB}{extended Miller Chang bound}
\acrodef{EME}{minimum eigenvalue ratio detector}
\acrodef{EMI}{electromagnetic interference}
\acrodef{ENP}{estimated noise power}
\acrodef{ESA}{European Space Agency}
\acrodef{EU}{European Union}
\acrodef{FAR}{false alarm rate}
\acrodef{FCC}{Federal Communications Commission}
\acrodef{FDMA}{frequency division multiple access}
\acrodef{FDMA}{frequency division multiple access}
\acrodef{FEC}{forward error correction}
\acrodef{FEC}{forward error correction}
\acrodef{FFD}{full function device}
\acrodef{FFR}{full function reader}
\acrodef{FF}{far-field}
\acrodef{FFT}{fast Fourier transform}
\acrodef{FG}{factor graph}
\acrodef{FH-SS}{frequency-hopping spread-spectrum}
\acrodef{FH}{frequency-hopping}
\acrodef{FIM}{Fisher information matrix}
\acrodef{FLL}{Frequency-locked loop}
\acrodef{FS}{frame synchronization}
\acrodef{GA}{Gaussian approximation}
\acrodef{GD}{gradient descent}
\acrodef{GDOP}{geometric dilution of precision}
\acrodef{GLR}{generalized likelihood ratio}
\acrodef{GLRT}{generalized likelihood ratio test}
\acrodef{GML}{generalized maximum likelihood}
\acrodef{GPRS}{general packet radio service}
\acrodef{GPS}{global positioning system}
\acrodef{HAP}{high altitude platform}
\acrodef{HCRB}{hybrid Cram\'{e}r-Rao bound}
\acrodef{HDSA}{high-definition situation-aware}
\acrodef{Hi-RADIAL}{High-accuracy RAdio Detection, Identification, And Localization}
\acrodef{HMM}{hidden Markov model}
\acrodef{HPA}{high-power amplifier}
\acrodef{HPBW}{half power beam width}
\acrodef{HW}{hardware}
\acrodef{i.i.d.}{independent, identically distributed}
\acrodef{ICT}{information and communication technologies}
\acrodef{ID}{integrate \& dump}
\acrodef{IE}{informative element}
\acrodef{IEEE}{Institute of Electrical and Electronics Engineers}
\acrodef{IF}{intermediate frequency}
\acrodef{IFFT}{inverse fast Fourier transform}
\acrodef{IMF}{ideal matched filter}
\acrodef{IMU}{inertial measurement unit}
\acrodef{INR}{interference-to-noise ratio}
\acrodef{INS}{inertial navigation system}
\acrodef{IoT}{Internet of things}
\acrodef{IIoT}{industrial Internet of things}
\acrodef{INS}{inertial navigation system}
\acrodef{IR-UWB}{impulse radio UWB}
\acrodef{IR}{impulse radio}
\acrodef{IRI}{inter-reader interference}
\acrodef{IRS}{intelligent reflecting surface} 
\acrodef{ISI}{inter-symbol interference} 
\acrodef{isi}{intra-symbol interference} 
\acrodef{ISM}{industrial, scientific and medical}
\acrodef{ISNR}{interference-plus-signal-to-noise-ratio}
\acrodef{IT}{interference temperature}
\acrodef{ITC}{information theoretic criteria}
\acrodef{JBSF}{jump back and search forward}
\acrodef{JF}{just forward}
\acrodef{KF}{Kalman filter}
\acrodef{LDC}{low duty cycle}
\acrodef{LDPC}{low density parity check}
\acrodef{LEO}{localization error outage}
\acrodef{LG}{Laguerre-Gaussian}
\acrodef{LIS}{large intelligent surface}
\acrodef{LLR}{log-likelihood ratio}
\acrodef{LLRT}{log-likelihood ratio test}
\acrodef{LRT}{likelihood ratio test}
\acrodef{LNA}{low-noise amplifier}
\acrodef{LOS}{line-of-sight}
\acrodef{LRT}{likelihood ratio test}
\acrodef{LS}{least square}
\acrodef{LS}{least squares}
\acrodef{M-PSK}{$M$-ary phase shift keying}
\acrodef{M-QAM}{$M$-ary quadrature amplitude modulation}
\acrodef{m.g.f.}{moment generating function}
\acrodef{MAC}{medium access control}
\acrodef{MAE}{mean absolute error}
\acrodef{MAI}{multiple access interference}
\acrodef{MAN}{metropolitan area network}
\acrodef{MAP}{maximum a posteriori}
\acrodef{MB-OFDM}{multi-band OFDM}
\acrodef{MB-UWB}{multi-band UWB}
\acrodef{MB}{multi-band}
\acrodef{MC}{multi-carrier}
\acrodef{MCB}{Miller Chang bound}
\acrodef{MCRB}{modified Cram\'{e}r-Rao bound}
\acrodef{MDD}{minimum distance distribution}
\acrodef{MDL}{minimum description length}
\acrodef{MF}{matched filter}
\acrodef{MGF}{moment generating function}
\acrodef{MI}{mutual information}
\acrodef{MIMO}{multiple-input multiple-output}
\acrodef{MISO}{multiple-input single-output}
\acrodef{ML}{maximum likelihood}
\acrodef{MM}{min-max}
\acrodef{MME}{maximum-minimum eigenvalue ratio detector}
\acrodef{MMSE}{minimum mean-square error}
\acrodef{mmWave}{millimeter wave}
\acrodef{MPC}{multipath component}
\acrodef{MRC}{maximal ratio combiner}
\acrodef{MS}{mobile station}
\acrodef{MSB}{most significant bit}
\acrodef{MSE}{mean square error}
\acrodef{MSE}{mean squared error}
\acrodef{MSK}{minimum shift keying}
\acrodef{MUI}{multi-user interference}
\acrodef{MUR}{multistatic radar}
\acrodef{MVU}{minimum variance unbiased}
\acrodef{MZZB}{modified Ziv-Zakai bound}
\acrodef{NB}{narrowband}
\acrodef{NBI}{narrowband interference}
\acrodef{NEO}{navigation error outage}
\acrodef{NFER}{near-Þeld electromagnetic ranging}
\acrodef{NF}{near-field}
\acrodef{NFF}{near-field focused}
\acrodef{NL}{nonlinear}
\acrodef{NLOS}{non-line-of-sight}
\acrodef{NP}{Neyman-Pearson}
\acrodef{NTIA}{National Telecommunications and Information Administration}
\acrodef{NTP}{network time protocol}
\acrodef{OAM}{orbital angular momentum} 
\acrodef{OC}{optimum combining}
\acrodef{OFDM}{orthogonal frequency division multiplexing}
\acrodef{OOK}{on-off keying}
\acrodef{OP}{outage probability}
\acrodef{OT}{optimum threshold}
\acrodef{P-Max}{$P$-Max}  
\acrodef{p.d.f.}{probability density function}
\acrodef{p.m.f.}{probability mass function}
\acrodef{PA}{power amplifier}
\acrodef{PAM}{pulse amplitude modulation}
\acrodef{PAN}{personal area network}
\acrodef{PAR}{peak-to-average ratio}
\acrodef{PD}{probability of detection}
\acrodef{PDP}{power delay profile}
\acrodef{PE}{probability of emulation}
\acrodef{PEB}{position error bound}
\acrodef{PEP}{packet error probability}
\acrodef{PF}{particle filter}
\acrodef{PFA}{probability of false alarm}
\acrodef{PHY}{physical layer}
\acrodef{PL}{path-loss}
\acrodef{PLL}{phase-locked loop}
\acrodef{PMD}{probability of missed detection}
\acrodef{PN}{pseudo-noise}
\acrodef{ppm}{part-per-million}
\acrodef{PPM}{pulse position modulation}
\acrodef{PR}{pseudo-random}
\acrodef{PRake}{partial rake}
\acrodef{PRF}{pulse repetition frequency}
\acrodef{PRP}{pulse repetition period}
\acrodef{PSD}{power spectral density}
\acrodef{PSEP}{pairwise synchronization error probability}
\acrodef{PSK}{phase shift keying}
\acrodef{PSWF}{prolate spheroidal wave function}
\acrodef{PU}{primary user}
\acrodef{QAM}{quadrature amplitude modulation}
\acrodef{QoS}{quality of service}
\acrodef{QPSK}{quadrature phase shift keying}
\acrodef{R.V.}{random variable}
\acrodef{RADAR}{radar}
\acrodef{RCS}{radar cross section}
\acrodef{RDL}{"random data limit"}
\acrodef{REM}{radio environment map}
\acrodef{REO}{ranging error outage}
\acrodef{RF}{radio-frequency}
\acrodef{RFID}{radio-frequency identification}
\acrodef{RFR}{reduced function reader}
\acrodef{RFT}{reduced function tag}
\acrodef{RII}{ranging information intensity}
\acrodef{RIS}{reconfigurable intelligent surface}
\acrodef{rms}{root mean square}
\acrodef{RMSE}{root-mean-square error}
\acrodef{ROC}{receiver operating characteristic}
\acrodef{RRC}{root raised cosine}
\acrodef{RSN}{radar sensor network}
\acrodef{RSS}{received signal strength}
\acrodef{RSSI}{received signal strength indicator}
\acrodef{RTLS}{real time locating systems}
\acrodef{RTT}{round-trip time}
\acrodef{S-V}{Saleh-Valenzuela}
\acrodef{SA}{simulated annealing}
\acrodef{SaG}{stop-and-go}
\acrodef{SBS}{serial backward search}
\acrodef{SBSMC}{serial backward search for multiple clusters}
\acrodef{SCM}{supply chain management}
\acrodef{SCR}{signal-to-clutter ratio}
\acrodef{SEP}{symbol error probability}
\acrodef{SIS}{small intelligent surface}
\acrodef{SFD}{start frame delimiter}
\acrodef{SIMO}{single-input multiple-output}
\acrodef{SINR}{signal-to-interference plus noise ratio}
\acrodef{SIR}{signal-to-interference ratio}
\acrodef{SISO}{single-input single-output}
\acrodef{SNR}{signal-to-noise ratio}
\acrodef{SoC}{system on chip}
\acrodef{SoO}{signal of opportunity}
\acrodef{SoP}{system on package}
\acrodef{SOT}{sub-optimum threshold}
\acrodef{SPAWN}{sum-product algorithm over a wireless network}
\acrodef{SPEB}{squared position error bound}
\acrodef{SPMF}{single-path matched filter}
\acrodef{SQNR}{signal-to-quantization-noise ratio}
\acrodef{SRE}{smart radio environment}
\acrodef{SS}{spread spectrum}
\acrodef{ST}{simple thresholding}
\acrodef{SU}{secondary user}
\acrodef{SVD}{singular value decomposition}
\acrodef{SW}{software}
\acrodef{SW}{sync word}
\acrodef{TDE}{time delay estimation}
\acrodef{TDL}{tapped delay line}
\acrodef{TDMA}{time division multiple access}
\acrodef{TDOA}{time difference-of-arrival}
\acrodef{TH}{time-hopping}
\acrodef{TNR}{threshold-to-noise ratio}
\acrodef{TOA}{Time-of-arrival}
\acrodef{TOF}{time-of-flight}
\acrodef{TPC}{transmit power control}
\acrodef{TR}{transmitted-reference}
\acrodef{TS}{tabu search}
\acrodef{UAV}{unmanned aerial vehicle}
\acrodef{UB}{union bound}
\acrodef{UCA}{uniform circular array}
\acrodef{UDP}{undetected direct path}
\acrodef{UHF}{ultra-high frequency}
\acrodef{ULP}{user location protocol}
\acrodef{UMP}{uniformly most powerful}
\acrodef{UMPI}{uniformly most powerful invariant}
\acrodef{UT}{user terminal}
\acrodef{UTC}{coordinated universal time}
\acrodef{UTM}{universal transverse Mercator}
\acrodef{UTRA}{UMTS terrestrial radio access}
\acrodef{UAV}{unmanned aerial vehicle}
\acrodef{UUV}{unmanned underwater vehicle}
\acrodef{UWB}{ultrawide-band}
\acrodef{UWBcap}[UWB]{Ultrawide band}
\acrodef{VFIL}{virtual force iterative localization}
\acrodef{VGA}{variable-gain amplifier}
\acrodef{VNA}{vector network analyzer}
\acrodef{WAF}{wall attenuation factor}
\acrodef{WB}{wideband}
\acrodef{WBI}{wideband interference}
\acrodef{WCL}{weighted centroid localization}
\acrodef{WED}{wall extra delay}
\acrodef{WiMAX} {worldwide interoperability for microwave access}
\acrodef{WLAN}{wireless local area network}
\acrodef{WLS}{weighted least squares}
\acrodef{WMAN}{wireless metropolitan area network}
\acrodef{WPAN}{wireless personal area networks}
\acrodef{WRAPI}{wireless research application programming interface}
\acrodef{WSN}{wireless sensor network}
\acrodef{WSR}{wireless sensor radar}
\acrodef{WSS}{wide-sense stationary}
\acrodef{WWB}{Weiss-Weinstein bound}
\acrodef{WWLB}{Weiss-Weinstein lower bound}
\acrodef{ZZB}{Ziv-Zakai bound}
\acrodef{ZZLB}{Ziv-Zakai lower bound}

\newcommand{\stx}{S_{\text{T}}}
\newcommand{\srx}{S_{\text{R}}}
\newcommand{\rt}{R_{\text{T}}}
\newcommand{\rr}{R_{\text{R}}}
\newcommand{\rhot}{\rho_{\text{T}}}
\newcommand{\phit}{\varphi_{\text{T}}}
\newcommand{\rhor}{\rho_{\text{R}}}
\newcommand{\phir}{\varphi_{\text{R}}}
\newcommand{\elln}{\ell_{n}}

\newcommand{\Et} {E_{\mathrm{t}}}
\newcommand{\Es}{E_{\mathrm{s}}}
\newcommand{\boldx} {{\bf x}}

\begin{document}

\title{Holographic MIMO Communications exploiting the Orbital Angular Momentum}

\author{
\IEEEauthorblockN{Giulia~Torcolacci\IEEEauthorrefmark{1},~\IEEEmembership{Student~Member,~IEEE},
Nicolò~Decarli\IEEEauthorrefmark{2},~\IEEEmembership{Member,~IEEE},
Davide~Dardari\IEEEauthorrefmark{1},~\IEEEmembership{Senior~Member,~IEEE}}

 \IEEEauthorblockA{\IEEEauthorrefmark{1}~\footnotesize{Department of Electrical, Electronic and Information Engineering "G. Marconi" (DEI) and WiLAB-CNIT, University of Bologna, Cesena (FC), Italy. \\  }}
    \IEEEauthorblockA{\IEEEauthorrefmark{2}~\footnotesize{National Research Council~-~Institute of Electronics, Computer and Telecommunication Engineering (CNR-IEIIT) and WiLab-CNIT, Bologna (BO), Italy. \\ \textit{Corresponding author}: Giulia Torcolacci (g.torcolacci@unibo.it) }}
  
   \IEEEcompsocthanksitem{Part of this article was presented at IEEE GLOBECOM 2022 \cite{TorDecDar:C22}.}}

\maketitle

\begin{abstract}
This study delves into the potential of harnessing the orbital angular momentum (OAM) property of electromagnetic waves in near-field and line-of-sight scenarios by utilizing large intelligent surfaces, in the context of holographic multiple-input multiple-output (MIMO) communications. The paper starts by characterizing OAM-based communications and investigating the connection between OAM-carrying waves and optimum communication modes recently analyzed for communicating with smart surfaces.
Subsequently, it proposes implementable strategies for generating and detecting OAM-based communication signals using intelligent surfaces and optimization methods that leverage focusing techniques. Then, the performance of these strategies is quantitatively evaluated through simulations. The numerical results show that OAM waves while constituting a viable and more practical alternative to optimum communication modes are sub-optimal in terms of achievable capacity.
\end{abstract}



\section{Introduction}
The ceaseless quest for high-speed, reliable, and ubiquitous wireless services is leading today’s radio networks to their utmost. Fifth-generation (5G) wireless networks, exploiting the most advanced radio technologies, e.g., \ac{MIMO} communications, have already been deployed in many countries and are establishing themselves as essential techniques to manage the ever-increasing number of devices’ connections. However, it is foreseen that sixth-generation (6G) wireless networks will push these requirements to the ultimate limit, introducing even more stringent requisites regarding user throughput, latency, scalability, and reliability, paving the way for massively novel applications and use cases. In this framework, the exploitation of higher operating frequencies, where a more considerable amount of spectral resources is available, in conjunction with the usage of large antenna arrays and antennas densification, has been envisaged as a potential, beneficial approach to tackle the challenging prerequisites that have been outlined. Nevertheless, the adoption of \ac{mmWave} communications and terahertz technologies entails larger path loss, increased susceptibility to atmospheric turbulence, and in some cases, failure of traditional \ac{EM} propagation models based on the far-field assumption. Indeed, when the antenna size becomes large or the frequency increases, the interaction between the transmitting and receiving antennas can occur in the radiating near-field region~\cite{GraEtAl:J21, BjoEtAl:J19}. Here the plane wave approximation of the wavefront becomes inapplicable, and the actual spherical-shaped wavefronts must be considered instead. This operating regime has been addressed as particularly favorable since it commences new and unexplored opportunities for future wireless systems, whose primary goal is to approach greater link-level spectrum efficiencies and unprecedented performance. From all the considerations above, the concept of \emph{holographic communications} emerged \cite{DarDec:J20,GongetAl:J22,SangetAl:J22}, which can be referred to as the capability to utterly manipulate the \ac{EM} field that is generated or received by antennas to exploit the entirety of the available channel’s \ac{DoF} and hence increase the communication capacity of the wireless link, even in \ac{LOS} channel conditions. Specifically, previous work targeted the exploitation of \ac{LIS} antennas as transmitting and receiving devices in the near field, thus presenting a more comprehensive description of the communication problem in terms of \textit{communication modes}, i.e., parallel channels that can be obtained between a couple of antennas at \ac{EM} level \cite{Mil:C00, Dar:J20}. For instance, in~\cite{Dar:J20}, the optimal basis functions determining the optimum communications modes and their number, i.e., the available \ac{DoF}, that can be established between two \acp{LIS} have been obtained from solving an eigenfunction problem \cite{Mil:C00}.

On another side, in the last decade, several works investigated the possibility of exploiting the \ac{OAM} property of the \ac{EM} waves to theoretically unfold further orthogonal radio channels. Initially, this was thought of as a new dimension to improve the \ac{DoF} in wireless communications that is entirely independent of concepts like time, space, frequency, and polarization.
Consequently, many relevant applications exploiting \ac{OAM} have been envisaged in the latest years, and an extensive number of possible technologies and deployments have been proposed. For instance, a novel type of space-division multiplexing technique based on modes’ orthogonality has been theorized, namely \ac{OAM} mode division multiplexing (\ac{OAM}-MD) \cite{WanEtAl:J12}. In addition, the possibility of managing the increasing users’ density through an \ac{OAM}-based multiple access scheme (\ac{OAM}-MDMA) was proposed as well \cite{CheWenEtAl:J18}. Several other use cases have been investigated \cite{LiaEtAl:C17, CheWenEtAl:18,ZhaEtAl:J17,YanEtAl:C16,YuaEtAl:J16, MarEtAl:J15} and, intuitively, many open issues and problems are still to be addressed since systems performance has been shown \cite{CheEtAl:J20} to be strongly affected by the kind of antenna that is used, the overall systems architecture, the operating frequency, and many other elements.
In spite of the initial enthusiasm in \ac{OAM}, various investigations highlighted that substantial advantages with regard to more traditional techniques, such as \ac{MIMO} communications, cannot be obtained by exploiting the \ac{OAM} property of the radio waves \cite{EdfJoh:J11,ZhaoEtAl:J15,TamaEtAl:J12, OldEtAl:J15, GafEtAl:J17}.

\ac{OAM}, whose employment was first studied in the optic domain \cite{AllEtAl:J92,BeiEtAl:J93} and then also proposed for radio communications \cite{ThiEtAl:J07,MohEtAl:J09}, is frequently characterized by the use of specific beams, such as that corresponding to \ac{LG} modes \cite{AllPadBab:B99,AllEtAl:J92}. However, these modes are often considered without accounting for the specific transmitting and receiving devices, i.e., without considering particular antenna structures. In fact, a plethora of radio \ac{OAM} generation methods have been identified \cite{CheEtAl:J20}, e.g., spiral phase plate (SPP) antennas, \acp{UCA}, and metasurfaces, and each of them was identified as optimal for a specific operating frequency. Indeed, it was shown that spiral reflectors and \acp{UCA} work best at \ac{RF} and are more suitable for long-range transmissions, while SPPs and metasurfaces are better for the \acp{mmWave} band \cite{CheEtAl:J20}. Nevertheless, the relationship among these \ac{LG} modes and the concept of optimum communication modes discussed in \cite{Mil:C00, Dar:J20} is often unclear, as well as the gap between the \ac{DoF} that can be obtained with \ac{OAM} and those corresponding to the optimum strategy.

Numerous studies to date on \ac{OAM}-based wireless systems contemplate the usage of \acp{UCA} as transmission and reception devices \cite{ChenWen:J20, CagGaf:J16, CanAll:J15, HuetAl:16}. However, these systems suffer from severe limitations in terms of hardware flexibility and phase quantization errors. In this regard, recent research \cite{AffEtAl:J21, WeitetAl:J17, RenetAl:J17, ZhaoetAl:J21, XuetAl:J23} has demonstrated, both theoretically and experimentally, the feasibility and benefits of exploiting \ac{OAM} in wireless systems by utilizing metasurfaces, especially in the terahertz band \cite{YanetAl:J23, RuietAl:J23, SuetAl:J23, ZhaoZheetAl:J21, WilletAl:J22} where most of the propagation occurs in the near-field region of the transmitter. To the best of the authors' knowledge, no analytical studies have addressed the use of \ac{OAM}, or more in general, arbitrary \ac{EM} phase and amplitude profiles, independently of the particular hardware implementation or antenna technology, as we instead perform by adopting homogenized, spatially continuous surfaces such as \acp{LIS}.

In this work, motivated by the recent advancements concerning the communication between \acp{LIS}, we revise the exploitation of \ac{OAM}-based techniques as orthogonal communication channels. The proposed approach allows the investigation of the \ac{OAM} capabilities in terms of capacity achievements and effectiveness, especially in relation to optimal strategies for defining the communication modes between a couple of antennas. The main contributions of the paper can be summarized as follows:
\begin{itemize}
    \item Review of the concept of \ac{OAM} communication in the framework of holographic MIMO and in relationship with communication modes;
    \item Characterization of the gap between \ac{OAM}-based strategies and optimum communication modes for different configurations in terms of \ac{DoF};
    \item Proposal of the exploitation of beam focusing for reducing the typical divergence effects of OAM beams within the near-field (Fresnel) region, showing the related benefits and drawbacks;
    \item Proposal and analysis of different detection strategies for the implementation of OAM-based receivers;
    \item Discussion of the main advantages and disadvantages concerning the exploitation of \ac{OAM} in the holographic MIMO scheme.
\end{itemize}

The remainder of the paper is organized as follows: Section~\ref{sec:CommModes} revises the concept of communication between \acp{LIS} and introduces the notion of communication modes, thus discussing the adoption of optimal transmitting and receiving basis function sets. Section~\ref{Sec:OAM} describes the characteristics of \ac{OAM} propagation. Then, the relationship between \ac{OAM} and optimum communication modes is illustrated, and the characteristics of \ac{OAM} modes are discussed. In Section~\ref{Sec:OAMnew} the overall \ac{OAM}-based communication scheme is presented, and beam focusing within the near-field is introduced as a way to improve the system performance. Then, in Section.~\ref{sec:DetectionStrategies} practical detection strategies are presented, while Section~\ref{Sec:Results} illustrates some numerical results to characterize the gap between \ac{OAM} communications and optimum strategies. Finally, a discussion concerning the main strengths and weaknesses of using \ac{OAM}-based radio communications, together with an implementation complexity analysis, is provided, and Section~\ref{Sec:Conclusion} concludes the paper.

\vspace{-0.3cm}
\section{Communications with LISs}\label{sec:CommModes}

\acp{LIS} can be denoted as active, re-configurable planar antennas whose sizes are much larger than the operating wavelength, that can control the amplitude and phase profiles of the \ac{EM} waves with high flexibility and resolution. Notably, metamaterials have been proposed \cite{Tre:J15, SilEtAl:J14, OveEtAl:J17} as a viable technology to create those kinds of antennas, thus enabling the design of predefined \ac{EM} waves’ features in terms of shape, polarization, and steering. From an analytical viewpoint, \acp{LIS} can be abstractly modeled as continuous surfaces composed of an infinite number of sub-wavelength antenna elements, regardless of their specific hardware implementation. Research in \cite{Dar:J20} has investigated the fundamental limitations of communication using \acp{LIS}, showing that by carefully designing the amplitude and phase profiles of the transmitting and receiving \acp{LIS}, multiple communication modes, known as channel's \ac{DoF}, can be achieved in \ac{LOS} near-field conditions. Particularly, these communication modes can be seen as a collection of parallel, orthogonal channels at \ac{EM} level, which allows for the capability to spatially multiplex signals within the wireless system.
In the following subsection, the formal definition of communication modes is briefly revised. The reader can refer to \cite{Mil:C00, Dar:J20} for a complete treatment.

\vspace{-0.3cm}
\subsection{Communication Modes}\label{Sec:LIS}
Abstracting in the first instance from the specific shape or antennas' configuration, let us generally consider a transmitting and a receiving \ac{LIS} of area $\stx$ and $\srx$.
 We indicate with $\textbf{s} \in \stx$ and $\textbf{r} \in \srx$ the vectors pointing from the origin of the contemplated reference system towards a point on the transmitting and the receiving \ac{LIS}, respectively.
A monochromatic source at frequency $f_0$, i.e., $\phi(\textbf{s})$, is used at the transmitting LIS. This source generates a wave function $\psi(\textbf{r})$ at the receiving antenna, which is a solution of the inhomogeneous Helmholtz equation
\begin{equation}\label{eq:Helmholtz}
\nabla^{2} \psi(\mathbf{r})+\kappa^{2} \psi(\mathbf{r})=-\phi(\mathbf{s})\, .
\end{equation}
A solution of \eqref{eq:Helmholtz} in free-space propagation condition is provided as
\begin{equation}\label{eq:psidef}
\psi(\textbf{r}) = \int_{\stx}  G(\textbf{r},\textbf{s})\phi(\textbf{s})\,\text{d}\textbf{s} 
\end{equation}
where the scalar Green function $G(\mathbf{r}_1,\mathbf{r}_2)$ between the points represented by the vectors $\mathbf{r}_1$ and $\mathbf{r}_2$ is given by
\begin{equation}\label{eq:GreenFunction}
G\left(\mathbf{r}_{1}, \mathbf{r}_{2}\right)=\frac{\exp \left(-\jmath \kappa\left|\left|\mathbf{r}_{1}-\mathbf{r}_{2}\right|\right|\right)}{4 \pi\left|\left|\mathbf{r}_{1}-\mathbf{r}_{2}\right|\right|}
\end{equation}
where $\kappa=2 \pi / \lambda$ is the free-space wavenumber and $\lambda=c / f_0$ indicates the wavelength, with $c$ being the light speed. It is important to note that in \eqref{eq:GreenFunction}, the reactive field components that typically vanish for link distances greater than a few wavelengths are neglected, as the system is supposed to operate in the radiating near-field region.

In the following, we will treat the \ac{EM} field as a complex-valued scalar quantity (e.g., by considering a single polarization), despite being a vectorial complex quantity. This simplification does not affect the generality or validity of the results, which can be extended to the vectorial case. Formally, communication modes can be described by the mean of specific orthonormal basis expansions of the transmitting function $\phi(\textbf{s})$ and receiving function $\psi(\textbf{r})$, respectively denoted by $ {\phi_n(\textbf{s}) }$ and $ {\psi_n(\textbf{r}) }$, where $n = 1, 2, ..., \infty$. 
If the basis sets are devised such that exists a bijective relationship among the $n$-th transmitting basis function \(\phi_n(\textbf{s})\) and the $n$-th receiving one \(\psi_n(\textbf{r})\), the existence of multiple communication modes is ensured. In this case, each transmitting function \(\phi_n(\textbf{s})\) induces an effect \(\xi_n\psi_n(\textbf{r})\) on the receiving antenna, with \(\xi_n\) identifying the largest feasible coupling coefficient. Precisely, large \(\xi_n\) coefficients indicate well-coupled modes, i.e., pair of functions (\( \phi_n(\textbf{s}) \), \( \psi_n(\textbf{r}) \)), between transmitting and receiving \acp{LIS} corresponding to practicable, parallel communication channels; small \(\xi_n\) instead, denotes the prevalence of the wave's dispersion away from the receiver. 
In addition, the linear combinations of $\phi_{n}(\mathbf{s})$ and $\psi_{n}(\mathbf{r})$ functions are proportional to the current spatial distribution impressed on the transmitting \ac{LIS}, and to the resulting electric field at the receiving LIS side, respectively, thus the latter is referred as $n$-th electric field component in the following.

In principle, infinite basis sets are realizable, but to achieve the highest number of communication modes with the largest coupling, the following coupled eigenfunction problem needs to be solved, thus leading to the optimal basis functions \cite{Mil:C00, Dar:J20}
\begin{equation}\label{eq:eigenfunc_phi}
\xi_n^2\, \phi_n(\textbf{s}) = \int_{\stx} K_{\text{T}}(\textbf{s},\textbf{s}^\prime)\,\phi_n(\textbf{s}^\prime)\,\text{d}\textbf{s}^\prime 
\end{equation}
\begin{equation} \label{eq:eigenfunc_psi}
\xi_n^2\, \psi_n(\textbf{r}) = \int_{\srx} K_{\text{R}}(\textbf{r},\textbf{r}^\prime)\,\psi_n(\textbf{r}^\prime)\,\text{d}\textbf{r}^\prime 
\end{equation}
where the kernels \(K_{\text{T}}(\textbf{s},\textbf{s}^\prime)\) and  \(K_{\text{R}}(\textbf{r},\textbf{r}^\prime)\) are given by
\begin{equation}\label{eq:KernelTX}
K_{\text{T}}(\textbf{s},\textbf{s}^\prime) = \int_{\srx} G^*(\textbf{r},\textbf{s})G(\textbf{r},\textbf{s}^\prime)\,\text{d}\textbf{r}
\end{equation}
\begin{equation}\label{eq:KernelRX}
K_{\text{R}}(\textbf{r},\textbf{r}^\prime) = \int_{\stx} G(\textbf{r},\textbf{s})G^*(\textbf{r}^\prime,\textbf{s})\,\text{d}\textbf{s} \; .
\end{equation}
Determining these solutions can be very challenging and may require a burdensome number of numerical calculations, particularly when dealing with very large LISs. In addition, from a practical standpoint, the knowledge of the system's geometry with a tolerance smaller than a wavelength is necessary, which may not be feasible for real-world applications. However, the optimal functions can serve as a baseline to measure the performance difference between the optimal strategy and any other practical approach that uses sub-optimal basis function sets. For this reason, sub-optimum strategies have been proposed, e.g., based on the approximation of the basis function sets with designated focusing/steering functions that depend on the specific operating conditions \cite{DecDar:J21}. 
\begin{figure*}[t]
    \centering
    \includegraphics[width=0.8\linewidth, keepaspectratio]{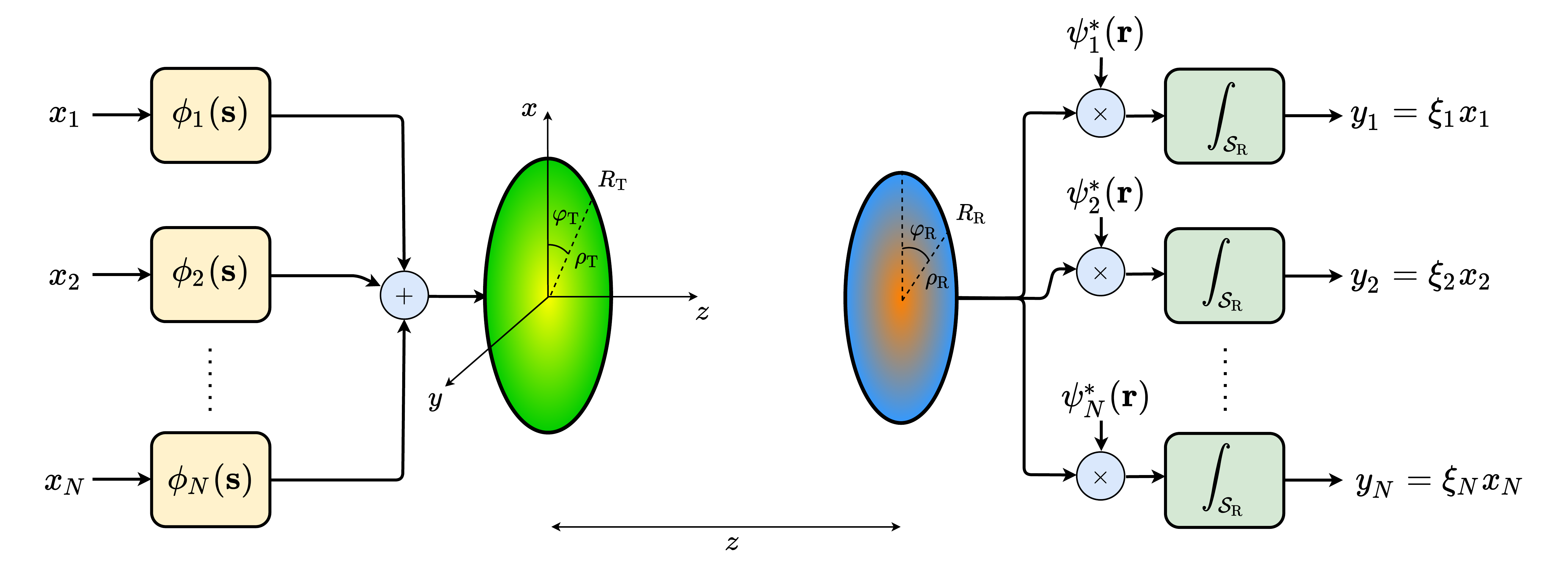}
    \caption{Schematic representation of communication modes between a couple of circular LISs in paraxial configuration and spaced of $z$.}
    \label{fig:Scenario}
\end{figure*}

Overall, the communication problem can be seen as a wireless interconnection of continuous apertures with finite extension in space, capable of generating and detecting the most suitable \ac{EM} waveforms to maximize the achievable channel capacity. When only a finite number $N$ of basis functions are strongly coupled (i.e., with significant values of the corresponding $\xi_n$), the equivalent communication scheme which is obtained can be visualized as in Fig.~\ref{fig:Scenario}. 
Remarkably, the adopted basis decomposition allows the definition of a system input-output representation in terms of $N$ parallel channels that is capacity-optimum, where $N$ identifies the available \ac{DoF} for the specific communication system. Indeed, by associating the $N$ input data streams $\{x_n\}$ to the transmitting basis set $\{\phi_n(\textbf{s}) \}$, $ \textbf{s} \in \stx$, and neglecting the presence of noise, we are able to directly recover the transmitted data by performing the correlation of the received \ac{EM} field with the corresponding basis functions $\{\psi_{n}(\mathbf{r})\}$, $\textbf{r} \in \srx$, thus obtaining $y_n=\xi_n x_n$, $n=1,\ldots, N$.

\vspace{-0.5cm}
\subsection{Optimum Communication Modes with Circular LISs}\label{Sec:OptModes}

Let us contemplate a paraxial geometry, which assumes that the transmitting and receiving \acp{LIS} are parallel to each other and aligned at their centers. 
Considering circular antennas in paraxial conditions, the optimal wave functions describing the communication modes are the so-called circular prolate spheroidal functions (CPSFs) \cite{Xu:J17}, extending the classical linear \acp{PSWF}, which are instead solutions under paraxial conditions for rectangular antennas \cite{Mil:C00}. 

As illustrated in Fig.~\ref{fig:Scenario}, let us consider two parallel, circular \acp{LIS} in \ac{LOS}. The coordinates $\rhot$ and $\phit$ in polar form represent a generic point on the transmitting \ac{LIS}, while coordinates $\rhor$ and $\phir$ represent a generic point on the receiving \ac{LIS}. The transmitting \ac{LIS} antenna is located at a $z=0$, and the two \acp{LIS} are separated by a distance of $z$. Moreover, the transmitting \ac{LIS} has a radius $\rt$, and the receiving LIS has a radius $\rr$.
The distance between two generic points on the transmitting and receiving \acp{LIS} is denoted as 
\begin{equation}\label{eq:r}
r= |\textbf{r}-\textbf{s}|= \sqrt{\rhot^{2}+\rhor^{2}- 2 \rhot \rhor \cos \left(\phir-\phit\right)+z^{2}}\, .
\end{equation}
According to this geometry, the basis functions at the transmitting \ac{LIS} are \cite{Xu:J17}
\begin{equation} \label{phi_CPSF}
\phi_n(\rhot,\phit) = \frac{1}{\sqrt{2\pi \rt \rhot}} f_{m,\ell}^{c} \left(\frac{\rhot}{\rt}\right)e^{- \jmath \kappa \frac{\rhot^{2}}{2z}}e^{\jmath \ell\phit}
\end{equation}
where $n=1,\ldots, N$ is the index spanning the elements in the set $\mathcal{S}=\{m,\ell\}$ corresponding to well-coupled communications modes, $\{f_{m,\ell}^{c}\}$ are the real-valued CPSFs that solve the band-limited Hankel-transform eigenvalue problem
\begin{equation}\label{eigen_problem_CPSF}
\gamma_{m,\ell}^{c} f_{m,\ell}^{c}(x)=\int_{0}^{1} J_{\ell}\left(c x x^\prime \right) \sqrt{c x x^\prime}  f_{m,\ell}^{c}\left(x^\prime\right) \text{d} x^\prime
\end{equation}
in which $\gamma_{m,\ell}^{c}$ are the eigenvalues, $c = \frac{\kappa \rt \rr}{z}$ is the bandwidth parameter, $x^\prime = \frac{\rhot}{\rt}$ and $x = \frac{\rhor}{\rr}$.
Correspondingly, at the receiving LIS side we have the basis functions \cite{Xu:J17}
\begin{equation} \label{psi_CPSF}
\psi_n(\rhor,\phir) =  \frac{1}{\sqrt{2\pi \rr \rhor}} \frac{e^{\jmath \kappa z}}{\jmath^{\ell}}     f_{m,\ell}^{c} \left(\frac{\rhor}{\rr}\right)e^{\jmath \kappa \frac{\rhor^{2}}{2z}}e^{\jmath \ell\phir} \, .
\end{equation}
According to \cite{Xu:J17}, the coupling coefficient for the communication modes are
\begin{equation}
\xi_n =  \gamma_{m,\ell}^{c}  \sqrt{\frac{\rt\rr}{4\kappa z}} \, .
\end{equation}
Due to the properties of eigenvalues $\gamma_{m,\ell}^{c}$ related to CPSFs, the number of communication modes (i.e., well-coupled modes) is given by \cite{Xu:J17}
\begin{equation}\label{eq:Nmodes}
N=\left(\frac{\pi\rt\rr}{\lambda z}\right)^2 = \frac{\stx\srx}{\lambda^2 z^2}  
\end{equation}
where the two areas of the \acp{LIS} are given by $\stx=\pi\rt^2$ and $\srx=\pi\rr^2$. It is worth noting that~\eqref{eq:Nmodes} has the same form as classical results related to rectangular surfaces, as reported in \cite{Mil:C00}. However, this result is accurate only when $z$ is greater than the dimensions of the LISs.\footnote{To this end, an extension of~\eqref{eq:Nmodes} for smaller values of $\rt$ and $\rr$ can be found in \cite{Dar:J20}.}

Notably, both the transmitting and receiving basis functions $\phi_n(\rhot,\phit)$ and $\psi_n(\rhor,\phir)$ are separable in the radial and angular components, that is, they can be written respectively as
\begin{align}\label{eq:SeparableBasis} 
\phi_n(\rhot,\phit) &= \phi_n^{\rho}(\rhot)\,\phi_n^{\varphi}(\phit) \\
\psi_n(\rhor,\phir) &= \psi_n^{\rho}(\rhor)\,\psi_n^{\varphi}(\phir)\, .
\end{align}

\section{Orbital Angular Momentum}\label{Sec:OAM}

\subsection{General Definition}

From the \ac{EM} point of view, angular momentum is a vectorial characteristic that quantitatively defines the intrinsic rotation of the \ac{EM} field \cite{AllEtAl:J92}. While propagating in the axial direction indeed, a photons fashion can simultaneously rotate around its axis. Moreover, there are two different modalities of rotation, and each of them is associated with a specific component of the total angular momentum.
Indeed, the total angular momentum can be computed as the sum of the spin angular momentum (SAM) and the orbital angular momentum (\ac{OAM}). The SAM is related to the dynamic rotation of the electromagnetic field around its direction of propagation, and it is associated with the wave’s polarization \cite{AllPadBab:B99}. \ac{OAM}, instead, concerns the spatial distribution of the electromagnetic field and its rotation around the main beam axis \cite{ThiEtAl:C14}.

From the physical point of view, the defining characteristic of \ac{EM} waves that possess non-zero \ac{OAM} is the emergence of wavefronts that deviate from the traditional far-field assumption of being parallel planes propagating in the axial direction. These wavefronts acquire a helical shape, twisted around the propagation direction. For this reason, \ac{OAM} is frequently described as a vortex or with the notion of helical/twisted/screwed wavefront.
Despite considering the scalar approximation as discussed in Sec.~\ref{sec:CommModes}, the particular helical structure of the wavefronts associated with \ac{OAM} is typically represented in the wave equations by an exponential phase term, in the form of $ e^{\jmath\ell\varphi}$, where $\ell \in\mathbb{Z}$ is referred to as the \textit{topological charge} and $\varphi$ is the transverse azimuthal angle, defined as the angular position on a plane perpendicular to the propagation direction. In \cite{AllPadBab:B99}, it has been demonstrated that this characteristic represents a sufficient condition for the existence of \ac{OAM}-carrying waves in a mostly paraxial propagation regime. Each vortex is thus characterized by an integer number $\ell$, whose magnitude represents the number of twists that a wavefront makes within a distance equal to the wavelength, and its sign determines the chirality or, equivalently, the direction of the twist. In fact, when $|\ell|>0$, the wavefront characterizing a specific topological charge has $|\ell|$ helices intertwined. Each of these states associated with a specific topological charge \(\ell\) takes the name of OAM mode. 
The most interesting feature of OAM modes is their intrinsic orthogonality. In fact, we have
\begin{equation} \label{eq:OAMortho}
 \int_{0}^{2\pi} e^{\jmath\ell_j\varphi}\left(e^{\jmath\ell_k\varphi}\right)^*\,\text{d}\varphi = \begin{cases} 0 & j \neq k \\
                      2\pi &  j = k
        \end{cases}
 \end{equation}
thus, in principle, beams exhibiting this characteristic could carry different data streams leading to spatial multiplexing capabilities.

Many types of beams carrying angular momentum have been investigated to impress the \ac{EM} waves with the desired vorticose phase profile, among which \ac{LG} beams are probably the most widely known \cite{AllBej:J92}. \ac{LG} beams are paraxial solutions of the wave equation in homogeneous media \cite{BarnetAl:J16}, and they are just an example of the plethora of existing waves able to carry \ac{OAM}. Gaussian beams, Airy beams, and many others were subjects of research \cite{LiuEtAl:J17,KadEtAl:J18}, representing possible solutions to the problem of determining the best field distribution for \ac{OAM} propagation.
Interestingly, the common element between these kinds of beams is that all of them are reasonably able to transport \ac{OAM}, and this characteristic is consistently described from the mathematical viewpoint through the exponential term \(e^{\jmath\ell\varphi}\).

\subsection{OAM Modes and Communication Modes}

The main difference concerning the definition of the \ac{OAM} modes as, for example, LG beams, and communication modes as for the definition in Sec.~\ref{sec:CommModes}, is that for the former neither concepts nor references related to the specific transmitting/receiving antennas are given. Differently, communication modes, as identified by the couples of functions $( \phi_n(\textbf{s}),\psi_n(\textbf{r}))$, are strictly related to the geometry of the transmitting and receiving antennas.
The fundamental dissimilarity between \ac{LG} beams and communication modes arises from their distinct definitions. Indeed, \ac{LG} beams are solutions of the wave equation in the absence of boundaries, i.e., with the source's transverse dimensions approaching infinity. Conversely, communication modes in Sec.~\ref{sec:CommModes} are obtained by taking into account finite, well-defined antenna sizes. As a result, \ac{LG} modes can be considered as approximate solutions of the actual propagating modes when the transmitting aperture is sufficiently large to generate the beams and the receiving aperture is adequate to capture them \cite{Xu:J17}.
However, when boundaries corresponding to the finite transmitting/receiving antennas are considered, a limited number $N$ of well-coupled communication modes is obtained, and it holds that \ac{LG} beams do not constitute proper solutions of~\eqref{eq:Helmholtz} \cite{Mil:J19}.

\section{OAM-based Communications using LISs}\label{Sec:OAMnew}

The adoption of CPSFs as basis functions according to~\eqref{phi_CPSF}-\eqref{psi_CPSF} leads to the highest number of communication modes. Unfortunately, such an approach results quite complex to be implemented, even in the paraxial scenario, for several reasons:
\begin{itemize}
    \item The transmitting and receiving \acp{LIS} should know exactly their mutual distance and sizes in order to shape the amplitude and phase profiles along the radial coordinates $\rhot$ and $\rhor$;
    \item A complex amplitude/phase profile must be drawn at the transmitting LIS side according to~\eqref{phi_CPSF}. Similarly, the received \ac{EM} field must be properly weighted with a complex amplitude/phase profile according to~\eqref{psi_CPSF}. These operations require high flexibility at both the \acp{LIS}.
\end{itemize}
In sight of this, there is an interest in designing ad-hoc transmitting and receiving basis functions of more straightforward implementation. 
To simplify the transmitting and receiving \acp{LIS} architectures and exploit, at the same time, the multiplexing capabilities offered by the \ac{OAM} helical wavefronts, the characteristic \ac{OAM} exponential term could be leveraged to construct basis functions in the form of~\eqref{eq:SeparableBasis} that satisfy~\eqref{eq:Helmholtz}. 
In particular, we hereby propose to use the following orthonormal basis functions at the transmitting LIS side in the form of~\eqref
{eq:SeparableBasis} where
\begin{align}
\phi_n^{\rho}(\rhot) &= \frac{1}{\rt }\Pi_{\rt}(\rhot) \label{basis_OAM1}\\
\phi_n^{\varphi}(\phit) & =\frac{1}{ \sqrt{\pi}} e^{\jmath \elln\phit} \label{basis_OAM2}
\end{align}
with 
\begin{equation}
    \Pi_a(\beta)=
        \begin{cases} 1 & \beta \in (0, a) \\
                      0 &  \text{otherwise}
        \end{cases}\, 
\end{equation}
being the rectangular function, which in our specific case identifies a disk of radius $\rt$, and
\begin{align}
   \elln= 
        \begin{cases} -\frac{n-1}{2} & \mod(n,2)=1 \\
                      \frac{n}{2} &  \mod(n,2)=0
        \end{cases}
\end{align}
for $n=1,\ldots, N$. Consequently, $\elln=0,+1,-1,+2,-2,\ldots, (N-1)/2$, assuming $N$ odd.\footnote{Note that the same notation for the basis functions $\phi_n(\rhot,\phit)$ in Sec.~\ref{sec:CommModes} is being used here for simplicity to indicate the OAM-based functions at the transmitting LIS, even though they do not coincide with the optimal ones. Similarly, we will use $\psi_{n}(\rhor,\phir)$ to denote the $n$-th EM field component when the $n$-th basis function $\phi_n(\rhot,\phit)$ in~\eqref{basis_OAM1}-\eqref{basis_OAM2} is employed.}
The transmitting functions proposed in~\eqref{basis_OAM1}-\eqref{basis_OAM2} are normalized such that they exhibit unitary energy, i.e.,
\begin{align}\label{eq:normalization}
\int_0^{\rt} \int_0^{2\pi}\phi_n(\rhot,\phit) \phi_m^*(\rhot,\phit)\rhot \,\text{d}\phit \,\text{d}\rhot  = \delta_{nm} 
\end{align}
being $\delta_{nm}$ the Kronecker delta, hence ensuring that the overall transmitted energy is equal to $\Et = N$.

These bases can be viewed as the extension to a circular surface of the typical phase tapering profiles required to generate \ac{OAM} waves with \acp{UCA}. In fact, a well-known technique for obtaining the helical wavefronts consists of feeding a circular array of $N_{\text{TX}}$ elements equally spaced over a ring, with a successive phase delay $\Delta=2 \ell \pi/N_{\text{TX}}$ such that, after a complete turn, the phase is incremented of a multiple integer $\ell$ of $2\pi$ (i.e., phase varying linearly with the azimuthal angle) \cite{MohEtAl:J10}.
Notably, functions \eqref{basis_OAM1}-\eqref{basis_OAM2} could be used to realize spatial multiplexing by requiring phase-tapering only at the transmitter, 
thereby significantly simplifying the implementation of the \ac{LIS}. These functions do not form a complete basis set, unlike the optimum solutions, hence a performance degradation in terms of \ac{DoF} is expected, as it will be investigated in the following. It is important to note that when $n=1$ (or, equivalently, $\elln=0$), the resulting beam is the same as a uniform circular aperture, while the subsequent modes are degenerate. This means that the same coupling is obtained for adjacent modes (e.g., for $n=2$ and $n=3$, which correspond to topological charges with the same absolute values but opposite chirality).

When the basis function $\phi_n(\rhot,\phit)$ specified in~\eqref{basis_OAM1}-\eqref{basis_OAM2} is employed at the transmitting antenna, the resulting field at a distance $z$ as described by~\eqref{eq:psidef} can be computed as 
\begin{align}\label{eq:OAMrx1}
\psi_{n} (\rhor,\phir)=& \frac{1}{ \rt\sqrt{\pi}}\int_0^{\rt} \int_0^{2\pi} e^{\jmath \elln\varphi_T} \frac{e^{-\jmath \kappa r}}{4 \pi r}\rhot \,\text{d}\phit \,\text{d}\rhot
\end{align}
Particularly, when waves propagate within the Fresnel zone, which allows for multiple communication modes to be established, the variable $r$ in the numerator of~\eqref{eq:OAMrx1} can be approximated with~\cite{Bal:B15}
\begin{equation}\label{eq:fresnel}
    r\approx z+ \frac{\rhot^2}{2z}+\frac{\rhor^2}{2z}-\frac{\rhot\rhor\cos(\phir-\phit)}{z}
\end{equation}
while in the denominator of \eqref{eq:OAMrx1} we consider $r\approx z$. This conventional Fresnel approximation leads to the expression of the received EM field as
\begin{align}\label{eq:OAMrx2}
\psi_{n}(\rhor,\phir) &\simeq \frac{1}{4 \pi z \rt \sqrt{\pi}} e^{-\jmath\kappa\frac{\rhor^2}{2z}}\nonumber e^{-\jmath\kappa z}\int_0^{\rt}\int_0^{2\pi}e^{\jmath \elln\phit}e^{-\jmath\kappa\frac{\rhot^2}{2z}}e^{\jmath\kappa\frac{\rhor\rhot\cos{(\phir-\phit)}}{z}} \rhot\,\text{d}\phit \,\text{d}\rhot \nonumber\\
&=\frac{(-1)^{\elln}}{2z \sqrt{2\pi \rt}}
e^{-\jmath\kappa\frac{\rhor^2}{2z}} e^{-\jmath \kappa z} e^{\jmath \elln \phir}\int_0^{\rt}  \rhot e^{-\jmath\kappa\frac{\rhot^2}{2z}} J_{\elln}\left(\frac{\kappa \rhor\rhot}{z}\right)\,\text{d}\rhot \, .
\end{align}
It can be noticed from~\eqref{eq:OAMrx1} that the exponential term $e^{\jmath \ell\phir}$ identifying \ac{OAM} is also present in the received \ac{EM} field expression. This shows that \ac{OAM} wavefronts propagate in free space with no changes and irrespectively of the selected transmitting profile along the radial direction, thus demonstrating how orthogonality is preserved at the receiver although modifying the OAM beams' shapes. 
In the remainder, we dub as \textit{OAM mode} each EM field component $\psi_{n}(\rhor,\phir)$ generated by the basis function in \eqref{basis_OAM1}-\eqref{basis_OAM2}.

\subsection{Focused OAM}
When considering the \ac{OAM} modes at the receiving LIS according to the discussion thus far, it can be noticed from the analytical expression in~\eqref{eq:OAMrx2} that, as the topological charge increases, higher-order modes exhibit maximum values at increasingly larger values of $\rhor$, which is a characteristic of the Bessel functions. Indeed, it is a widely acknowledged fact that the divergence of the OAM-carrying beams increases with the 
topological charge \cite{XieEtAl:J15}. Therefore, a large receiving \ac{LIS} is required to collect most of the beam and obtain a significant coupling intensity. 
Moreover, the beam-widening effect is more pronounced when operating in the near field, as the diffraction pattern of an aperture is inherently larger than that observed in the far field. This is a consequence of the convolution with a quadratic-phase term, as reported in~\eqref{eq:OAMrx2}. A widely adopted method for restricting the angular spread of a beam in the near field is through the use of focusing techniques \cite{NepBuf:J17}. This approach consists in adjusting the phase of a radiating source so that it compensates for the propagation delay towards a specific point (the focal point), where all the \ac{EM} field contributions add coherently, thus enhancing the EM field intensity in a limited-area region. Achieving this desired outcome, by utilizing the functions presented in~\eqref{basis_OAM1}-\eqref{basis_OAM2}, requires a modification of the transmitting phase profile $\phi_n(\rhot,\phit)$ along the radial coordinate $\rhot$, such that the phase converges at the focal point. In the context of paraxial geometry, where the focal point is located at the center of the receiving \ac{LIS} at a distance $z$, this can be accomplished by adopting
\begin{equation}\label{eq:focusing}
\phi_n^{\rho}(\rhot)\!=\!\frac{\Pi_{\rt}(\rhot)}{\rt}\left.e^{\jmath \kappa r}\right|_{\rhor=0}\!\approx\! \frac{\Pi_{\rt}(\rhot)}{\rt} e^{\jmath \kappa z  \left(1+\frac{\rhot^2}{2z^2}\right)}
\end{equation}
where the right-hand approximation is derived from the Fresnel approximation outlined in~\eqref{eq:fresnel}.

From a practical perspective, focusing can be achieved by utilizing the phase profile $\phi_n^{\rho}(\rhot)=\Pi_{\rt}(\rhot) \exp{(\jmath \kappa {\rhot^2}/{2z})}/\rt$, as the inclusion of a constant phase term does not affect the focusing behavior.
As the separability described in~\eqref{eq:SeparableBasis} highlights, the application of the phase profile along the transverse radial coordinate $\rhot$ does not alter the helical phase front, thus maintaining orthogonality at both the transmitter and receiver sides.
Indeed, any variation of the phase profile along the radial coordinate $\rhot$ does not entail any loss in terms of beams' separation. A limitation of this method is that the distance between the antennas must be known at the transmitting LIS side in order to properly focus the EM field. Nevertheless, this strategy requires phase-tapering only, thus leading to a simple implementation. As a consequence, the overall system complexity in the presence of focusing is in any case reduced with respect to the optimum strategy in Sec.~\ref{Sec:OAM}.B.

When the phase profile~\eqref{eq:focusing} is adopted, at the receiving LIS side the resulting EM field can be expressed as
\begin{align}\label{eq:FocOAMrx}
\psi_{n}(\rhor,\phir)&\simeq \frac{1}{4 \pi z \rt \sqrt{\pi}} e^{-\jmath\kappa\frac{\rhor^2}{2z}}\int_0^{\rt}\int_0^{2\pi}  e^{\jmath \elln\phit}e^{\jmath\kappa\frac{\rhor\rhot\cos{(\phir-\phit)}}{z}} \rhot\,\text{d}\phit \,\text{d}\rhot \nonumber\\
&=\frac{(-1)^{\elln}}{2z \rt \sqrt{\pi }}
e^{-\jmath\kappa\frac{\rhor^2}{2z}} e^{\jmath \elln \phir} \int_0^{\rt}  \rhot J_{\elln}\left(\frac{\kappa \rhor\rhot}{z}\right)\,\text{d}\rhot \, .
\end{align}
Interestingly, the expression in~\eqref{eq:FocOAMrx} corresponds to the beam that would be obtained with the far-field (Fraunhofer) approximation, even though the field is being observed within the Fresnel zone of the transmitting LIS \cite{DecDar:J21,BjoEtAl:J21}. Indeed, by focusing along $\rhot$, the quadratic phase term responsible for the diffraction effects typical of the Fresnel zone is compensated, hence resulting in a more concentrated beam at the focal point. Notice that, when $\elln=0$ (i.e., $n=1$), we have
\begin{align}\label{eq:FocOAMrx2}
\psi_{1}(\rhor,\phir)=& \frac{1}{2z \rt \sqrt{\pi}} e^{-\jmath\kappa\frac{\rhor^2}{2z}}  \int_0^{\rt} \rhot  J_0\left(\frac{\kappa \rhor\rhot}{z}\right)\,\text{d}\rhot \nonumber\\
=&  \frac{1}{2z\sqrt{\pi}} e^{-\jmath\kappa\frac{\rhor^2}{2z}}J_1\left(\frac{\kappa \rhor \rt}{z}\right)\, .
\end{align}
The intensity profile corresponding to~\eqref{eq:FocOAMrx2} is the classical far-field pattern of a circular aperture antenna (i.e., the \textit{Airy disk} \cite{Bal:B15}). Remarkably, higher topological charges produce a beam divergence partially compensated by focusing, as it will be investigated in the following numerical example. 

\subsection{Numerical Example}

\begin{figure}[h]
  \centering
  \begin{subfigure}[b]{0.5\linewidth}
    \includegraphics[width=\linewidth]{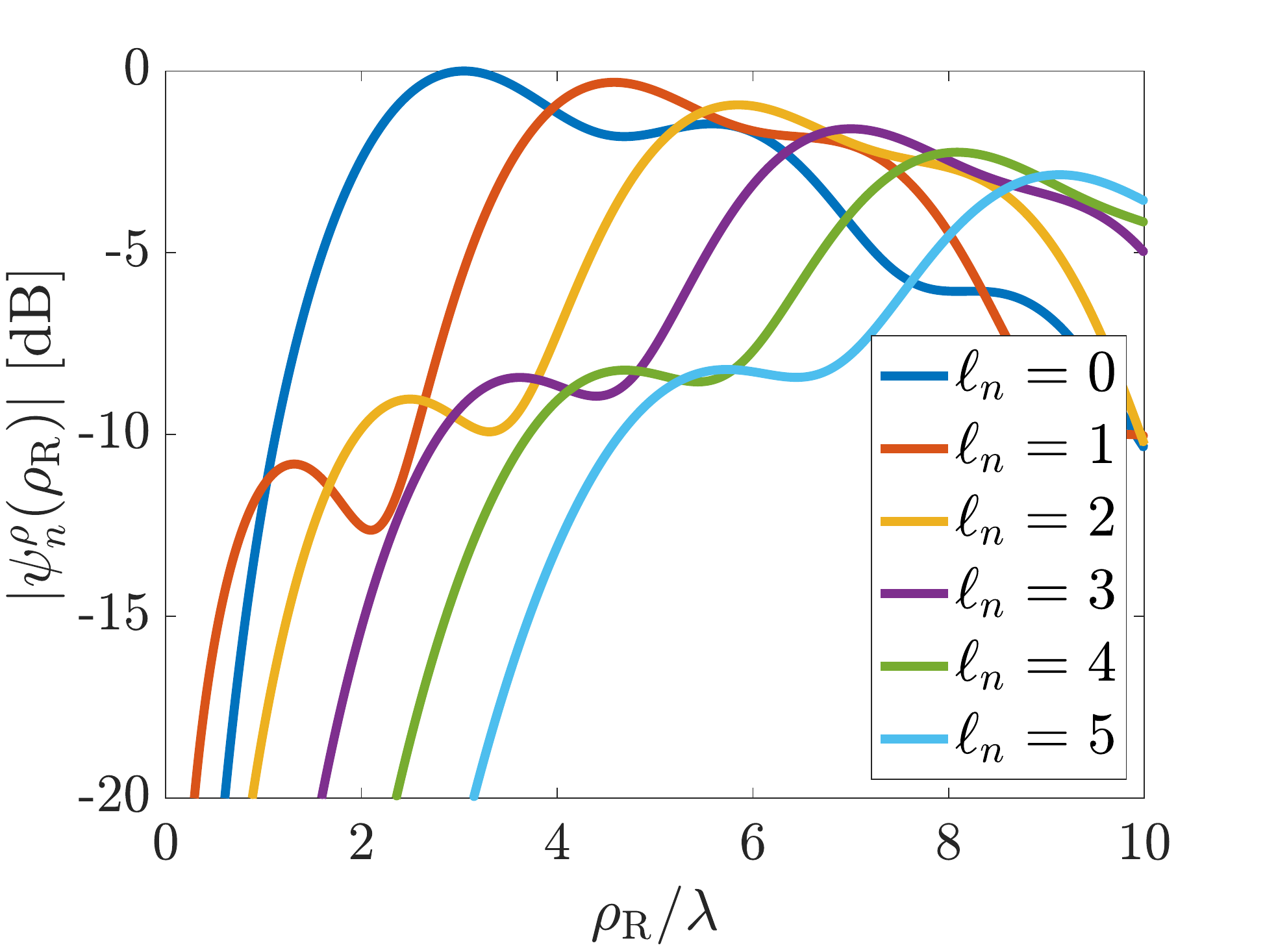}
    \caption{\textit{Unfocused case}}
    \label{fig:OAMRX}
  \end{subfigure}%
  \begin{subfigure}[b]{0.5\linewidth}
    \includegraphics[width=\linewidth]{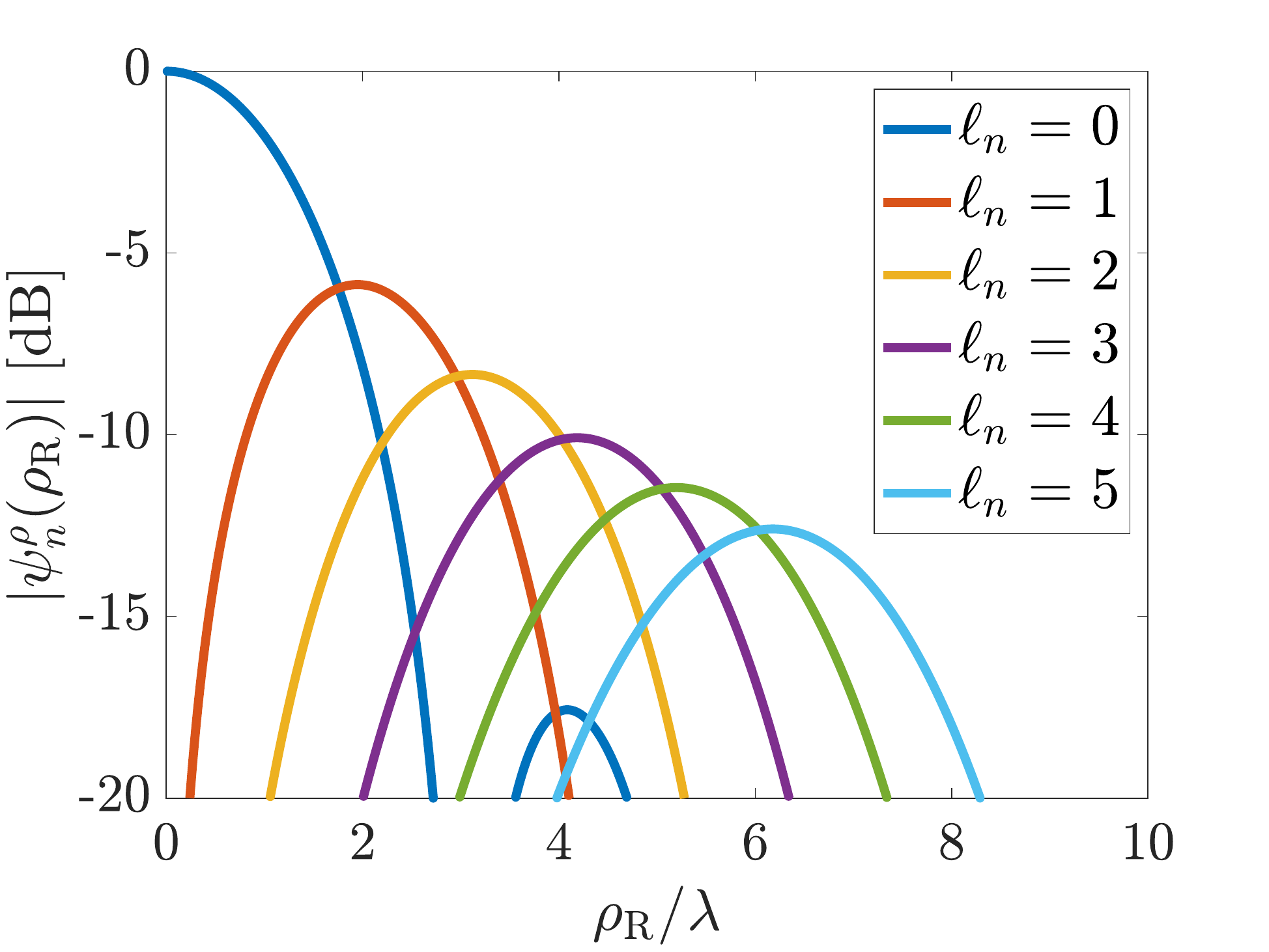}
    \caption{\textit{Focused case}}
    \label{fig:OAMRXfoc}
  \end{subfigure}
  \caption{Received EM field  along the radial coordinate as a function of the ratio $\rhor / \lambda$ for different topological charges $\ell_n$ (a) in the absence and (b) in the presence of a focusing transmitting LIS (normalized to $\operatorname{max}\{|\psi_1^{\rho}(\rho_{\mathrm{R}})|\})$.}
\end{figure}

As an example, let us consider two parallel, circular \acp{LIS} placed at a distance $z=D\lambda$.
The transmitting LIS has a radius $\rt=T\lambda$, while the receiving LIS has a radius $\rr=R\lambda$. Fig.~\ref{fig:OAMRX} illustrates the normalized amplitude of the received electric field's radial component $\psi_{n}^\rho(\rhor)$, where $\psi_{n}(\rhor,\phir)=\psi_{n}^\varphi(\phir)\psi_{n}^\rho(\rhor)$ and $\psi_{n}^\varphi(\phir)=e^{\jmath \ell_n \phir}$, as a function of the ratio $\rhor / \lambda$ for the case $R=T=10$ and $D=50$. It can be observed that the beam divergence increases with higher values of the topological charge $\ell_n$.  Furthermore, the beam widening effect, which is a characteristic of the received \ac{EM} field within the Fresnel zone, is also visible. In fact, within the chosen numerical setup, the boundary between the near-field and far-field (i.e., the Fraunhofer distance $d_{\mathrm{ff}}$) is in our case \cite{Bal:B15} 
\begin{equation}
d_{\mathrm{ff}}=\frac{8\rt^2}{\lambda}
\end{equation}
and it can be written in terms of a multiple of the wavelength $\lambda$ as $D_{\mathrm{ff}}=d_{\mathrm{ff}}/\lambda=8T^2$. It should be noted that  $D_{\mathrm{ff}}> D$ in the suggested numerical example.

In contrast, when a focusing transmitting LIS side is adopted according to \eqref{eq:focusing} with the same system configuration, the amplitude of the radial component of the normalized received electric field is that as reported in Fig.~\ref{fig:OAMRXfoc}. The focusing effect is clearly visible, producing much more concentrated beams, thus maximizing the energy collected by the receiving \ac{LIS}. 
As it is possible to notice from the figure, each beam is characterized by a different level of divergence, increasing with the mode order. 
Additional examples of the amplitude and phase profiles obtained both at the transmitting and receiving \ac{LIS} for different topological charges are reported in Appendix~\ref{app:OAMprofiles}, either in the presence or in the absence of focusing at the transmitting LIS.

\section{Detection Strategies}\label{sec:DetectionStrategies}
When $N$ data streams are transmitted simultaneously to carry the symbols $\boldx=\{x_1,\ldots, x_N\}$, i.e., spatially multiplexed according to the basis functions $\phi_n(\rhor,\phir)$ in~\eqref{basis_OAM1}-\eqref{basis_OAM2}, the information-carrying \ac{EM} field resulting from the modes' superposition at a distance $z$ (i.e., at the receiving LIS) is given by
\begin{align}\label{eq:signalmodel}
y(\rhor,\phir) = \psi(\rhor,\phir) + n(\rhor,\phir)=\sum_{n=1}^{N}x_n\psi_{n}(\rhor,\phir) + n(\rhor,\phir)
\end{align}
where $y(\rhor,\phir)$ identifies the overall received \ac{EM} field, the first addend represents the useful \ac{EM} field carrying the data, and $n(\rhor,\phir)$ denotes a  noise  field affecting the \ac{OAM}-based system. In the absence of further assumptions (e.g., regarding electromagnetic interference \cite{deJesetAl:J21}), the noise field, when expressed in Cartesian coordinates, is considered as \ac{AWGN} within the useful signal's spatial bandwidth, as further characterized in Appendix~\ref{app:AWGNNoise}.

The total energy received per symbol is
\begin{equation}\label{eq:ESdef} 
\Es = \int_{0}^{\rr} \int_{0}^{2\pi}  \left| \psi(\rhor,\phir)\right|^2 \rhor \,\text{d}\phir \,\text{d}\rhor
\end{equation}
so that we define the \ac{SNR} as $\mathsf{SNR}=\Es/N_0$. 
Consequently, the \ac{SNR} associated with the $n$-th \ac{OAM} mode can be written as
\begin{equation}\label{eq:SNRdef}
\mathsf{SNR}_{n} = \mathsf{SNR} \, \frac{E_n}{\Es}
\end{equation}
being $E_n=\int_0^{2\pi}\int_0^{\rhor}\left|\psi_{n}(\rhor,\phir)\right|^2\rhor \,\text{d}\phir \,\text{d}\rhor$ the energy of the $n$-th received \ac{OAM} mode. In addition, we define the parameter $\eta$ as the ratio between the overall received energy and the transmitted energy as 
\begin{align}\label{eq:eta}
    \eta &= \frac{\Es}{\int_{0}^{\rt} \int_{0}^{2\pi} \left| \sum_{n=1}^{N} \phi_n(\rhot,\phit) \right|^2 \rhot  \,\text{d}\phit \,\text{d}\rhot} =\frac{\Es}{\Et}\, . 
\end{align}
This quantity is conceptually equivalent to the inverse of the free-space path loss (i.e., path gain), and it will be further discussed in Sec.~\ref{Sec:Results}.

\subsection{OAM Demultiplexing}\label{sec:OAMdemux}

At the receiving \ac{LIS} side, the overall \ac{EM} field $y(\rhor,\phir)$ is processed to extract the information data $\boldx$. Remarkably, thanks to the helical wavefront propagating with no changes along the azimuthal coordinate, a correlation at the receiving LIS side with a conjugate factor $e^{-\jmath \elln \phir}$ enables to isolate the $n$-th contribution $\psi_{n}(\rhor,\phir)$ from the total received field  in~\eqref{eq:signalmodel}, thus performing \ac{OAM} modes' demultiplexing. Formally, it holds
\begin{align}\label{eq:RXbranch}
 y_n(\rhor) &= \int_{0}^{2\pi}y(\rhor,\phir) e^{-\jmath \elln \phir}\,\text{d}\phir \nonumber \\
 &=  \int_{0}^{2\pi}\sum_{n=1}^{N}x_n\psi_{n}(\rhor,\phir)
 e^{-\jmath \elln \phir}\,\text{d}\phir +\int_{0}^{2\pi} n(\rhor,\phir) e^{-\jmath \elln \phir}\,\text{d}\phir \nonumber \\
 &= x_n \psi_{n}^\rho(\rhor)   + n_n(\rhor)\, 
\end{align}
where
\begin{align}\label{eq:fn}
 \psi_{n}^\rho(\rhor)= \frac{(-1)^{\elln}}{2z \rt \sqrt{\pi}}
e^{-\jmath\kappa\frac{\rhor^2}{2z}} e^{-\jmath \kappa z}\int_0^{\rt}  \rhot e^{-\jmath\kappa\frac{\rhot^2}{2z}} J_{\elln}\left(\frac{\kappa \rhor\rhot}{z}\right)\,\text{d}\rhot\, 
\end{align}
when focusing is not performed the transmitting LIS side and
\begin{align}\label{eq:fnfoc}
\psi_{n}^\rho(\rhor)= \frac{(-1)^{\elln}}{2z \rt \sqrt{\pi}}
e^{-\jmath\kappa\frac{\rhor^2}{2z}} e^{-\jmath \kappa z}\int_0^{\rt}  \rhot  J_{\elln}\left(\frac{\kappa \rhor\rhot}{z}\right)\,\text{d}\rhot\, 
\end{align}
when focusing is instead employed at the transmitting LIS. The function $n_n(\rhor)$ denotes the radial noise portion after demultiplexing affecting the $n$-th \ac{OAM} mode that is being detected. 
\begin{figure}[t!]
    \centering
    \includegraphics[width=0.7\linewidth,trim={6mm 5mm 0mm 5mm}, clip, keepaspectratio]{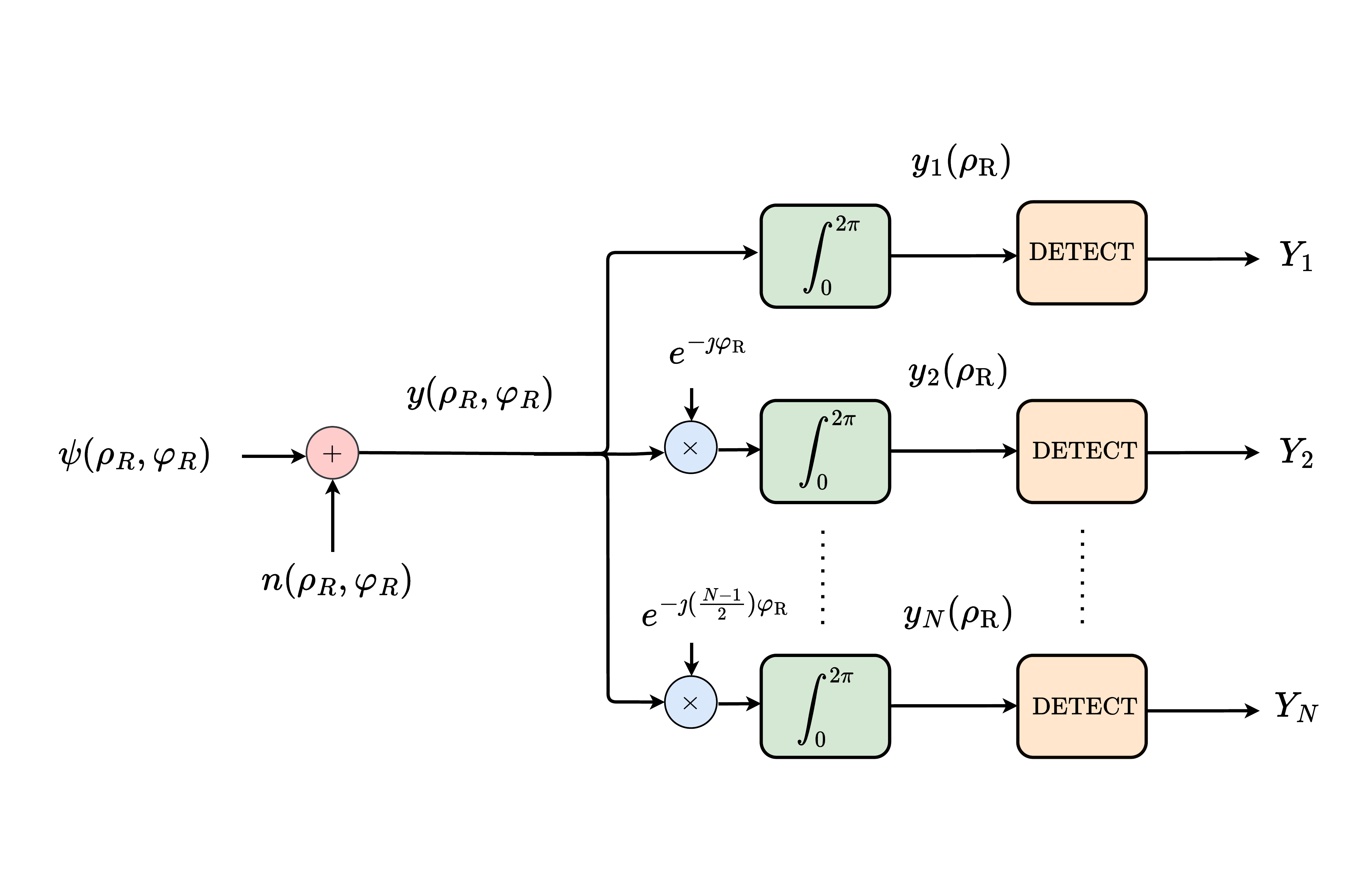}
    \caption{Equivalent block diagram of the OAM modes demultiplexing.}
    \hfill
    \label{fig:noisyRXscheme}
\end{figure}
Then, different approaches can be employed to build a scalar decision variable $Y_n$ by processing the signal over the radial component $\rhor$ of the \ac{LIS}, thus detecting the transmitted symbol $x_n$ from the $n$-th EM field component. An overview of the equivalent receiver block diagram performing OAM demultiplexing is reported in Fig.~\ref{fig:noisyRXscheme}. In the following, we illustrate three different detection strategies corresponding to different levels of implementation complexity and performance.

\subsection{Matched Filter}

The optimum strategy for maximizing the energy at the $n$-th branch of the receiver is that of performing matched filtering, i.e., computing the correlation also along the radial coordinate $\rhor$ using a template function $f_{n}(\rhor)$ matched to \eqref{eq:fn} or \eqref{eq:fnfoc}, i.e., $ f_{n}(\rhor)= \psi_{n}^\rho(\rhor)$. Notably, due to the assumption of spatially white noise and the adoption of a polar reference system, the correlation-based reception scheme requires the introduction of the determinant of the Jacobian matrix associated with the Cartesian to polar coordinates' transformation as in \eqref{eq:normalization}-\eqref{eq:OAMrx1}, that is, $\rhor$. Thus, the decision variable $Y_n^{\text{(MF)}}$ used for signal demodulation at the $n$-th branch is
\begin{align}\label{eq:MF}
    Y_n^{\text{(MF)}} & = \int_{0}^{\rr} y_n(\rhor) \,  f_{n}^*(\rhor)  
   \rhor\,\text{d}\rhor    \nonumber \\
   & = x_n \int_{0}^{\rr} \left|\psi_{n}^\rho(\rhor)\right|^2 \rhor\,\text{d}\rhor + \int_{0}^{\rr} n_n(\rhor) f_{n}^*(\rhor) \rhor \,\text{d}\rhor \nonumber \\
    &= x_n h_n^{\text{(MF)}} + w_n
\end{align}
where $h_n^{\text{(MF)}}$ denotes the $n$-th channel coefficient describing the effects of \ac{OAM} \ac{LOS} propagation. Precisely, it holds $h_n^{\text{(MF)}}=E_n$, with
\begin{align} \label{eq:h-coef_MF_nofoc}
E_n = \frac{1}{4 \pi z^2 \rt^2} \int_{0}^{\rr}\!\left| \int_0^{\rt} \rhot e^{-\jmath\kappa\frac{\rhot^2}{2z}} J_{\elln}\left(\frac{\kappa \rhor\rhot}{z}\right)\text{d}\rhot\right|^2  \,\rhor\text{d}\rhor
\end{align}
being the received energy related to the $n$-th OAM mode when focusing is not performed by the transmitting LIS, and
\begin{align}\label{eq:h-coef_MF_foc}
E_n = \frac{1}{4 \pi z^2 \rt^2} \int_{0}^{\rr}\left| \int_0^{\rt} \rhot  J_{\elln}\left(\frac{\kappa \rhor\rhot}{z}\right)\text{d}\rhot\right|^2  \rhor\,\text{d}\rhor
\end{align}
when accounting for the presence of focusing instead. In addition, $w_{n}$ denotes the noise component affecting the decision variable at the output of the $n$-th correlation-based device. We have that, $w_n \sim \mathcal{CN}(0,\, \sigma_{w, \text{MF}}^2)$, $n=1,2, \ldots, N$, are \ac{i.i.d.} circularly-symmetric complex Gaussian random variables, whose variance is derived in Appendix~\ref{app:AWGNNoise}.
\begin{figure}[t]
\centering
 \includegraphics[trim={2mm 2cm 0mm 3cm},clip, width=0.65\columnwidth]{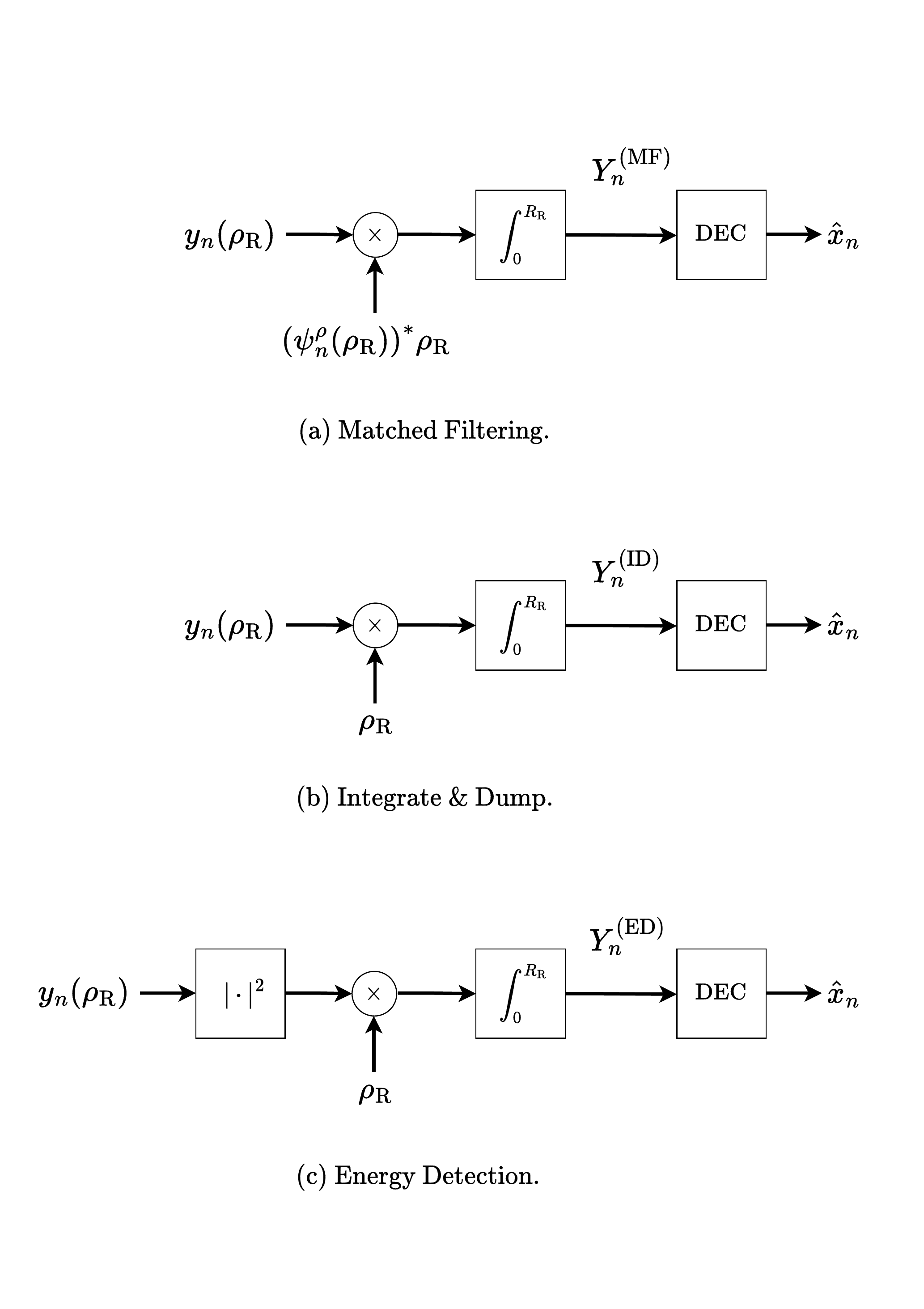}
\caption{Proposed radial processing schemes for symbol detection after OAM modes demultiplexing.}
\label{fig:Schemes}
\end{figure}
Unfortunately, such a strategy, whose equivalent processing scheme is reported in Fig.~\ref{fig:Schemes}a, despite maximizing the \ac{SNR} at its output, requires high flexibility at the receiving \ac{LIS} side. In fact:
\begin{itemize}
    \item The spatial correlation with a complex amplitude/phase pattern must be performed;
    \item The distance between the transmitter and receiver, as well the transmitting \ac{LIS}'s size, must be known at the receiver side to build the template function $f_n(\rhor)$.
\end{itemize}

\subsection{Integrate \& Dump} 

Considering that the total received field comprises a set of well-defined beams, especially in the focused case (as in Fig.~\ref{fig:OAMRXfoc}), a lower-complexity detection scheme could be achieved by means of a simple \ac{ID} approach (i.e., accumulation over the radial coordinate $\rhor$), thus obviating the need for point-wise processing along $\rhor$ involving a complex amplitude and phase spatial correlation. 

Specifically, we have
\begin{align}\label{eq:Acc}
Y_n^{\text{(ID)}} &=\int_{0}^{\rr} y_n(\rhor) \rhor \,\text{d}\rhor \nonumber \\ &= x_n \int_{0}^{\rr} \psi_n^{\rho}(\rhor)\rhor\,\text{d}\rhor  +  \int_{0}^{\rr} n_n(\rhor) \rhor \,\text{d}\rhor \nonumber \\
&= x_n h_n^{\text{(ID)}}+w_n^{\text{(ID)}}
\end{align}
with $h_n^{\text{(ID)}}$ being the complex channel coefficent, and $w_n^{\text{(ID)}} \sim \mathcal{CN}(0,\, \sigma_{w, \text{ID}}^2)$, $n=1,\ldots, N$, being the Gaussian noise variables with variance $\sigma_{w, \text{ID}}^2$, derived in Appendix~\ref{app:AWGNNoise}.
This strategy corresponds to assuming $ f_{n}(\rhor)= \Pi_{\rr}(\rhor)$, hence integrating on the totality of the receiving radial dimension.  The equivalent processing scheme is reported in Fig.~\ref{fig:Schemes}b.
It is worth noticing that, in this case, differently from the  \ac{MF} strategy, the lack of the square module operation in~\eqref{eq:Acc}, which is instead present in~\eqref{eq:MF}, makes the complex output variable $Y_n^{\text{(ID)}} $ dependent in its sign on the positive/negative result of the integral of the radial component $\psi_n(\rhor)$. Thus, the final decision criterion must account for this. Moreover, channel equalization is required to compensate for the phase term in the decision variable.

\subsection{Energy Detection}

In order to further simplify the receiver structure, a partially non-coherent scheme can be considered, by collecting the energy along the radial component $\rhor$ exploiting a square-law device, thus leading to an \ac{ED} scheme. After demultiplexing the $n$-th \ac{OAM} mode as previously described, the square module is then computed and, subsequently, the integration in the radial direction is performed to obtain the scalar decision variable $Y_n^{\text{(ED)}}$, i.e.,
\begin{align}\label{eq:ED}
    Y_n^{\text{(ED)}} &= \int_{0}^{\rr}  \left| y_n(\rhor) \right|^2  \rhor\,\text{d}\rhor \nonumber\\
    &= \int_{0}^{\rr}  \left| x_n \psi_n^{\rho}(\rhor)   +n_n(\rhor)  \right|^2  \rhor\,\text{d}\rhor .
\end{align}
The equivalent processing scheme is reported in Fig.~\ref{fig:Schemes}c. Conversely to the other detection strategies, energy detection cannot be realized with the general communication scheme of Fig.~\ref{fig:Scenario}, since integration over the angular coordinate $\phir$ only is required and, afterward, non-linear processing along the radial coordinate $\rhor$ is performed. 

Differently from the \ac{MF} and the \ac{ID} schemes, this detection strategy introduces some additional constraints to be considered. For instance, due to the presence of the square law device in the reception scheme, any information regarding the signal's phase is lost, thus requiring non-coherent modulation schemes such as \ac{OOK}, as we will assume in the following. This particular choice leads to two possible outcomes: a first case in which the decision variable $Y_n^{\text{(ED)}}$ is composed of the noise energy only, and a second one in which $Y_n^{\text{(ED)}}$ comprises both the transmitted signal and the noise energies. The final symbol's decision can be performed by comparing the energy level measured at the radial integration block's output, i.e., $Y_n^{\text{(ED)}}$, with an appropriate threshold value. Therefore, the decision criterion becomes
\begin{equation}\label{eq:DecisionCritED}
    \hat{x}_n =
        \begin{cases} +1,  &Y_n^{\text{(ED)}} \geq \zeta_n \\
                      0,  & Y_n^{\text{(ED)}} < \zeta_n
        \end{cases}  
\end{equation}
where $\zeta_n$ identifies the decision threshold that has been suitably defined for the $n$-th \ac{OAM} mode. In Sec.~\ref{Sec:Results}, appropriate threshold values for $\zeta_n$, $n = 1,\ldots,N$, will be numerically evaluated for a given \ac{SNR} at the receiver.\footnote{The definition of a proper threshold can be avoided by considering different signaling schemes, such as pulse position modulation, at the expense of the spectral efficiency \cite{DecetAl:J14}.} 

\subsection{Smart Integration}\label{sec:SmartInteg}

In Sec.~\ref{sec:OAMdemux}, it has emerged how, after performing a first correlation of the total received \ac{EM} field with the conjugate phase factor $e^{-\jmath \elln \phir}$ of the mode of interest to be demultiplexed, integration in the radial variable $\rhor$ needs to be performed. 
When the \ac{MF} scheme is employed, this operation results in maximizing the \ac{SNR} at the receiving \ac{LIS}, being the selected receiver template functions matched to the received waveforms. 
However, when the \ac{ID} or \ac{ED} strategies are contemplated, the absence of a radial processing matched to the received signal $\psi_n^{\rho}(\rhor)$ might lead to a subtle deterioration in performance. This is because when integrating along the totality of the radial domain, i.e., $ \rhor \in [0, \rr]$, a consistent amount of noise is accumulated, which comes from those locations on the receiving LIS where the radial signal component $ \psi_n^{\rho}(\rhor)$ has negligible values, as shown in Figs.~\ref{fig:OAMRXfoc}. Indeed, both with focused or unfocused beams, the received field amplitude tends to be concentrated on a given radial interval, hence rapidly vanishing outside of it. 

To counteract this phenomenon, a viable solution is that of computing the integral only where the received \ac{EM} field has a significant intensity. This operation has been dubbed as \emph{smart integration} since we assume to have an intelligent receiving device that, given the knowledge of the OAM order that has been demultiplexed, can identify a suitable domain on which to perform the integration operation. 
The procedure is similar to the problem of integration time determination for non-coherent receivers, for which several solutions have been proposed in the literature \cite{DecetAl:J14, DecetAl:C11, NemetAl:C06}. Examples will be given in the numerical results of Sec.~\ref{Sec:Results}.

\section{Numerical Study}\label{Sec:Results}

\subsection{Degrees of Freedom}

\begin{figure}[t]
    \centering
    \includegraphics[width=0.6\columnwidth]{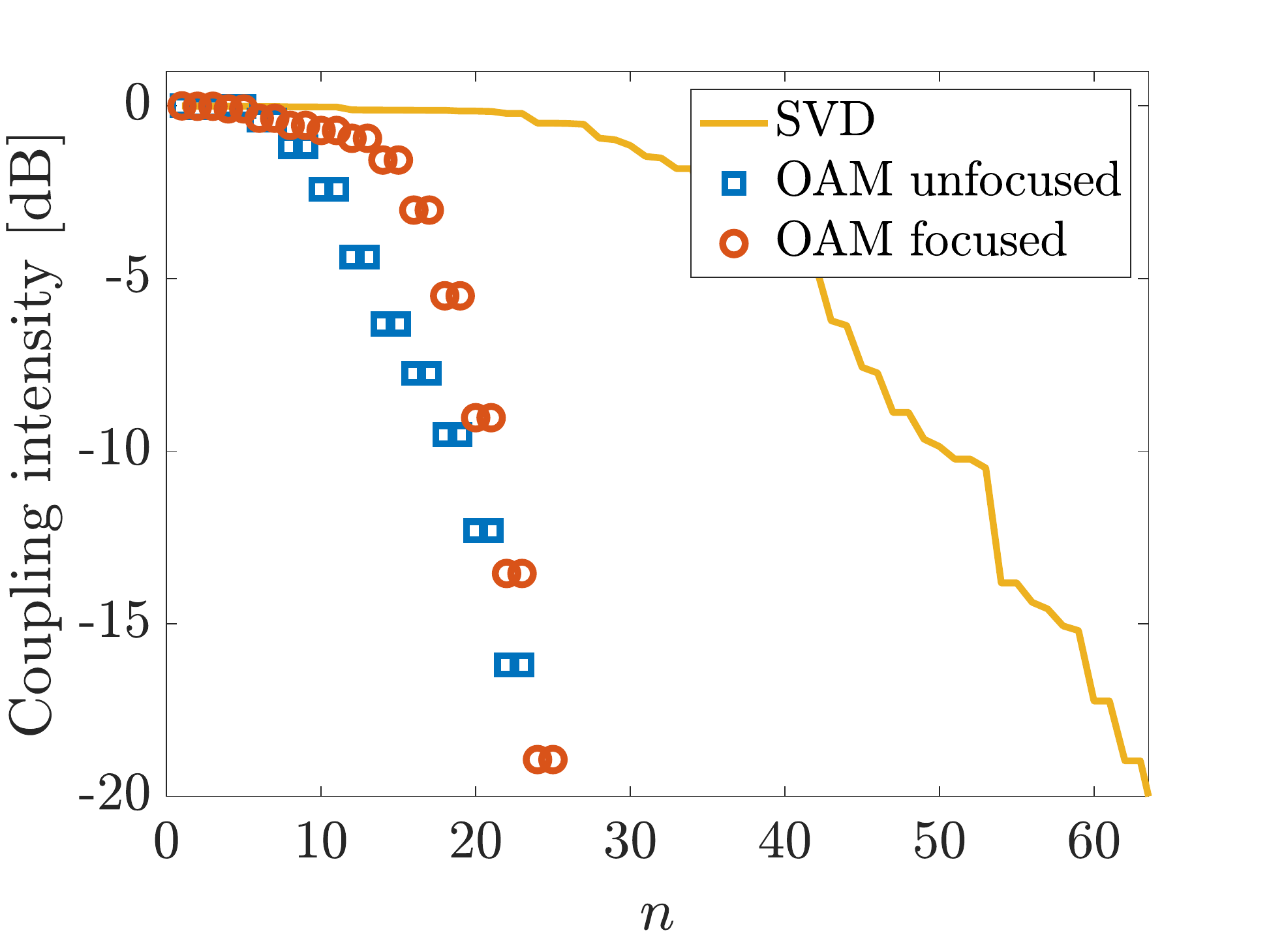}
    \caption{Coupling intensity for the unfocused and focused OAM modes in comparison with the coupling intensity of optimum communication modes (SVD).}
    \hfill
    \label{fig:SVD1}
\end{figure}

\begin{figure}
     \centering
     \begin{subfigure}[b]{0.48\linewidth}
         \centering
         \includegraphics[width=\linewidth]{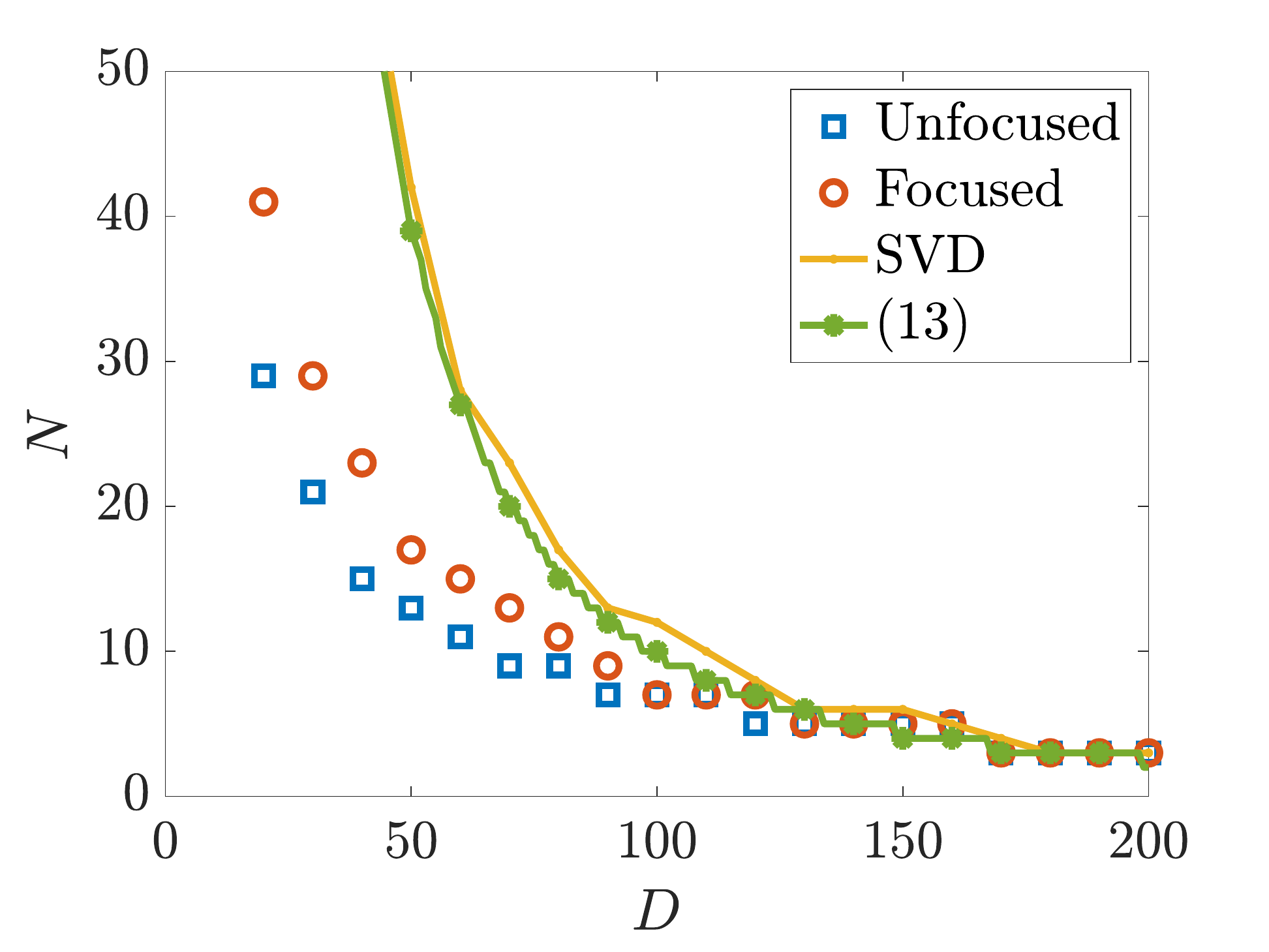}
         \caption{Antennas of the same size ($R=T=10$).}
         \label{fig:N_vs_z}
     \end{subfigure}
     \hfill
     \begin{subfigure}[b]{0.48\linewidth}
         \centering
         \includegraphics[width=\linewidth]{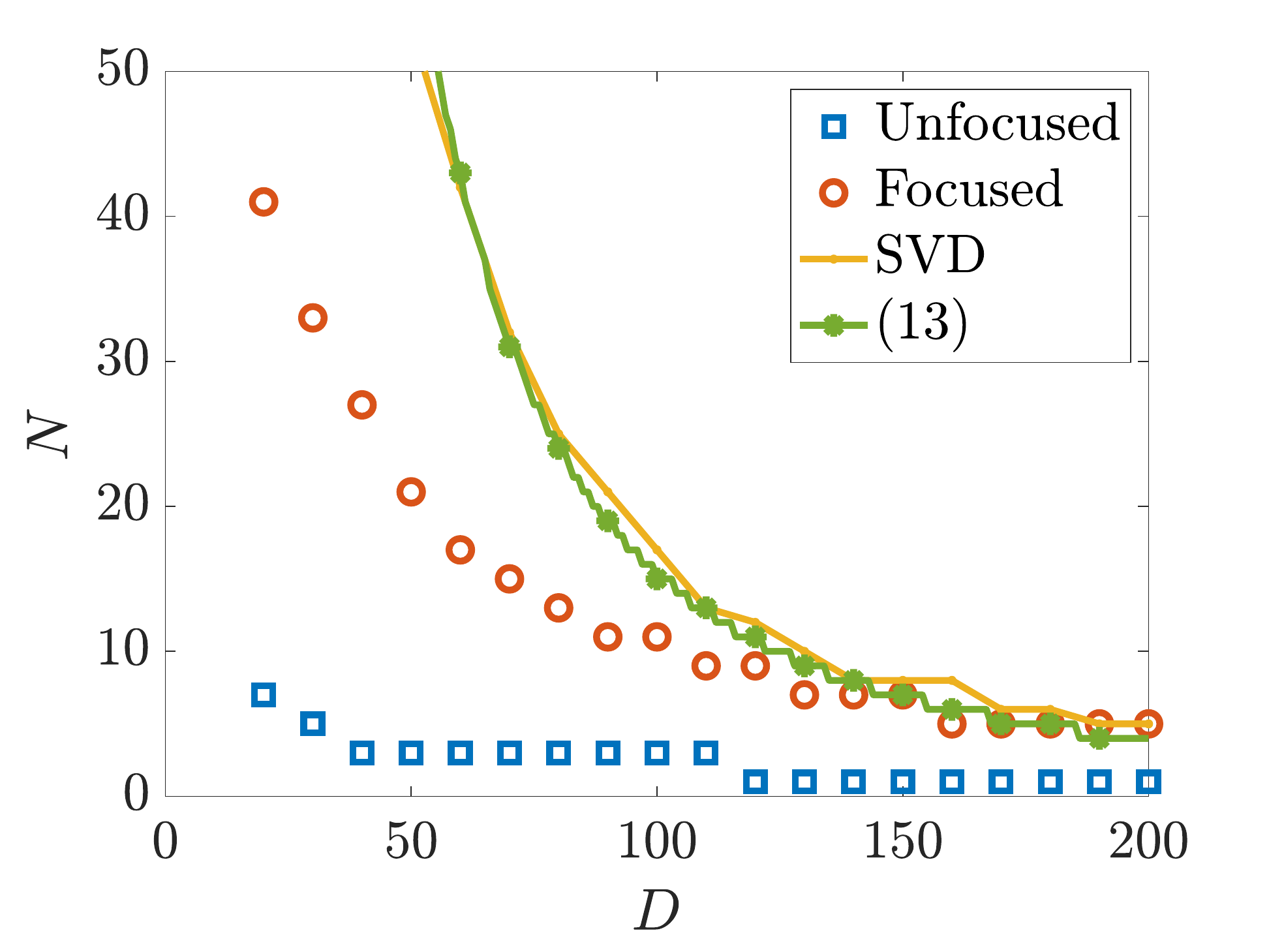}
         \caption{Downlink case ($R=5, T=25$).}
         \label{fig:N_vs_downlink}
     \end{subfigure}
     \hfill
     \begin{subfigure}[b]{0.48\linewidth}
         \centering
         \includegraphics[width=\linewidth]{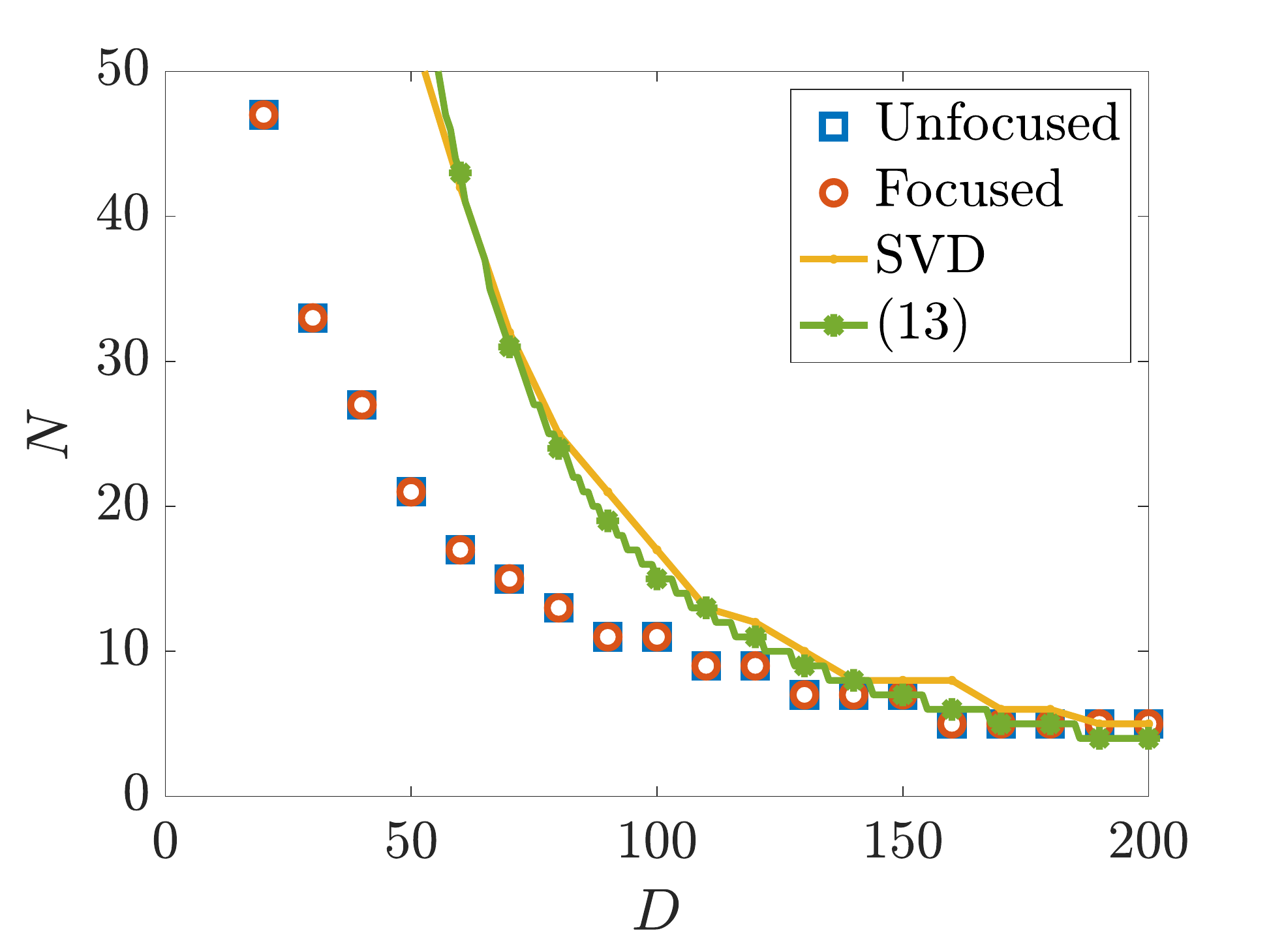}
         \caption{Uplink case ($R=25, T=5$).}
         \label{fig:N_vs_uplink}
     \end{subfigure}
        \caption{Number of OAM modes as a function of the link distance  $D$ (multiples of $\lambda$).}
        \label{fig:N_vs_Distance}
\end{figure}

To evaluate the effectiveness of the \ac{OAM} approach, we compare the coupling intensity obtained with \ac{OAM} modes to the coupling intensity obtained from optimum communication modes (i.e., the intensity of the singular values $\xi_n$ as from the optimum basis sets).
The optimal bases are obtained by discretizing the transmitting and receiving \acp{LIS} and computing the \ac{SVD} of the Green function~\eqref{eq:GreenFunction} between the surfaces. In Fig.~\ref{fig:SVD1}, the behavior of singular values is reported for the configuration $R=T=10$ and $D=50$, with $\lambda=0.1\,$m. It can be seen that they are almost constant, then drop quickly after a specific value. The number of effective communication modes $N$, i.e., the \ac{DoF}, can be defined as the number of singular values of intensity no smaller than a certain value with respect to the largest one.
The coupling intensity of the \ac{OAM} modes corresponds to the normalized values $E_n/E_1$, being the denominator coincident with the energy of the fundamental mode $n=1$ (i.e., $\ell_n=0$). As anticipated, OAM-based modes are degenerate, hence the same coupling is obtained for two consecutive indexes $n$, except for the case $n=1$. 
Due to the presence of beam divergence, OAM modes coupling falls off quite rapidly compared to the optimal case. However, it can be noticed how beneficial the adoption of focusing techniques is in terms of coupling strength.
In this setup, by fixing a threshold to $-5 \, \text{dB}$ from the best-connected communication mode, about 40 communication modes are obtained with SVD; differently, only 10 can be exploited with OAM and about 18 with OAM when performing focusing at the transmitting \ac{LIS}. Therefore, it is evident that the number of parallel channels is lower than that corresponding to the optimum strategy. However, system complexity is significantly reduced. In particular, the coupling intensity obtained for focused OAM-based communication modes with low topological charge is anyhow acceptable, so this could be reasonably exploited for enhancing the link capacity with small additional complexity in terms of hardware implementation. A detailed analysis of the trade-off between ease of deployment and the required level of system knowledge will be thoroughly provided in Sec.~\ref{sec:Discussion}.
In Fig.~\ref{fig:N_vs_z}, the number of modes $N$ is depicted as a function of the distance parameter $D$ for the scenario of $R=T=10$. For comparison, the number of communication modes obtained from~\eqref{eq:Nmodes} is also illustrated. In particular, it is convenient to rewrite \eqref{eq:Nmodes} as a function of the normalized quantities considered in the numerical results, which gives
\begin{equation}\label{eq:NmodesRes}
N=\frac{\pi^2 R^2 T^2}{D^2}  \, 
\end{equation}
and which depends only on geometrical quantities reported to the wavelength. Furthermore, the number of modes obtained by considering the 
singular values from the \ac{SVD} is also illustrated. In all cases, a threshold of $-5 \,\text{dB}$ is applied to determine the number of well-coupled modes.
As expected, regardless of the adopted strategy, the number of modes diminishes rapidly for increasing distance, reaching unity as the far-field limit is approached. The number of modes corresponding to significant singular values from the \ac{SVD} is accurately predicted by \eqref{eq:NmodesRes}. While the use of focused \ac{OAM}  results in a greater number of well-coupled modes, both the focused and unfocused \ac{OAM} transmissions have inferior performance compared to optimum communication modes when the distance is small. However, when the geometric setting does not allow for a large number of modes (e.g., for larger distances), \ac{OAM} can be used as a viable solution to exploit channel spatial multiplexing, with limited additional complexity and without the need to perform \ac{SVD}.

Furthermore, the following plots show what happens when the sizes of the \acp{LIS} greatly differ. In particular, in Fig.~\ref{fig:N_vs_downlink} \ac{LIS} antennas with $R=5$ and $T=25$ are considered. Since the transmitting antenna is much larger than the receiving one, we refer to this scenario as the downlink case. Thanks to the large dimensions of the transmitting \ac{LIS}, a high focusing gain can be obtained, thus drastically increasing the number of well-coupled OAM-based channels. Differently, if focusing is not adopted in this downlink case, the detrimental effect of beam divergence, exasperated by the large transmitting aperture, makes OAM-based strategies quite ineffective. 
The opposite case, corresponding to $R=25$ and $T=5$ (i.e., uplink case) is reported in Fig.~\ref{fig:N_vs_uplink}. In this case, the number of modes obtained with \ac{SVD} or corresponding to~\eqref{eq:NmodesRes} is equivalent to that from Fig.~\ref{fig:N_vs_downlink}, since it is related to optimum complete bases. Differently, when considering the simpler \ac{OAM}-based approach, the situation is asymmetric due to beam divergence. However, in this case, since the receiving \ac{LIS} is large enough to properly collect the received \ac{EM} energy for increasing topological charge values, a large number of \ac{OAM}-based channels can be realized, thus making the approach appealing. On the other hand, employing a focusing transmitter does not bring advantages due to the reduced size of the transmitting \ac{LIS}.

\begin{figure}[t]
    \centering
    \includegraphics[width=0.6\columnwidth]{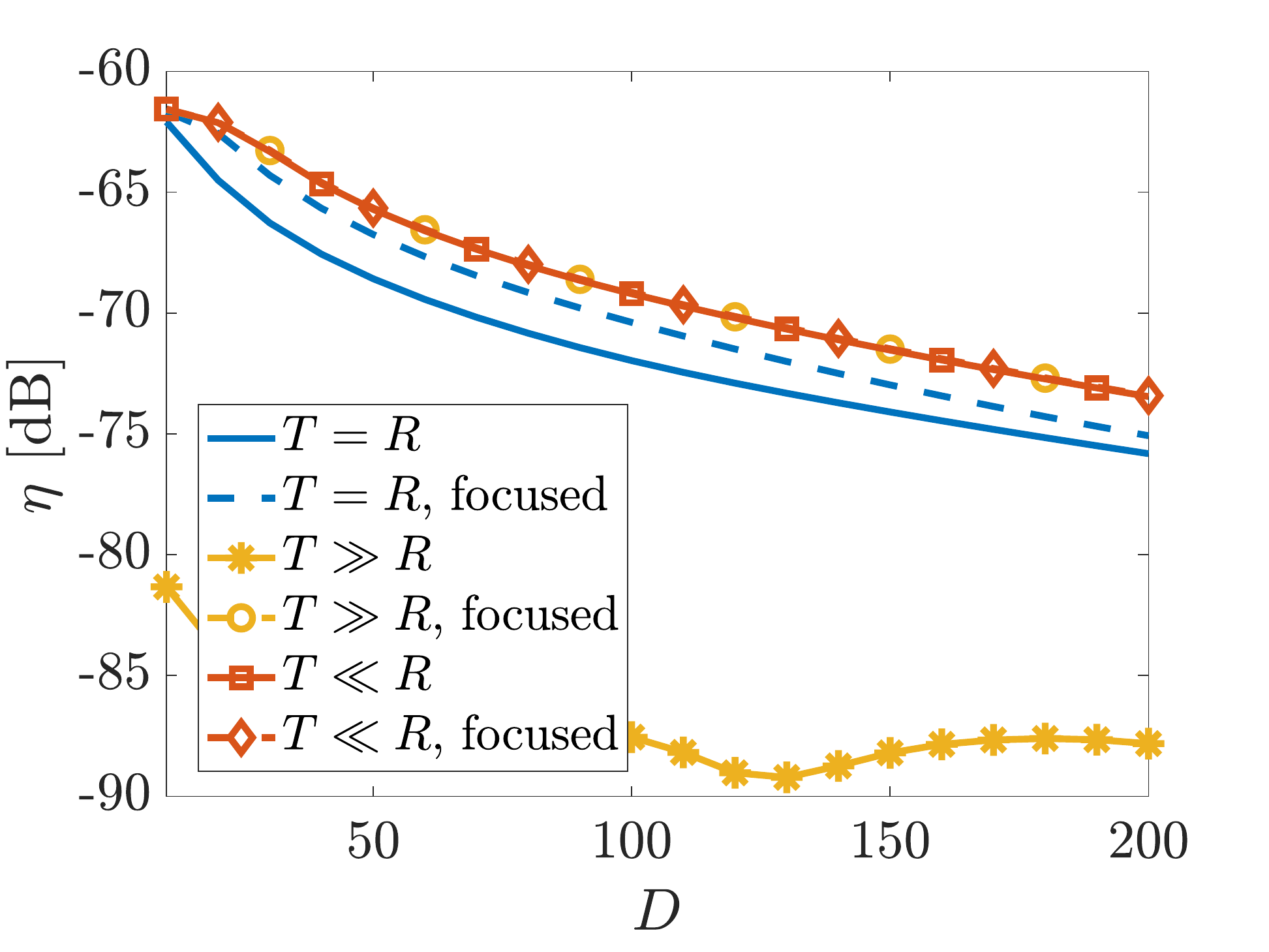}
    \caption{Path gain  $\eta$ as a function of the link distance $D$ (multiples of $\lambda$).}
    \label{fig:EtaPlot}
\end{figure}

\subsection{OAM Detection}
In this subsection, we characterize the \ac{OAM}-based communication scheme in terms of path gain and \ac{BER}. 
Fig.~\ref{fig:EtaPlot} reports the path gain $\eta$ for both the focused and unfocused cases as a function of the normalized link distance $D$. In particular, the same three configurations relative to Fig.~\ref{fig:N_vs_z}, Fig.~\ref{fig:N_vs_downlink}, and Fig.~\ref{fig:N_vs_uplink} are considered, corresponding to $T=R$, $T\gg R$ (i.e., downlink) and $T\ll R$ (i.e., uplink). Here we adopt $N=51$ to obtain the various curves since, given the information presented in Figs.~\ref{fig:N_vs_uplink}, it can be inferred that this value serves as an upper bound for the number of well-coupled OAM modes. When antennas with identical radii are assumed (case $T=R=10$), the path gain is more favorable with focused OAM, which does not impact the amount of transmitted energy but allows for concentrating a larger amount of it on the receiving LIS. The gap between the unfocused and focused case is clearly visible in downlink (i.e., $T=25$, $R=5$), where the large transmitting LIS produces a great divergence of the OAM modes. If such a divergence is not properly compensated by focusing, a significant performance loss in terms of energy collected by the receiver is experienced (e.g., around $17\,$dB at $D=100$). Differently, in uplink (i.e., $T=5$, $R=25$) no difference is experienced between the focused and unfocused cases. In this configuration, \textit{(i)} the small transmitting LIS is not capable of focusing the transmitted energy, \textit{(ii)} the OAM beam divergence is limited by the small transmitting LIS, and \textit{(iii)} the large receiving LIS is intrinsically more efficient in collecting the energy spread by the higher-order OAM modes.

\begin{figure}
\hspace{-0.6cm}
     \centering
     \begin{subfigure}[b]{0.48\linewidth}
         \centering
        \includegraphics[width=\linewidth]{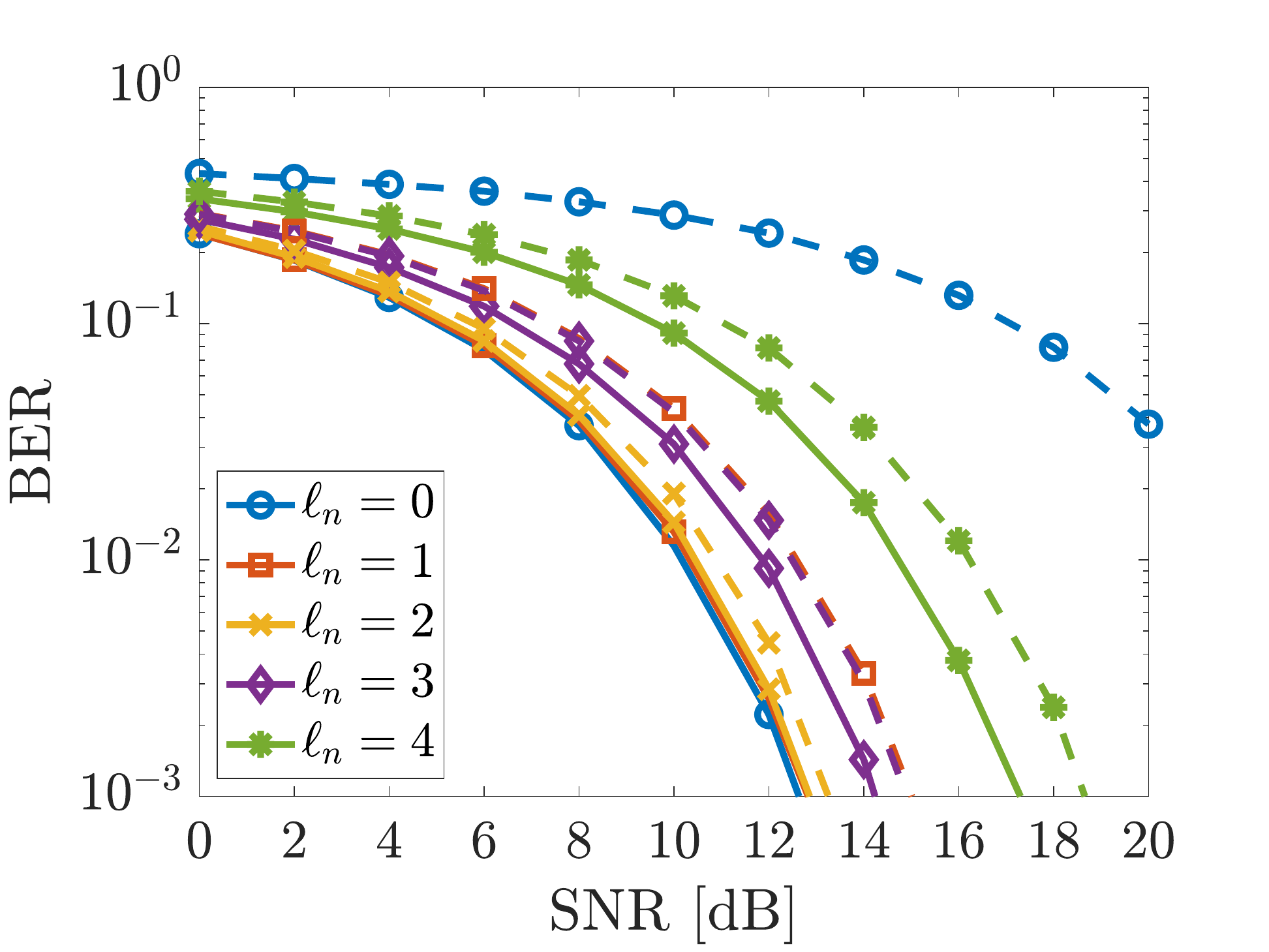}
    \caption{MF versus ID}
    \label{fig:BER_MF}
     \end{subfigure}
     \hfill
     \begin{subfigure}[b]{0.48\linewidth}
         \centering
         \includegraphics[width=\linewidth]{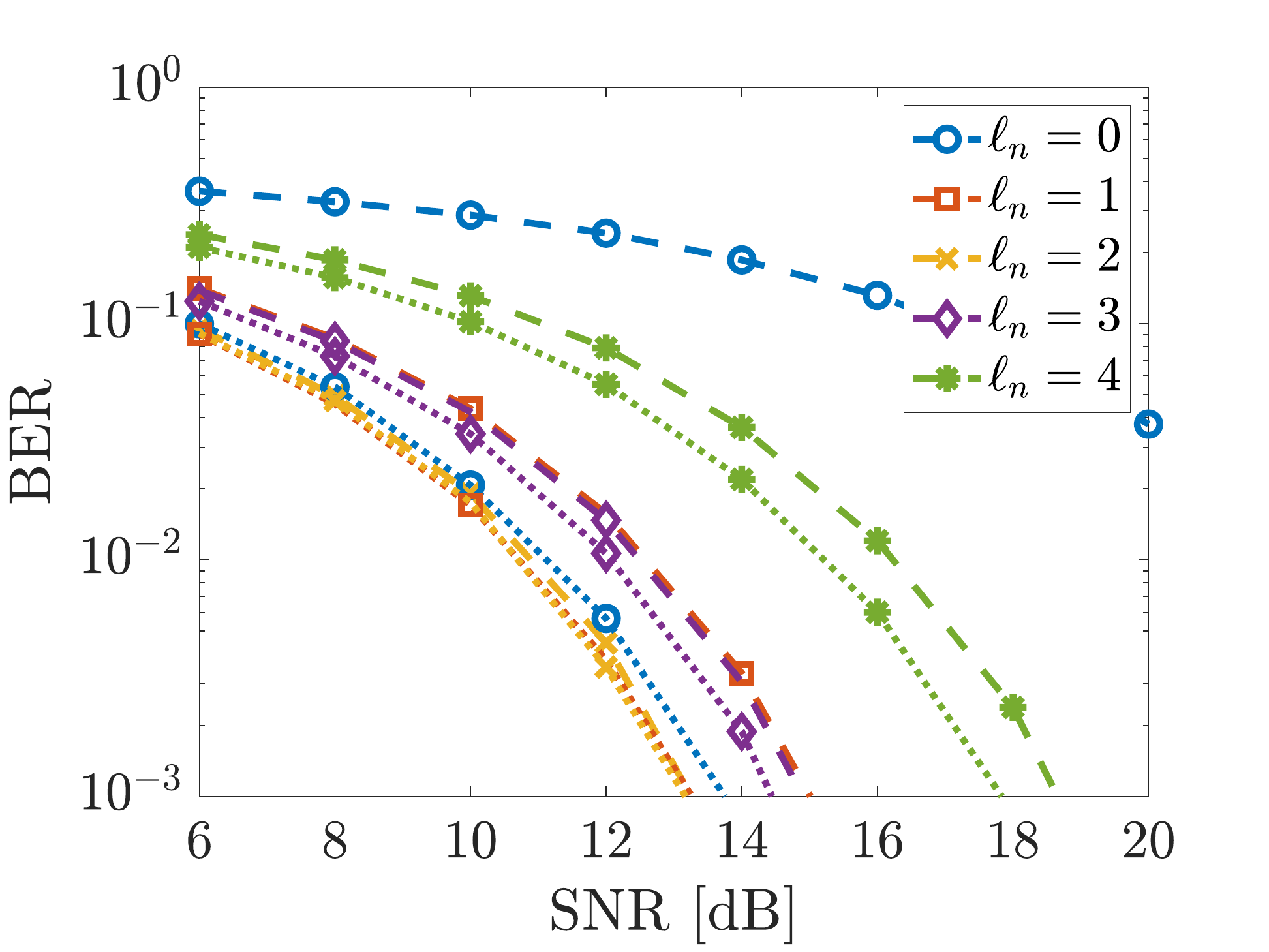}
     \caption{ID versus ID with smart integration.}
     \label{fig:BER_ID}
     \end{subfigure}
     \hfill
     \begin{subfigure}[b]{0.48\linewidth}
         \centering
         \includegraphics[width=\linewidth]{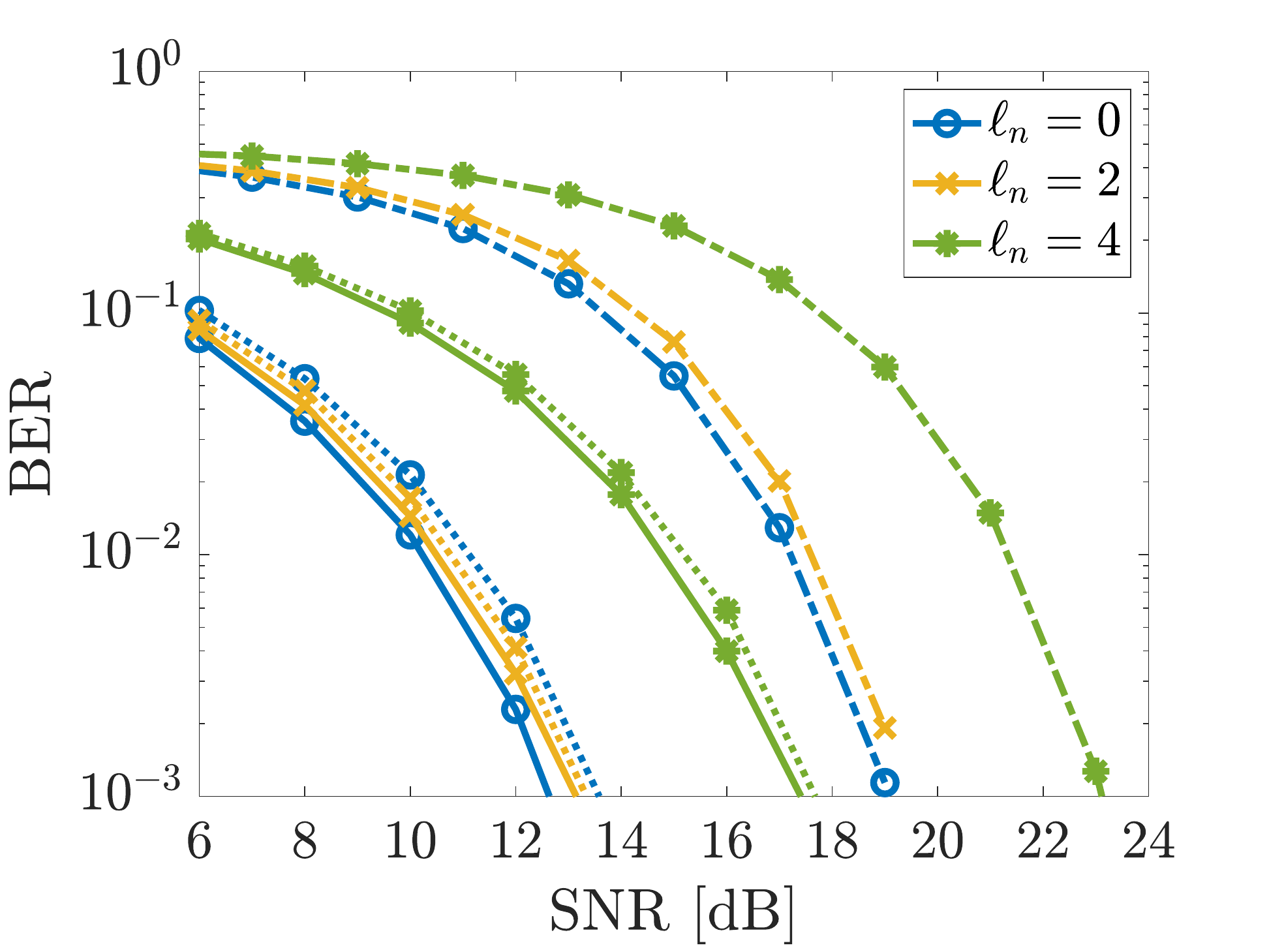}
    \caption{MF versus ID versus ED}
    \label{fig:BER_Comparison}
  \end{subfigure}
  \caption{BER as a function of the SNR for different OAM modes. Continuous lines ($-$) are for the MF; dashed lines ($-\,-$) are for the ID; dotted lines ($\cdot\cdot$) are for the ID (smart integration); dot-dashed lines ($\cdot-$) are for the ED (smart integration and optimum detection threshold).}
\end{figure}

\begin{figure}[t]
    \centering
    \includegraphics[width=0.6\columnwidth]{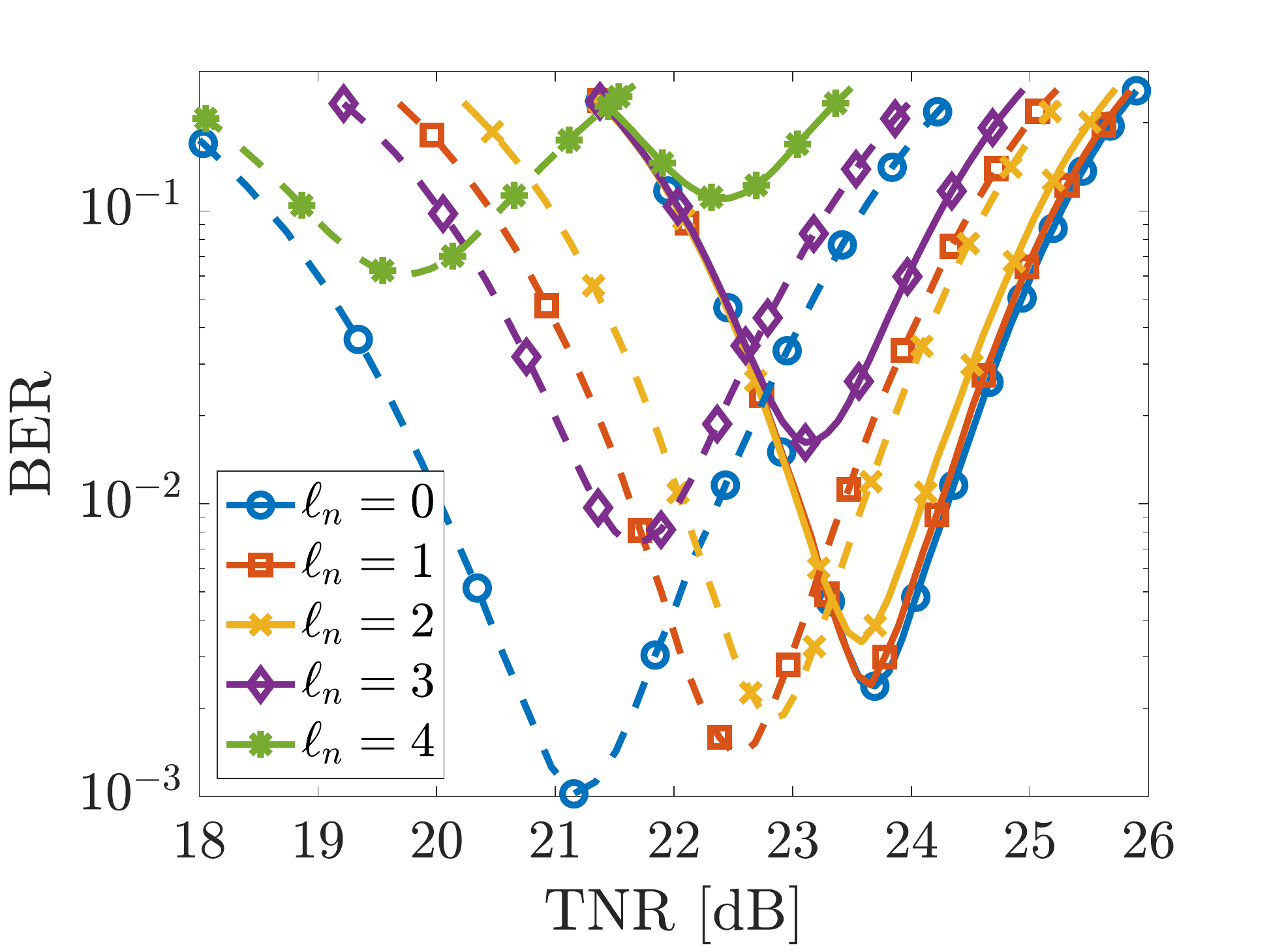}
    \caption{BER as a function of the TNR for different OAM modes and a \ac{SNR} of $19\,$dB. Continuous lines ($-$) are without smart integration; dashed lines ($-\,-$) are with smart integration.}
    \label{fig:BERvsTNR}
\end{figure}

Let us now concentrate on the performance of the different detection schemes that have been proposed.
Fig.~\ref{fig:BER_MF} depicts the \ac{BER} as a function of the \ac{SNR}, as defined in~\eqref{eq:SNRdef}, for $R=T=10$ and $D=100$ when focused OAM and a \ac{BPSK} modulation scheme are adopted. In particular, results are reported for a subset of OAM mode orders $\ell_n$, as indicated in the legend (i.e., modes with positive signs only are considered). Thanks to the \ac{MF} strategy, the continuous curves coincide with the bit error probability of an optimum uncoded BPSK system, considering the fraction of energy allocated to each OAM mode. In fact, as it is possible to notice, the best performance is experienced by the OAM modes with lower indexes, which are better-coupled thanks to their limited divergence. In particular, the OAM modes for $\ell_n=0,1,2$ experience almost the same coupling, thus resulting in a similar performance. In the same figure, the \ac{MF} performance is compared to that of the sub-optimum \ac{ID} scheme. This has been implemented considering the compensation of both the quadratic phase terms at the receiving LIS and the additional phase terms with an ideal single-tap equalizer. It can be noticed that, in this case, the performance loss of the \ac{ID} scheme can change drastically depending on the OAM mode order. In particular, for modes with index $\ell_n=2,3,4$, there is only a slight decrease in performance (below approximately $1\,$dB), while the fundamental mode for $\ell_n=0$ is severely impacted. This is due to the diverse waveform shapes associated with the various OAM modes and the adoption of the sub-optimum ID strategy, thus making the performance dependent on the specific signal shape, in contrast to the \ac{MF} method where the performance is independent of the signal shape.
In Fig.~\ref{fig:BER_ID}, the impact of the smart integration strategy for the \ac{ID} detection, as described in Sec.~\ref{sec:SmartInteg}, is reported. In this case, an integration window corresponding to the radial coordinates where the beams exhibit a reduction of $10\,$dB from their corresponding peak was considered. The results demonstrate that the smart integration technique enhances the performance for all OAM modes, albeit to varying degrees. The amelioration is small for modes that already exhibit minimal loss compared to the \ac{MF} (as observed in Fig.~\ref{fig:BER_MF}). However, a substantial improvement is evident for the fundamental mode, which experiences significant performance degradation when employing the standard ID strategy. 
The smart integration approach facilitates the enhancement of the \ac{BER} by balancing the useful and noise energies that are accumulated. Specifically, employing large integration windows results in an increased useful energy level but also a higher noise energy level. On the contrary, small integration windows limit the amount of useful signal energy but diminish the accumulated noise. Using smart integration, the performance degradation of ID with respect to the MF is highly reduced, making such a strategy a viable solution for OAM detection with significantly lower complexity. The primary limitation of the ID method is the requirement for channel estimation to compensate for the phase terms. This drawback can be circumvented by utilizing the ED strategy, whose performance assessment is reported in Fig.~\ref{fig:BER_Comparison}.
In this case, the smart integration technique is applied, as for the ID approach, and the optimum detection threshold for \ac{OOK} demodulation, i.e., the threshold $\zeta_n$ that ensures the minimum \ac{BER} for each \ac{OAM} mode and \ac{SNR} value, is utilized. The figure displays the results for three \ac{OAM} modes only to improve readability. It is evident that the performance of the \ac{ED} method is inferior to that of both the MF and the ID approach due to the non-linear processing and the use of the OOK modulation. Specifically, a degradation of approximately $5\,$dB is observed. However, this approach eliminates the need for amplitude correlation or channel estimation at the receiving LIS. 
In Fig.~\ref{fig:BERvsTNR}, the impact of the detection threshold to discriminate between the hypotheses of \textit{noise only} (i.e., $x_n =0$) and that of \textit{signal plus noise} (i.e., $x_n =1$) in the OOK demodulation is reported. In the figure, the \ac{BER} is reported as a function of the \ac{TNR} for different OAM mode orders $\ell_n$, considering a \ac{SNR} of $19\,$dB, where we defined $\mathsf{TNR}=\zeta_n/N_0$, $n = 1,2,\ldots,N$. It is possible to note that an optimum threshold value exists for each OAM mode. The fundamental mode (i.e., $\ell_n=0$), which has a higher coupling value, demonstrates the best BER. Conversely, higher-order modes exhibit degraded performance. In particular, the minimum values in Fig.~\ref{fig:BERvsTNR}, when considering the smart integration (dashed lines), corresponding to the BER values for an SNR of $19\,$dB in Fig.~\ref{fig:BER_Comparison}, when accounting for the ED curves. In the same figure, the ED performance as a function of the TNR, but without smart integration, is reported. As for the ID approach, the smart integration allows for improving the performance, by balancing the amount of energy and noise accumulated at the receiving LIS. Interestingly, when smart integration is not considered, the optimum threshold decreases monotonically when increasing the mode order $\elln$; in fact, the useful energy decreases accordingly due to the lower coupling, thus it is convenient to decrease the decision threshold in \eqref{eq:DecisionCritED}. Differently, when the smart integration approach is considered, the optimum threshold is still mode-dependent, although not exhibiting a monotonic behavior with respect to the mode order. This is reasonable since the corresponding beams are placed at different points along the radial coordinate, and the width of the integration window changes for each mode (also with a noise power $n_n(\rhor)$ which is not flat after de-multiplexing); thus, the optimum threshold must account for either the useful fraction of energy or the amount of noise energy in the specific integration window selected.

It is worth noticing that these curves assume that the same power is associated with the different OAM modes (i.e., equal power allocation at the transmitting LIS). In order to maximize the system capacity, water-filling power distribution at the transmitter side, accounting for the different coupling intensities of the OAM modes, is required.

\subsection{Discussion and Implementation Complexity Analysis}\label{sec:Discussion}

The exploitation of the \ac{OAM} property has both practical advantages and ineluctable disadvantages. Concerning the benefits of the OAM characteristics, it has been shown that this transmission technique allows multi-modal communications, hence enabling spatial multiplexing in LOS \ac{MIMO} communications. Specifically, the basis functions proposed in~\eqref{basis_OAM1}-\eqref{basis_OAM2} introduce the possibility to perform spatial multiplexing with a considerably lower complexity at the transmitting LIS side with respect to the optimum \ac{SVD} case.
In fact, with the \ac{SVD}/CPWFs approach, the basis functions depend on the overall system geometry (i.e., sizes of both circular LISs and relative distance), which needs to be known at both sides and requires a certain degree of hardware flexibility. With \ac{OAM}, instead,  such knowledge is not required at the transmitting LIS side (or distance only in the case of focused OAM) when operating in a paraxial regime. 

On the other side, the detection of OAM-multiplexed EM fields with a receiving LIS results quite demanding in terms of complexity and knowledge of the system geometry if optimal MF processing is adopted. Differently, the complexity can be reduced, and system geometry knowledge can be relaxed if \ac{ID} and \ac{ED} strategies are adopted.
A summary of the flexibility requirements at the transmitting (TX) and receiving (RX) LIS as well as of the required degree of knowledge at both sides is reported in Table~\ref{tab:Complexity}.

\renewcommand{\arraystretch}{1.3}
\begin{table*}[t!]
\centering
        \caption{Implementation requirements for the different detection strategies.}
	\label{tab:Complexity}
        \begin{tabular}{|c|c|c|c|c|c|}
        \hline
        \ &\cellcolor{yellow!50} TX LIS flexibility &\cellcolor{yellow!50} RX LIS flexibility & \cellcolor{yellow!50} Knowledge at TX & \cellcolor{yellow!50} Knowledge at RX \\ \hline
        
        \cellcolor{CornflowerBlue!50} Optimal &  Phase tapering,  & Phase tapering, & RX LIS size, & TX LIS size, \\ 
        \cellcolor{CornflowerBlue!50} SVD/CPWFs &  amplitude tapering & amplitude tapering & distance & distance \\ \hline

        \cellcolor{Red!30}TX: OAM & Phase tapering & Phase tapering, & N/A & TX LIS size, \\ 
        \cellcolor{Red!30}RX: MF & & amplitude tapering  &   & distance\\ \hline

        \cellcolor{Red!30}TX: Focused OAM & Phase tapering & Phase tapering,  & Distance  & TX LIS size, \\ 
        \cellcolor{Red!30}RX: MF & & amplitude tapering  &   & distance \\ \hline

        \cellcolor{GreenYellow!50}TX: OAM & Phase tapering & Phase tapering,  & Distance & N/A \\ 
        \cellcolor{GreenYellow!50}RX: ID & & channel equalization  &   &  \\ \hline
        
        \cellcolor{GreenYellow!50}TX: Focused OAM & Phase tapering & Phase tapering,  & Distance & N/A \\ 
        \cellcolor{GreenYellow!50}RX: ID & & channel equalization  &   & \\ \hline
        
        \cellcolor{violet!30}TX: OAM & Phase tapering & Phase tapering,  & N/A &  N/A \\ 
        \cellcolor{violet!30}RX: ED & & non-linear processing &   & \\ \hline

     \cellcolor{violet!30}TX: Focused OAM & Phase tapering & Phase tapering,  & Distance & N/A \\ 
        \cellcolor{violet!30}RX: ED & & non-linear processing  &   & \\ \hline
        \end{tabular}
\end{table*}

Apart from these considerations, an element of fundamental importance that is worth to be mentioned deals with the employment of large \ac{LIS} antennas (i.e., holographic MIMO configuration) rather than \acp{UCA} \cite{ChenWen:J20}. Suppose, in fact, to exploit a large \ac{UCA} to receive the \ac{OAM}-multiplexed \ac{EM} field, whose radius is even collimated to the radial coordinate $\rhor$ corresponding to the peak of a specific mode (for example, $\ell_n=3$ in Fig.~\ref{fig:OAMRXfoc}). In such a case, the energy that can be collected from the other modes (e.g., that for $\ell_n=1$ or $\ell_n=6$ in Fig.~\ref{fig:OAMRXfoc}) will be inevitably low, due to the different divergence behavior spreading the beams at different radii. In this sense, the use of a large receiving LIS enables much more flexibility and larger coupling intensities for an increased number of OAM modes.

Be that as it may, OAM communications also display several, not negligible negative aspects that necessarily need to be considered. Fig.~\ref{fig:SVD1} highlighted that the number of the effectively well-coupled OAM modes is much smaller than that of the optimum communication modes, both in the presence and absence of focusing. Focusing has been shown to partially compensate for this OAM deficiency and is particularly important when large LISs are employed, especially at the transmitting side. 
Behind all that, it is essential to note that the system under consideration assumes a paraxial \ac{LOS} scenario. Any misalignment between the transmitter and the receiver, such as lateral displacements or receiver angular errors, or the presence of multipath propagation, will result in power loss and potential crosstalk at the receiver side. Therefore, suitable strategies must be implemented to address these effects in more general scenarios characterized by mobility and multipath.

\section{Conclusion}\label{Sec:Conclusion}
The paper discussed holographic \ac{MIMO} communications that exploit the \ac{OAM} property of EM waves, thus investigating the relationship between optimum communication modes and \ac{OAM} modes. It proposed \ac{OAM}-related basis functions that offer a simpler implementation than the optimum solution based on eigenfunction decomposition, although reducing the channel \ac{DoF}. Various detection strategies at the receiver have been proposed and analyzed to investigate performance and complexity trade-offs. The paper also introduced focusing to improve the performance in terms of \ac{DoF} and coupling strength, showing its effectiveness in different configurations. The numerical analysis demonstrated the feasibility of a significant number of orthogonal \ac{OAM} channels, particularly when focusing is used to mitigate the beam divergence characteristic of \ac{OAM} modes in the near field, by characterizing the performance of the proposed OAM demultiplexing and detection schemes.


\appendices
\section{Example of OAM Amplitude and Phase Profiles}\label{app:OAMprofiles}
Table~\ref{tab:OAMprofiles} displays the transmitting phase profiles and the corresponding OAM modes (both amplitude and phase patterns) obtained at the reciving LIS for different topological charges, both with and without employing focusing at the transmitting LIS when $T=R=5$ and $D=20$.
In particular, results for $\ell_n=0$ are the beams obtained with unfocused/focused constant phase profiles along $\phit$, thus corresponding to classical diffraction patterns of circular apertures. The increasing beam divergence, partially compensated by focusing, can be seen in the subsequent rows for increasing values of $\ell_n$, as well as the corresponding helical-shaped phase profiles at the receiver side.

\begin{table*}[t!]
\raggedright
        \caption{Example of transmitting and receiving \ac{OAM} amplitude/phase profiles, with and without focusing, for $\ell_n=0, 1,3$ ($T=R=5$, $D=20$).}
	\label{tab:OAMprofiles}
\begin{tabular}{|>{\small}C{1.1cm}|C{2.2cm}|C{2.2cm}|C{2.2cm}|C{2.2cm}|C{2.2cm}|C{2.2cm}|}

\hline
\rowcolor{green!30}
\cellcolor{white} & \multicolumn{2}{|c|}{\textbf{TX phase}} & \multicolumn{2}{|c|}{\textbf{RX amplitude}} & \multicolumn{2}{|c|}{\textbf{RX phase}}  \\

\cline{2-7}
\rowcolor{yellow!50} 
\cellcolor{white} & \textbf{Unfocused} & \textbf{Focused} & \textbf{Unfocused} & \textbf{Focused} & \textbf{Unfocused} & \textbf{Focused} \\

\cline{1-7}
\cellcolor{CornflowerBlue!50} $\ell_n = 0$ & \includegraphics[width=2.8cm, trim={30mm 5mm 1mm 0mm}, clip, keepaspectratio]{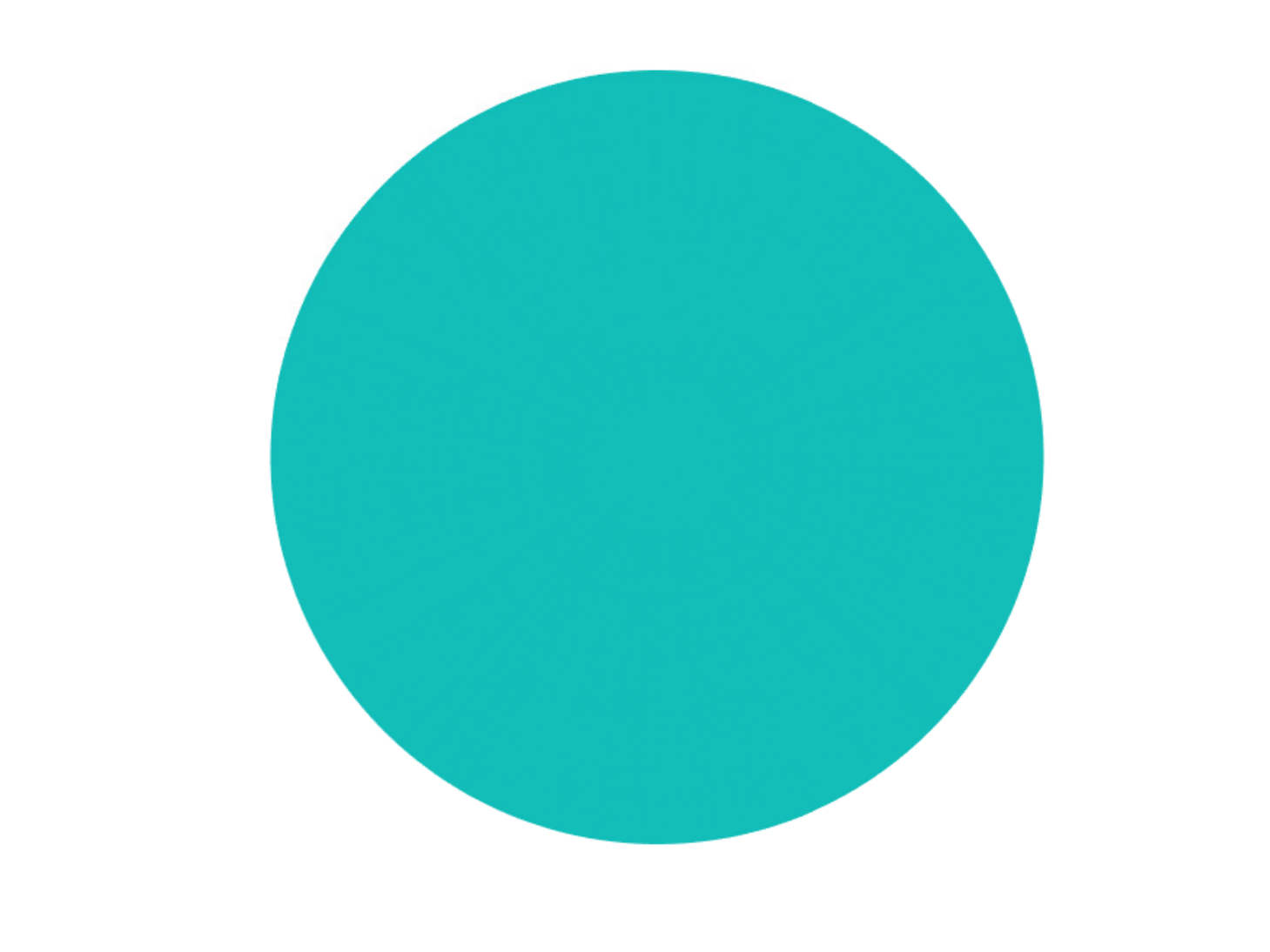} & \includegraphics[width=2.8cm, trim={30mm 5mm 1mm 0mm}, clip, keepaspectratio]{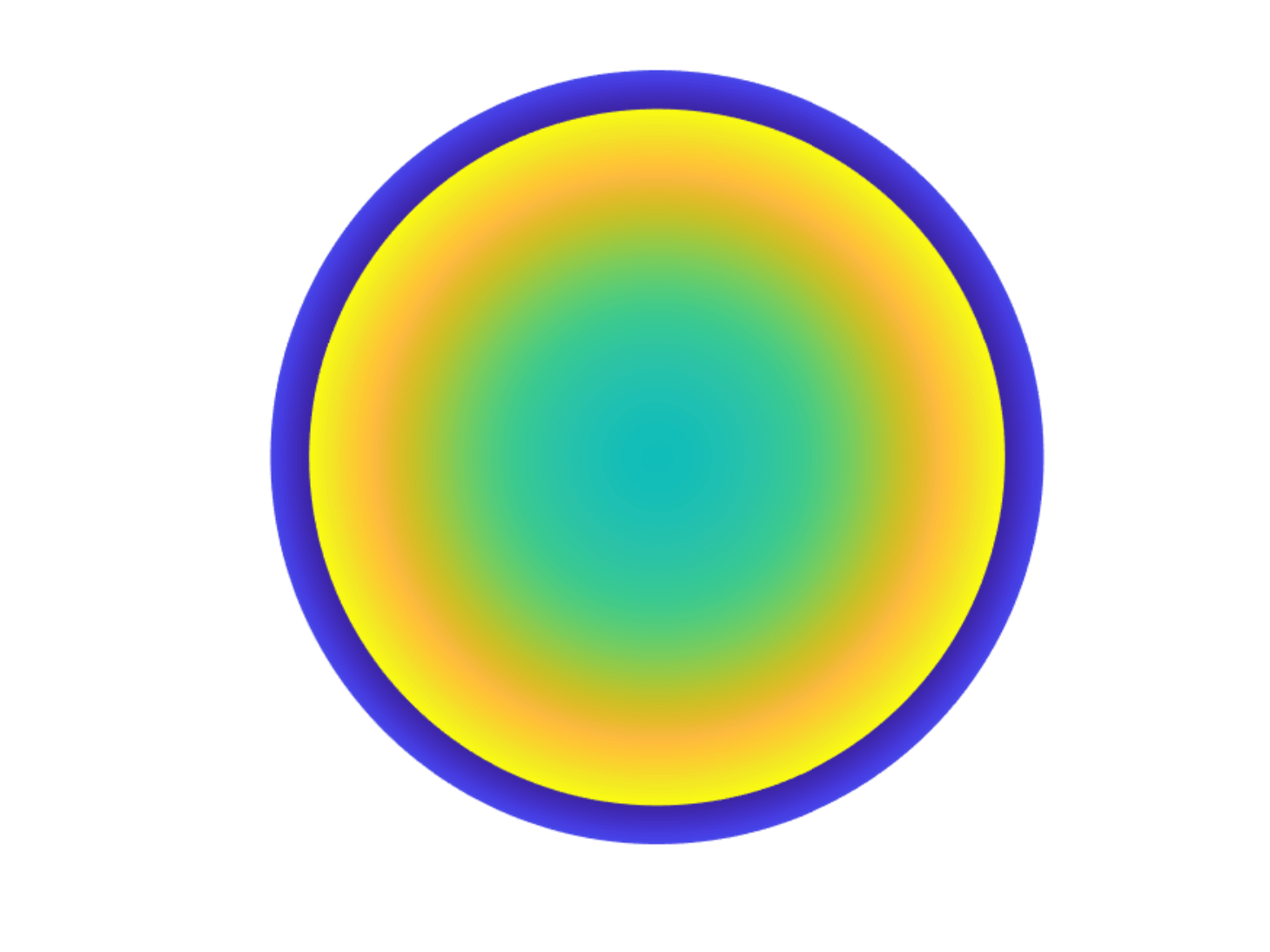} & \includegraphics[width=2.8cm, trim={30mm 5mm 1mm 0mm}, clip, keepaspectratio]{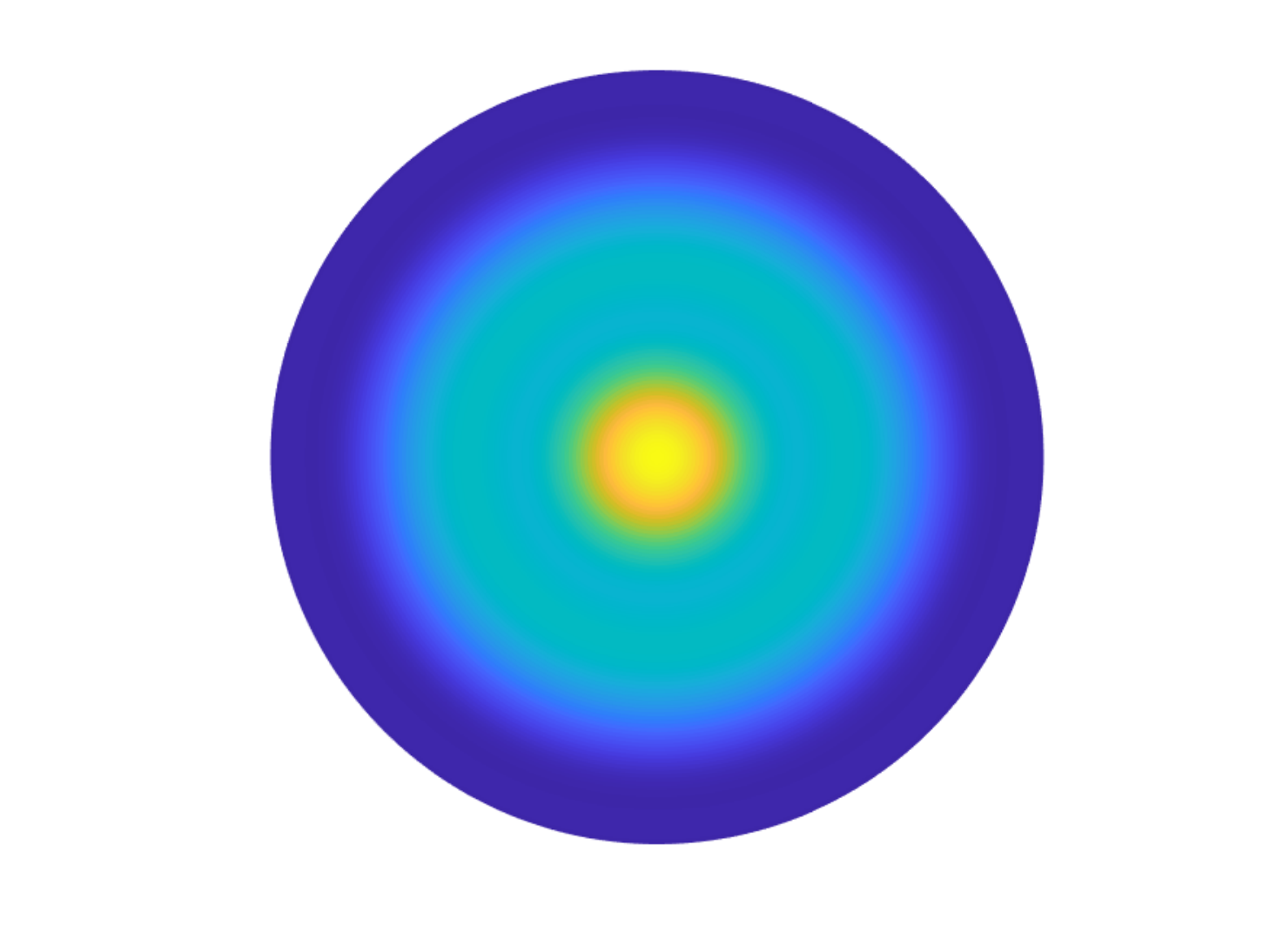} & \includegraphics[width=2.8cm, trim={30mm 5mm 1mm 0mm}, clip, keepaspectratio]{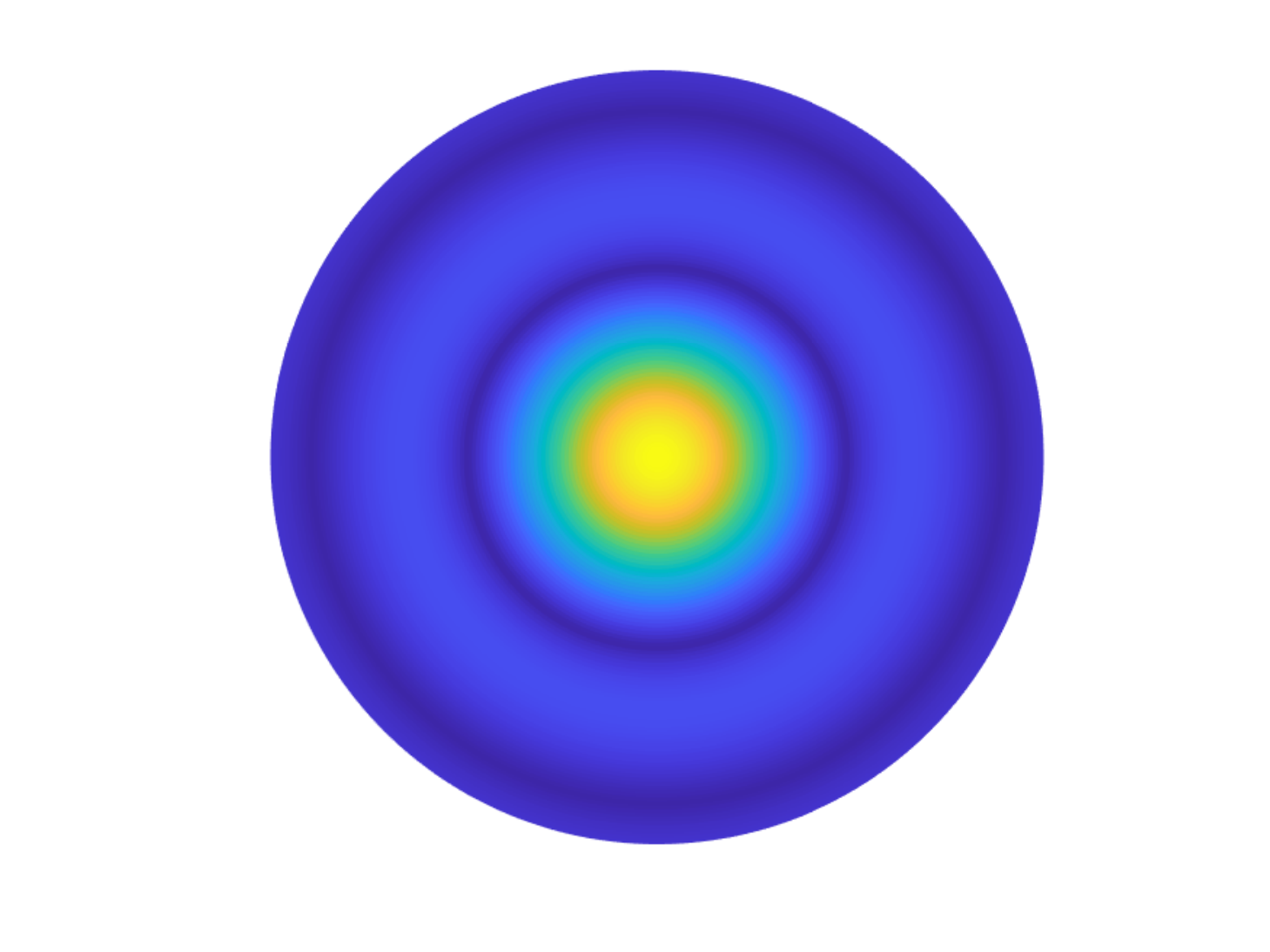} & \includegraphics[width=2.8cm, trim={30mm 5mm 1mm 0mm}, clip, keepaspectratio]{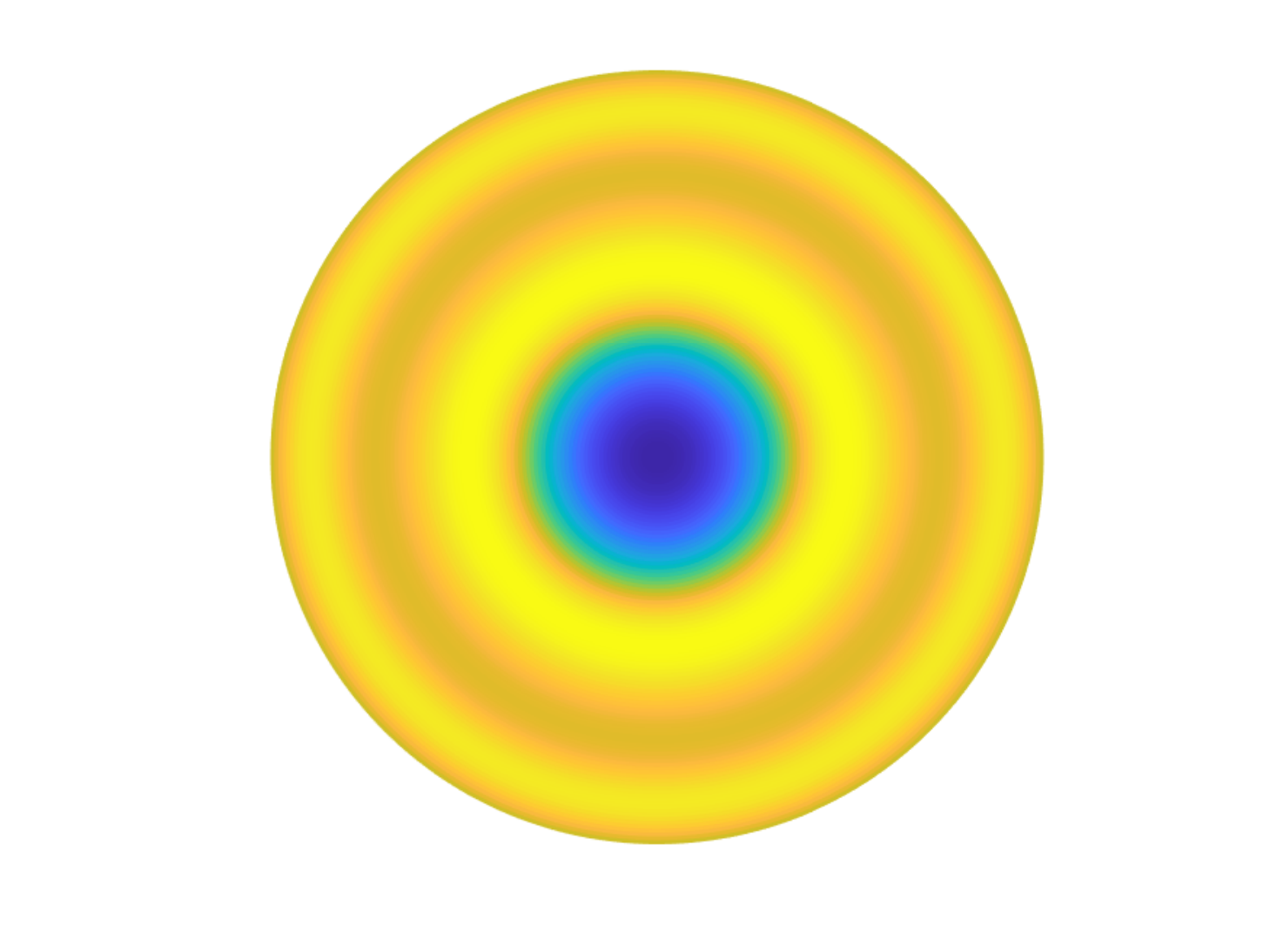} & \includegraphics[width=2.8cm, trim={30mm 5mm 1mm 0mm}, clip, keepaspectratio]{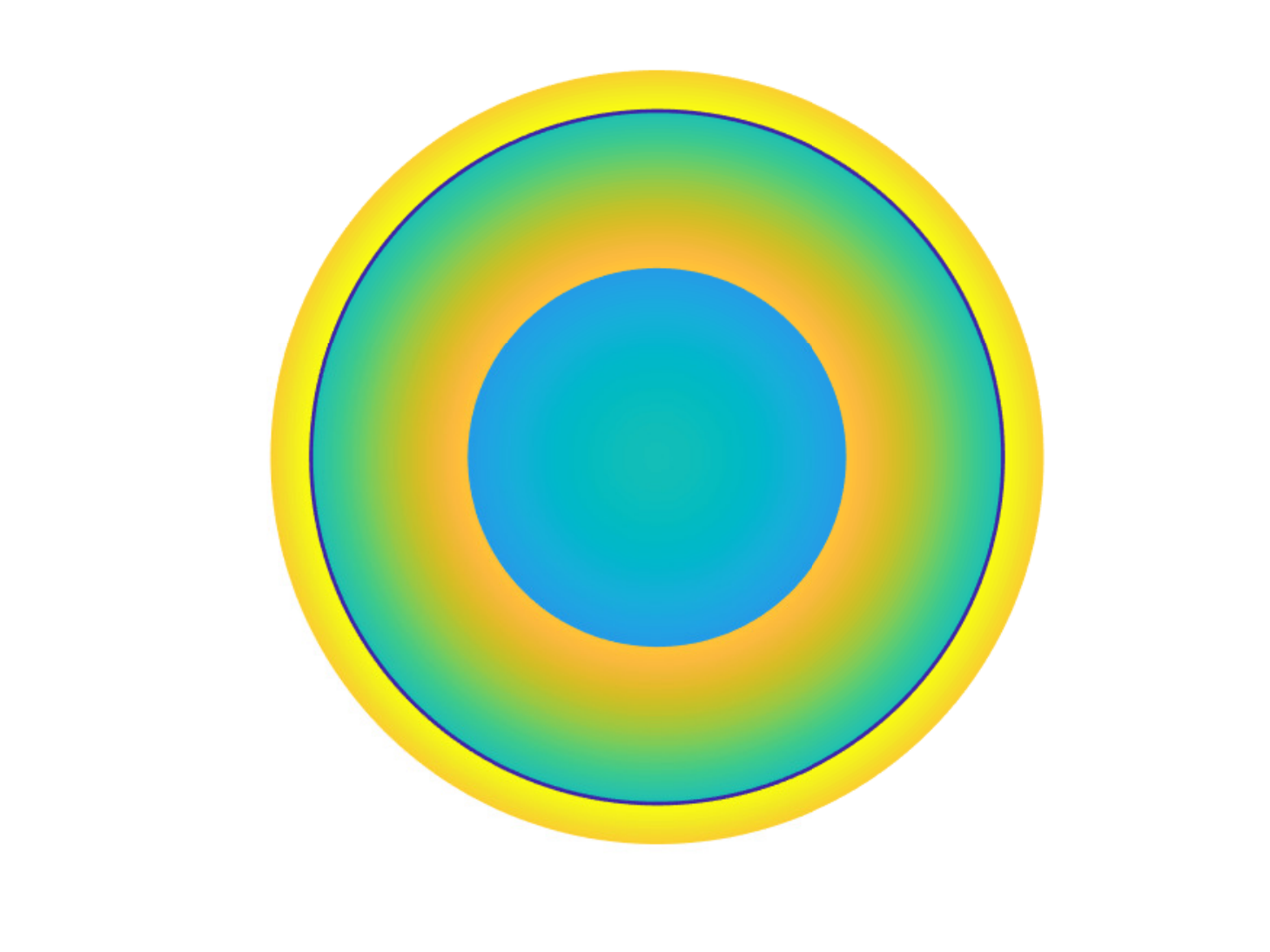} \\
\hline

\cellcolor{CornflowerBlue!50}$\ell_n = 1$ & \includegraphics[width=2.8cm, trim={30mm 5mm 1mm 0mm}, clip, keepaspectratio]{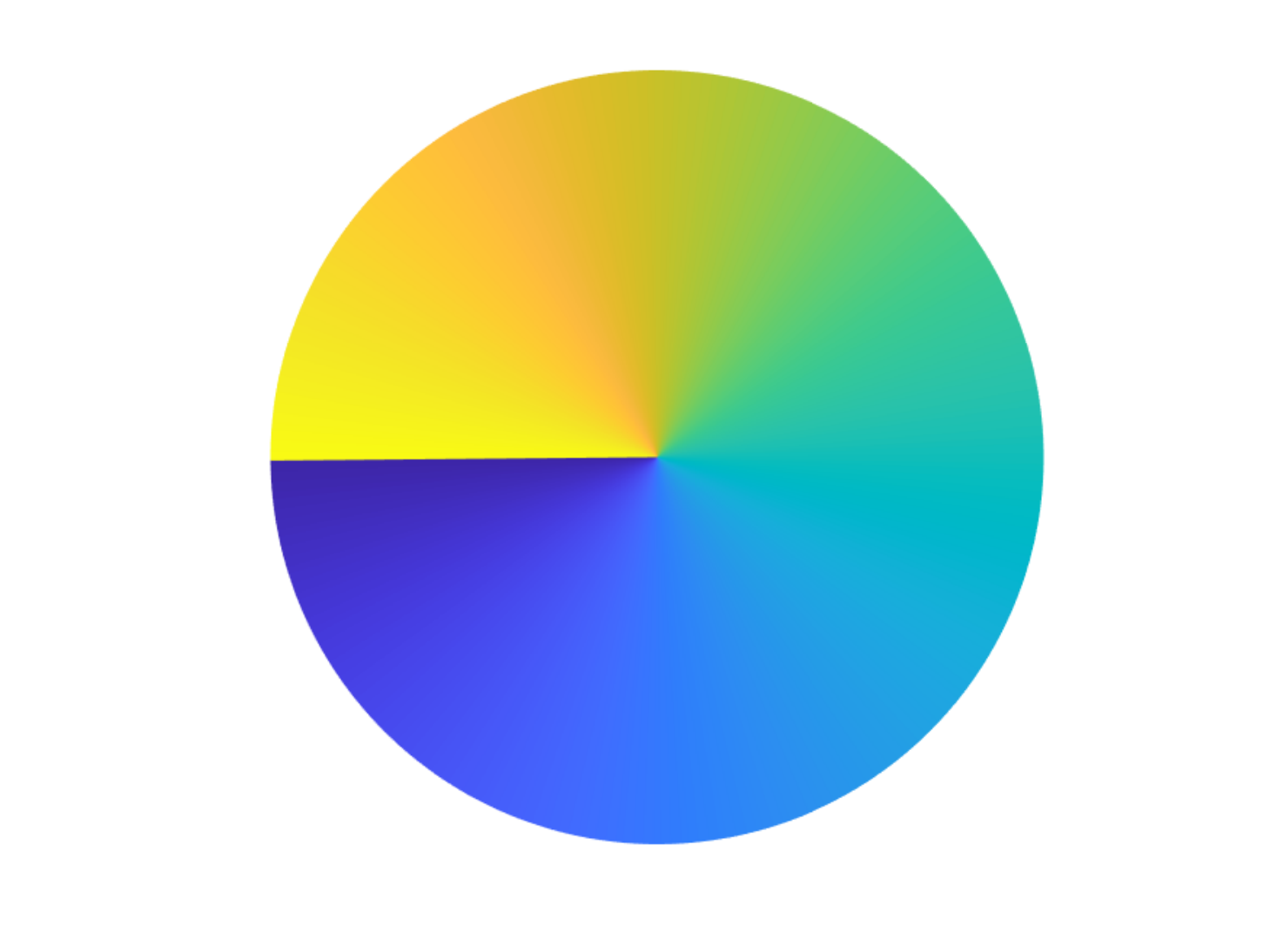} & \includegraphics[width=2.8cm, trim={30mm 5mm 1mm 0mm}, clip, keepaspectratio]{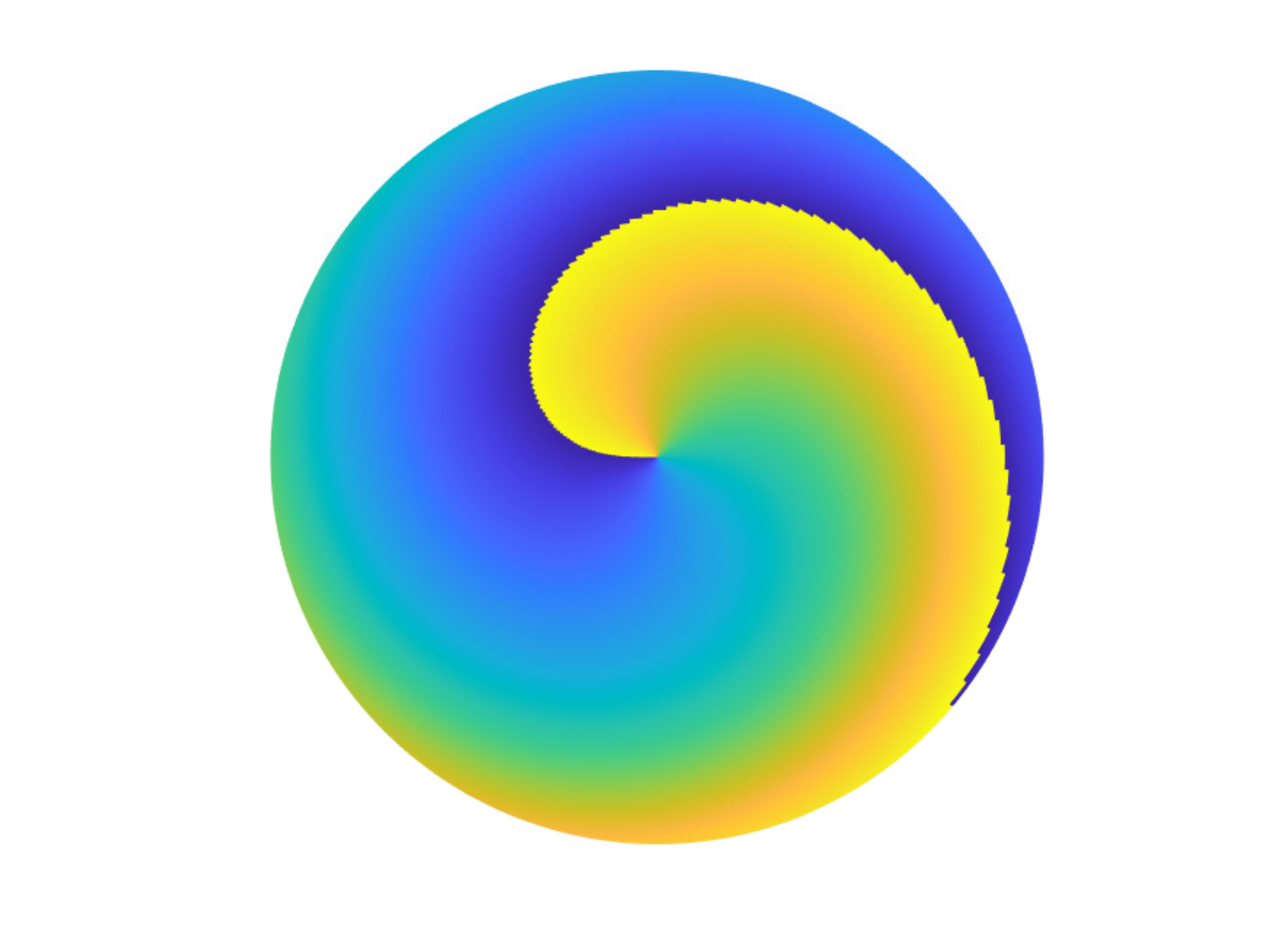} & \includegraphics[width=2.8cm, trim={30mm 5mm 1mm 0mm}, clip, keepaspectratio]{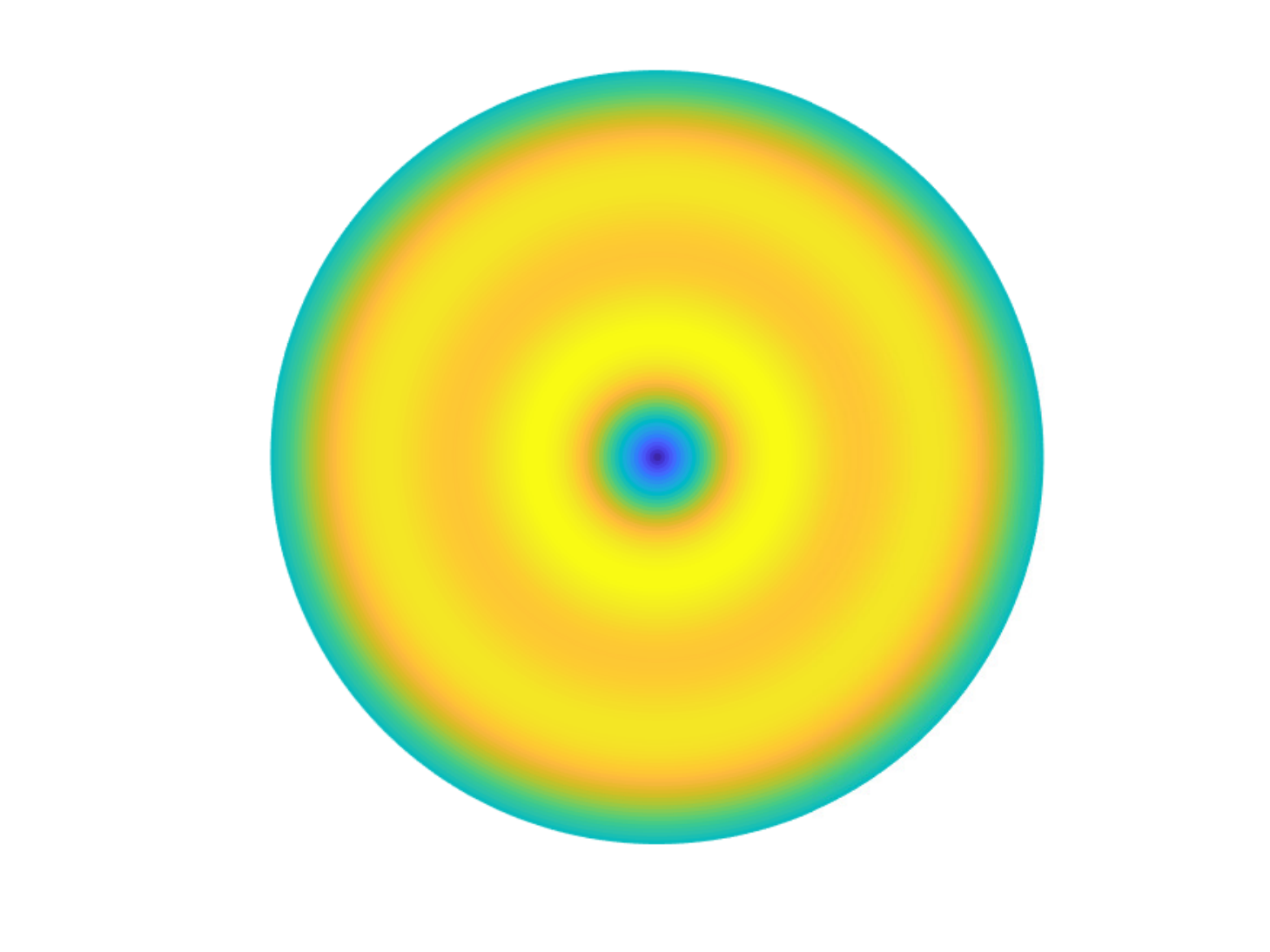} & \includegraphics[width=2.8cm, trim={30mm 5mm 1mm 0mm}, clip, keepaspectratio]{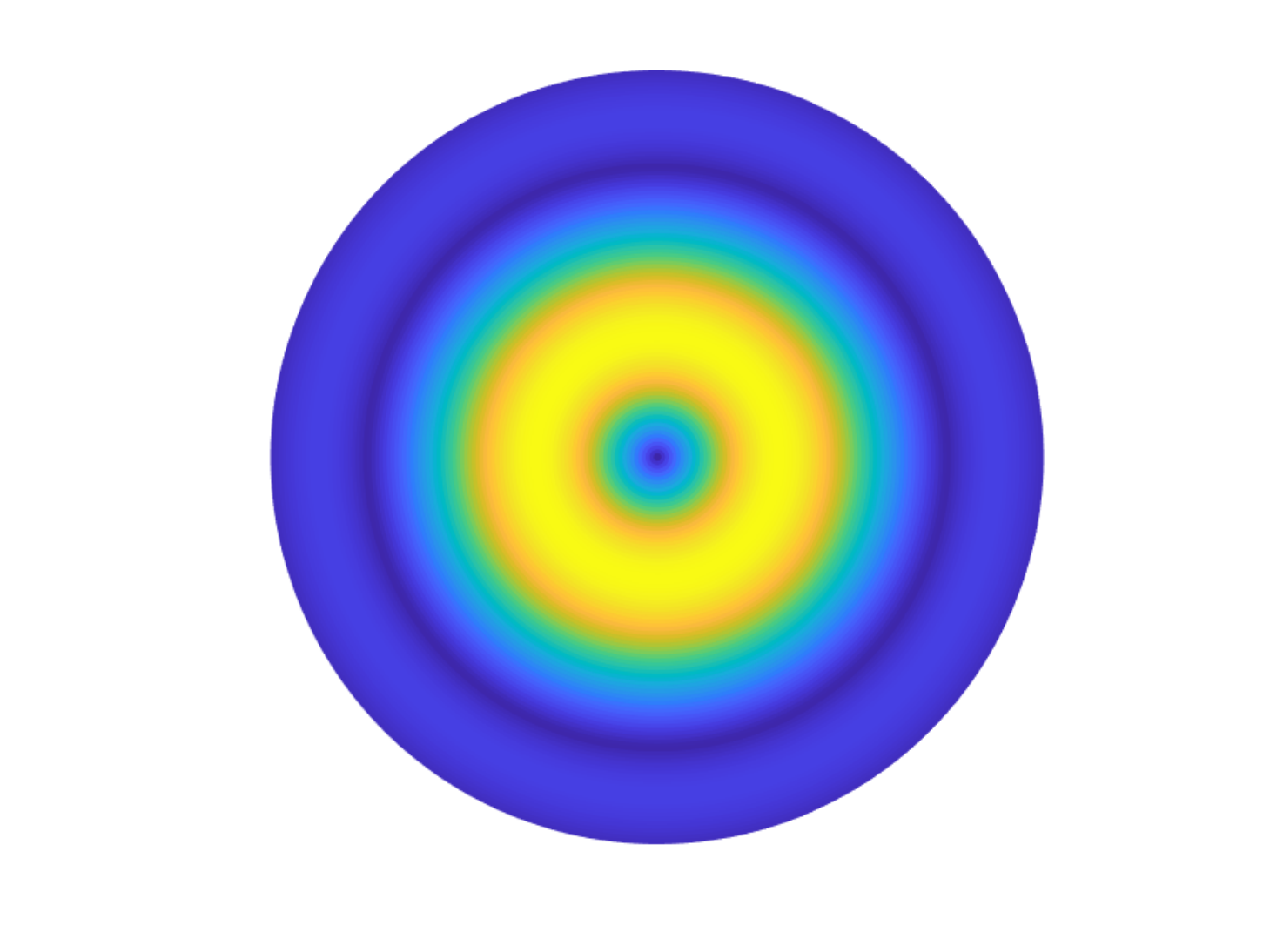} & \includegraphics[width=2.8cm, trim={30mm 5mm 1mm 0mm}, clip, keepaspectratio]{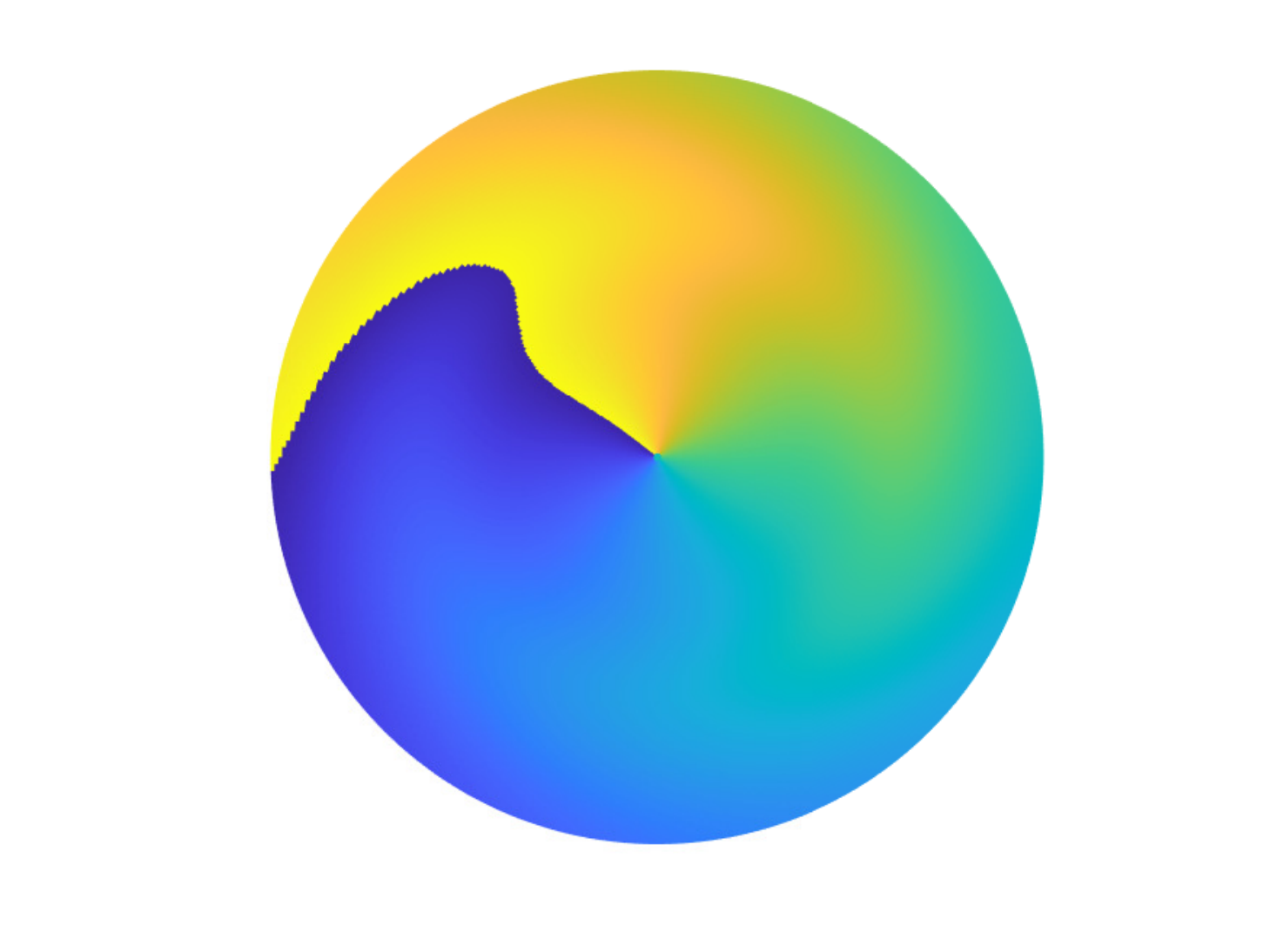} & \includegraphics[width=2.8cm, trim={30mm 5mm 1mm 0mm}, clip, keepaspectratio]{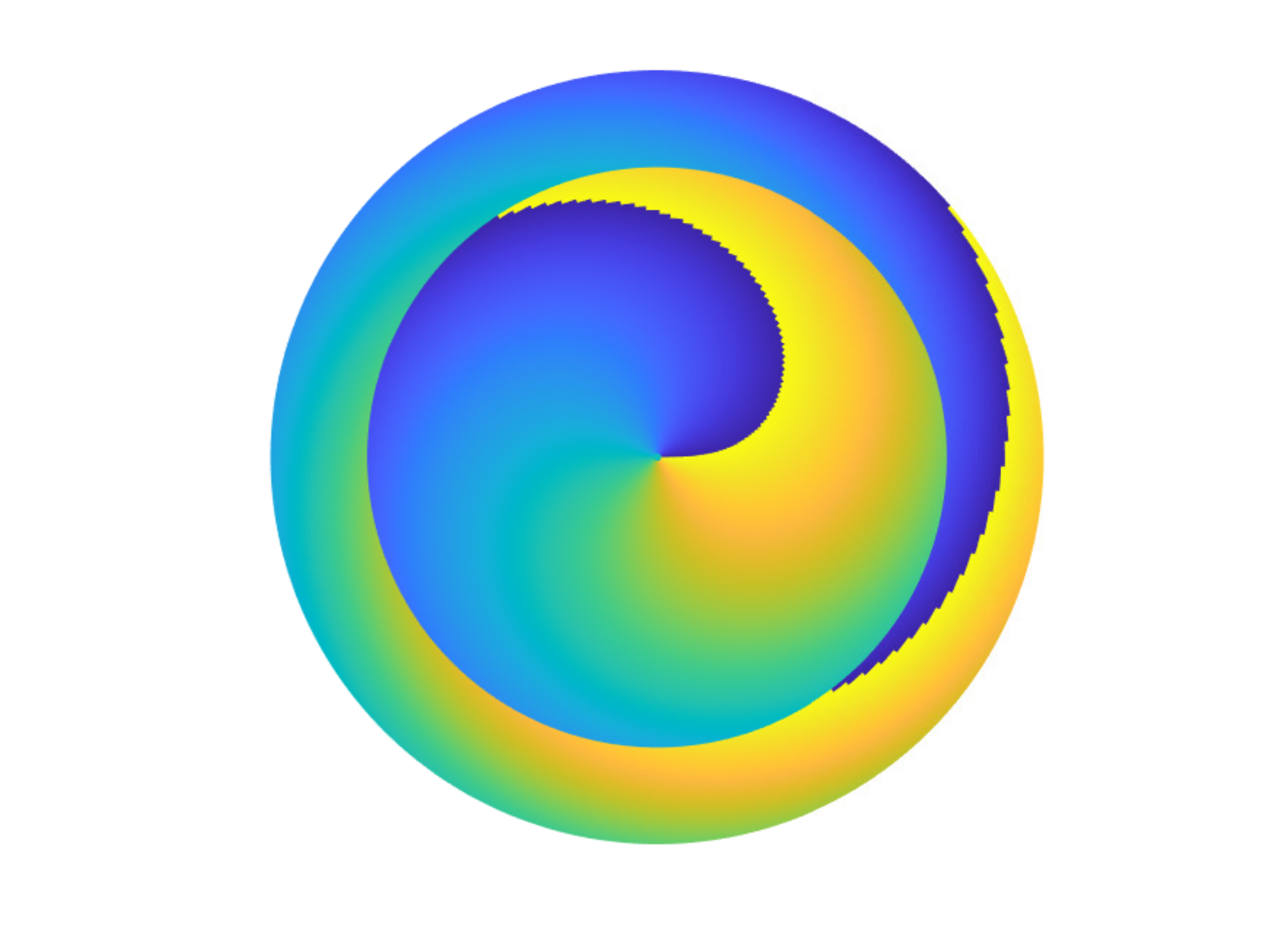} \\
\hline

\cellcolor{CornflowerBlue!50}$\ell_n = 3$ & \includegraphics[width=2.8cm, trim={30mm 5mm 1mm 0mm}, clip, keepaspectratio]{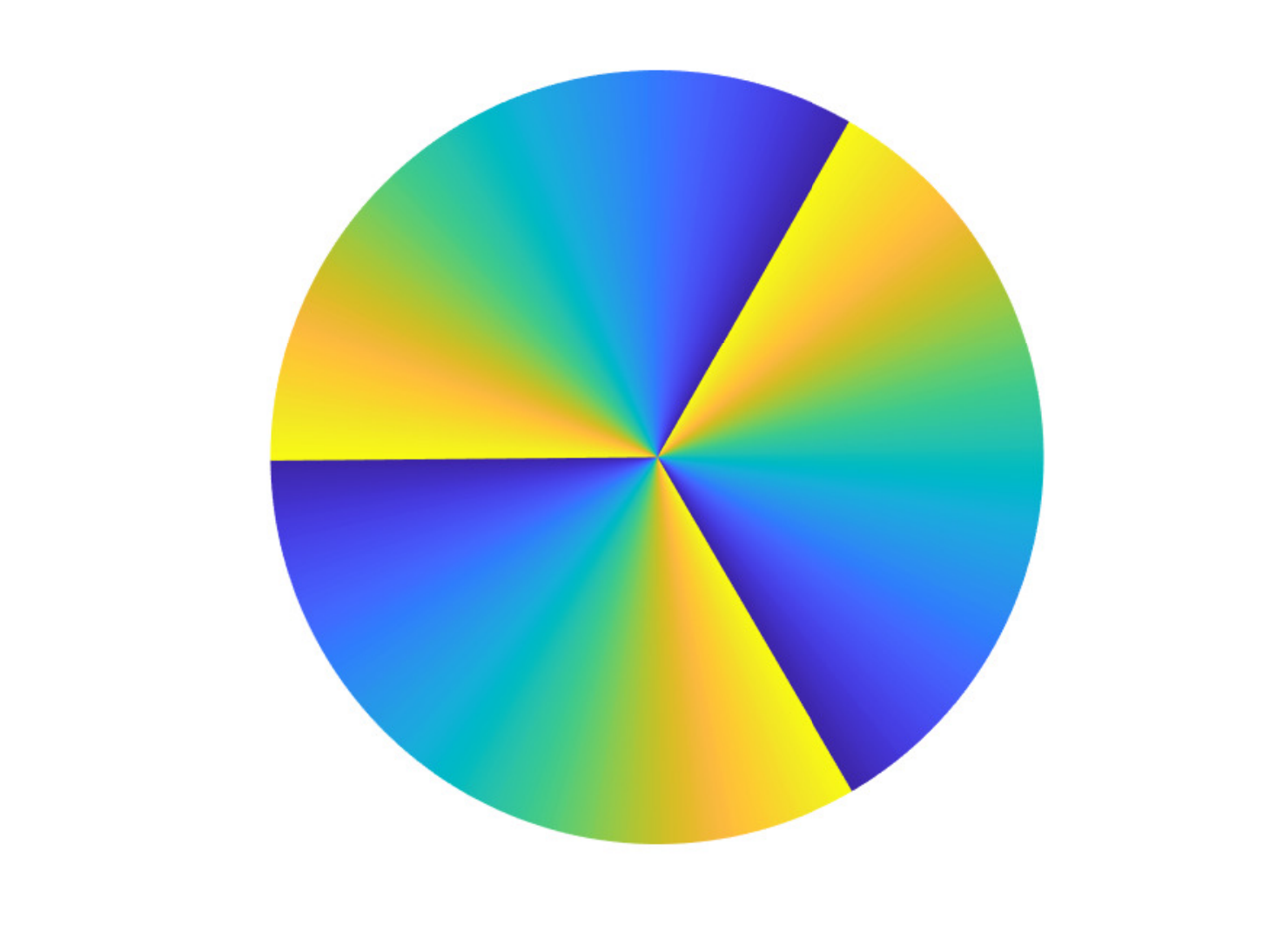} & \includegraphics[width=2.8cm, trim={30mm 5mm 1mm 0mm}, clip, keepaspectratio]{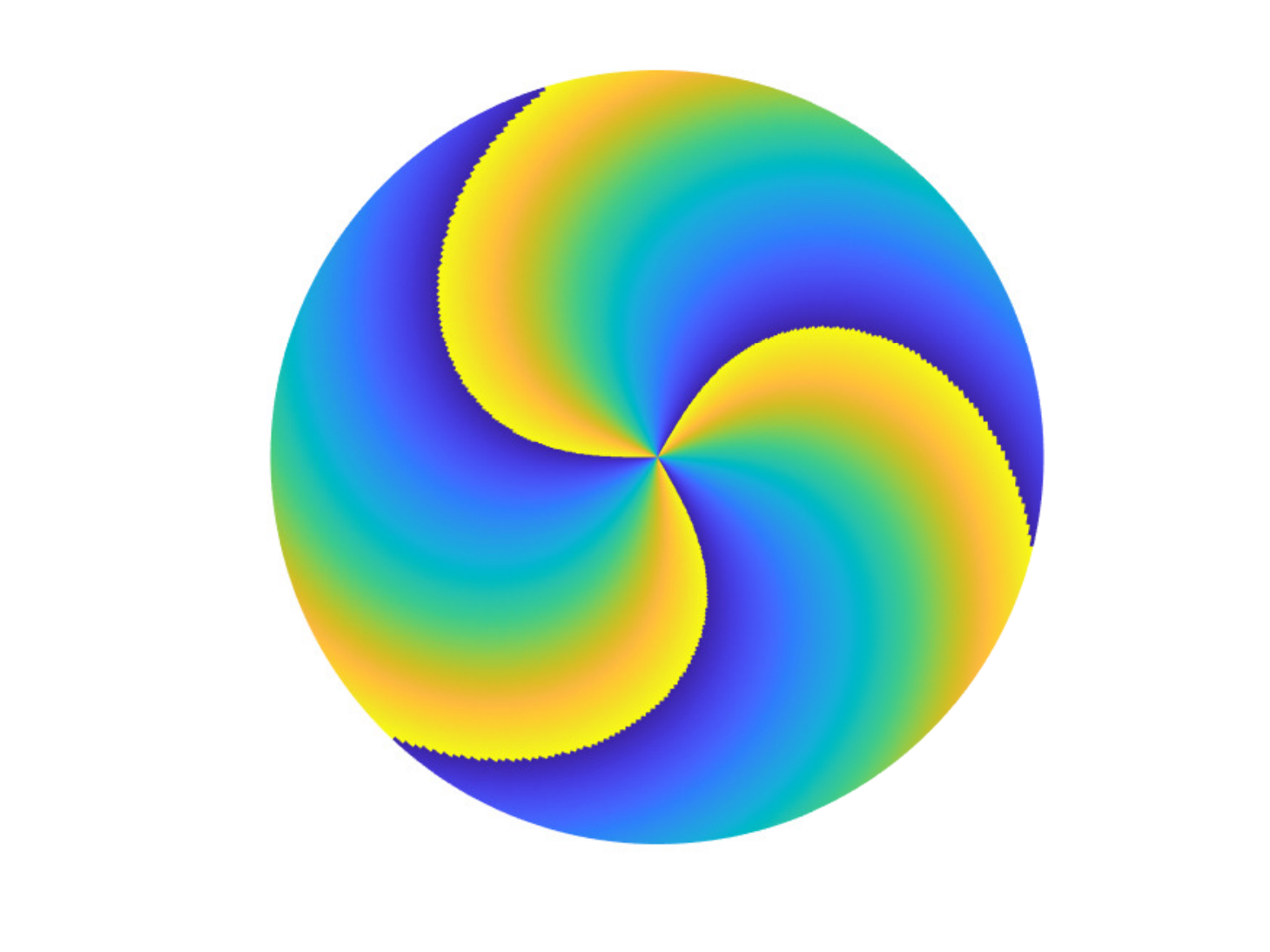} & \includegraphics[width=2.8cm, trim={30mm 5mm 1mm 0mm}, clip, keepaspectratio]{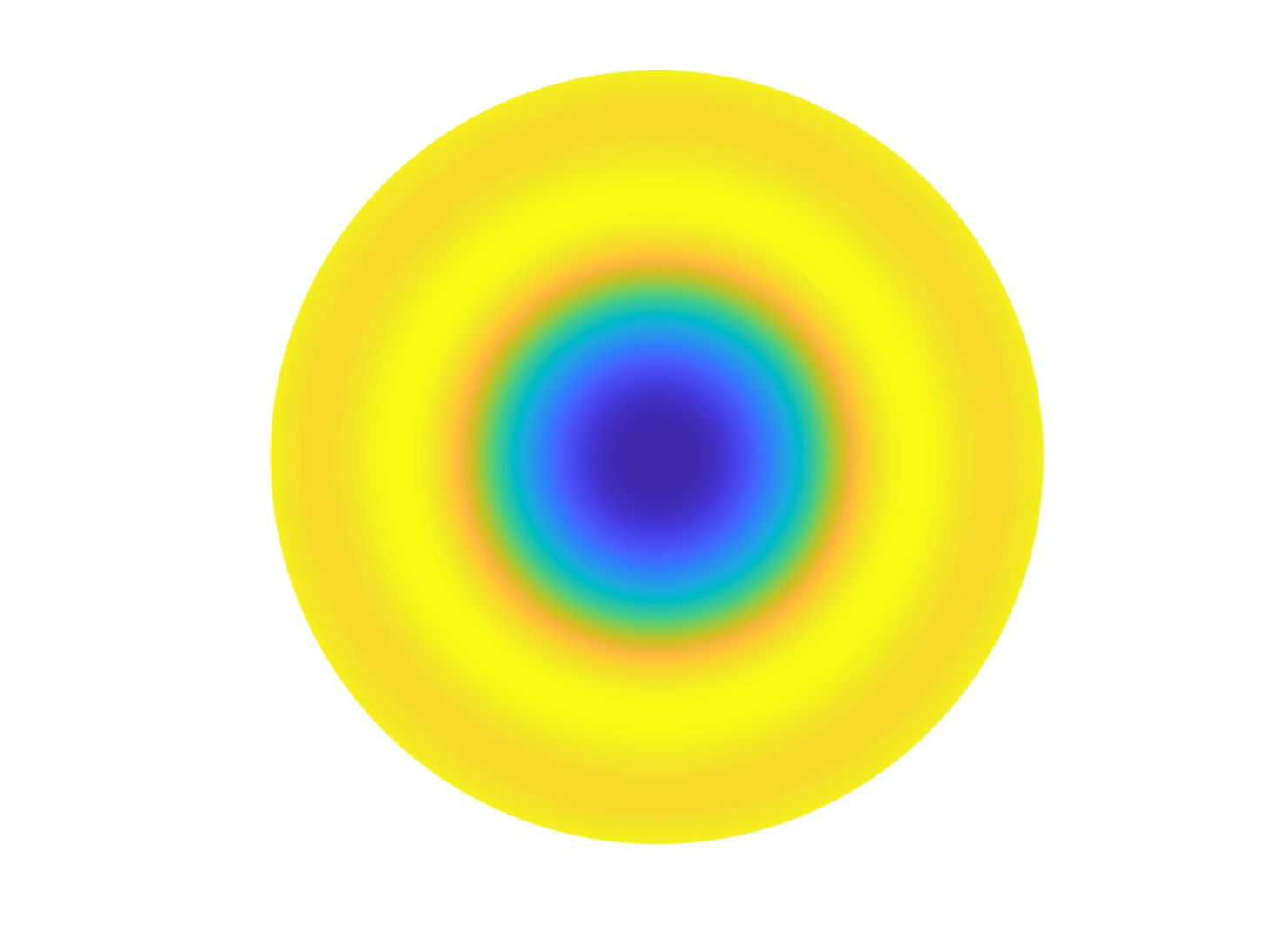} & \includegraphics[width=2.8cm, trim={30mm 5mm 1mm 0mm}, clip, keepaspectratio]{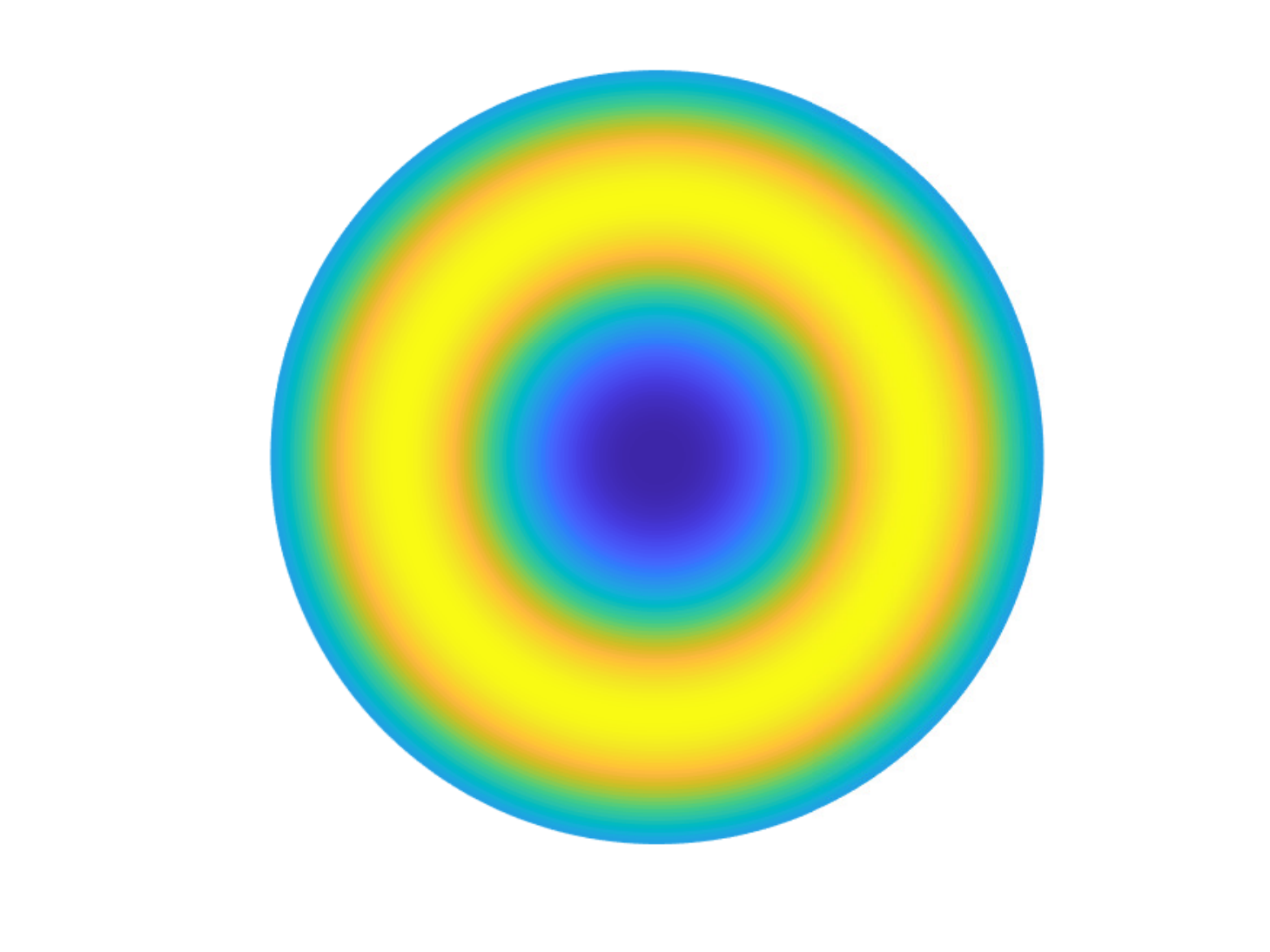} & \includegraphics[width=2.8cm, trim={30mm 5mm 1mm 0mm}, clip, keepaspectratio]{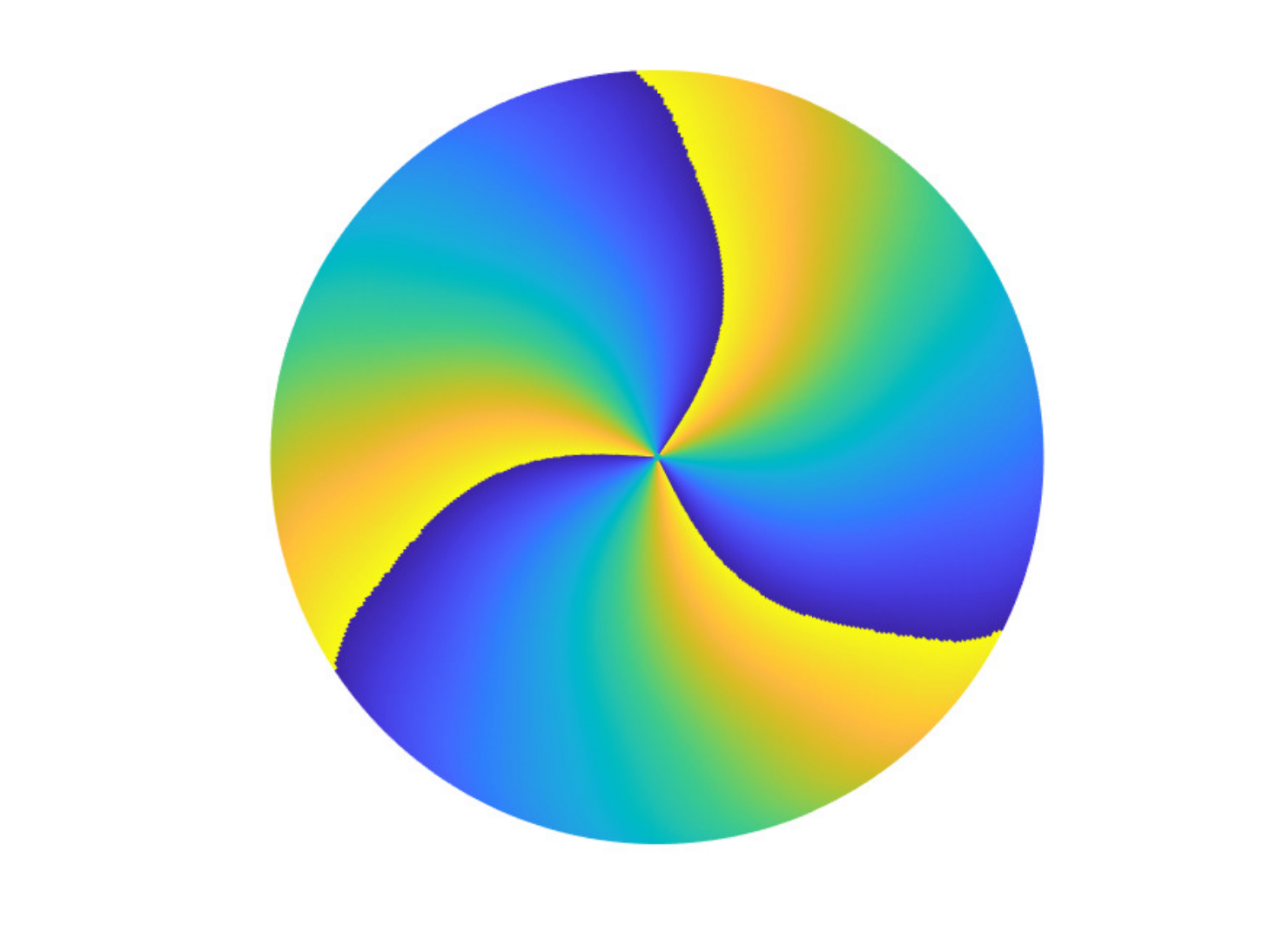} & \includegraphics[width=2.8cm, trim={30mm 5mm 1mm 0mm}, clip, keepaspectratio]{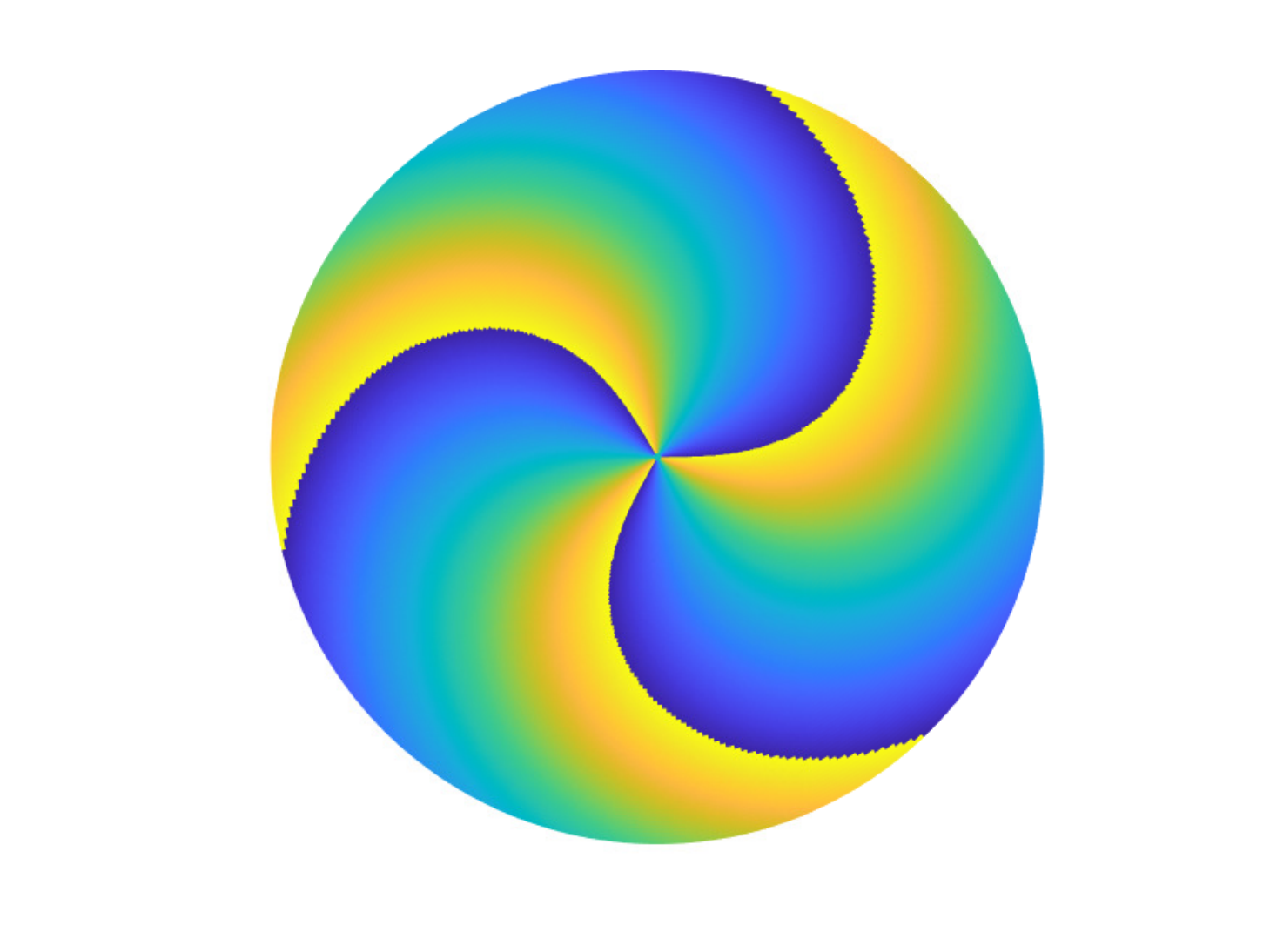} \\

\hline
\end{tabular}
\end{table*}

\section{Noise Characterization}\label{app:AWGNNoise}
Let us contemplate the noise statistical characterization by expressing the noise field $n(\rhor, \phir)$ in Cartesian coordinates, i.e., $\tilde{n}(x,y)$. Specifically, it is
\begin{align}
    w_{n} &=  \int_{\srx}   \tilde{n}(x,y)\tilde{F}_n^{*}(x,y) \, \text{d}x\text{d}y
\end{align}
where $\tilde{n}(x,y)$ is the complex \ac{AWGN} with one-sided power spectral density $N_0$ and $\tilde{F}_n(x,y)$ identifies the generic $n$-th receiving template function adopted at the receiving \ac{LIS} in Cartesian coordinates. We assume that this template function is in the same form of~\eqref{eq:SeparableBasis} when expressed in polar coordinates, i.e., $\tilde{F}_n(x,y) = F_n(\rhor,\phir) = f_n (\rhor) e^{\jmath \elln \phir}$, meaning that it is separable in the radial and angular coordinates and comprises the OAM exponential term. The radial component $f_n (\rhor)$ instead is an arbitrary function that depends on the selected detection strategy.
Due to the AWGN assumption in the spatial domain, the noise variance can be computed as
\begin{align}\label{eq:noisevariance}
\sigma_{w}^2 &= \mathbb{E} \left[\left|w_n\right|^2\right] = \int_{\srx}\int_{\srx} \tilde{F}_n^*(x,y)\tilde{F}_n(x',y') \mathbb{E} [\tilde{n}(x,y) \tilde{n}^*(x', y')]\, \text{d}x\text{d}y \, \text{d}x'\text{d}y' \nonumber \\
&= N_0 \int_{\srx} \left| \tilde{F}_n(x,y) \right|^2 \, \text{d}x\text{d}y = N_0 \int_0^{2 \pi} \int_0^{\rr} \left|f_n(\rhor)e^{\jmath \elln \phir} \right|^2 \rhor \, \text{d}\rhor \text{d}\phir \nonumber \\
&= 2 \pi N_0 \int_0^{\rr} \left| f_n(\rhor) \right|^2 \rhor\,  \text{d}\rhor
\end{align}
where $\rhor$ is the Jacobian determinant of the transformation from Cartesian to polar coordinates. Notably, the radial function $f_n(\rhor)$ adopted at the receiving LIS plays a fundamental role in determining the amount of noise power collected by the receiving \ac{LIS}. 
In particular, for the \ac{MF} case it holds $f_n (\rhor) =  \psi_n^{\rho} (\rhor)$, thus leading to 
\begin{align}\label{eq:sigmaqMF}
\sigma_{w, \text{MF}}^2 &= 2 \pi N_0 \int_0^{\rr} \left| \psi_n^{\rho} (\rhor) \right|^2 \rhor \, \text{d}\rhor = 2 \pi N_0 h_n^{\text{(MF)}} 
\end{align}
where $h_n^{\text{(MF)}}=E_n$ is given in~\eqref{eq:h-coef_MF_nofoc}-\eqref{eq:h-coef_MF_foc}. Similarly, for the \ac{ID} strategy, being $f_n (\rhor) =  \Pi_{\rr}(\rhor)$, it results
\begin{align}\label{eq:sigmaqID}
\sigma_{w, \text{ID}}^2 &= 2 \pi N_0 \int_0^{\rr} \left|  \Pi_{\rr}(\rhor) \right|^2 \rhor \, \text{d}\rhor = \pi N_0  \rr^2 \, .
\end{align}
While the optimal \ac{MF} approach maximizes the \ac{SNR} at the receiver, by perfectly equalizing the propagation effects, the noise variance of the \ac{ID} scheme increases with the observation interval, i.e., the integration domain in the radial direction enlarges. For this reason, proper countermeasures such as the smart integration technique discussed in Sec.~\ref{sec:SmartInteg} should be adopted.

\bibliographystyle{IEEEtran}
\bibliography{IEEEabrv,StringDefinitions,Biblio_LIS,Biblio_OAM}

\vfill

\end{document}